\documentstyle[sprocl,psfig]{article}
\input{epsf}
\def\be{\begin{equation}}
\def\ee{\end{equation}}
\def\bea{\begin{eqnarray}}
\def\eea{\end{eqnarray}}
\def\bq{\begin{quote}}
\def\eq{\end{quote}}
\def\bseq{\begin{subequation}}
\def\eseq{\end{subequation}}
\def\bsea{\begin{subeqnarray}}
\def\esea{\end{subeqnarray}}

\def\c2bb{\cos^2 2 \beta}

\def\s2bb{\sin^2 2 \beta}

\def\simlt{\stackrel{<}{{}_\sim}}
\def\simgt{\stackrel{>}{{}_\sim}}

\def\st{\tilde{t}}

\newcommand{\ptp}{{\em Prog. Theor. Phys.} }
\newcommand{\prc}{{\em Phys. Rep.} }
\newcommand{\rmp}{{\em Rev. Mod. Phys.} }
\newcommand{\mpl}{{\em Mod. Phys. Lett} }

\newcommand{\pl}{{\em Phys. Lett.} }
\newcommand{\prl}{{\em Phys. Rev. Lett.} }
\newcommand{\pr}{{\em Phys. Rev.} }
\newcommand{\np}{{\em Nucl. Phys.} }

\newcommand{\jmp}{{\em J. Math. Phys.} }
\newcommand{\nc}{{\em Nuovo Cimento} }
\newcommand{\ncl}{{\em Nuovo Cimento Letters} }

\def\NPB#1#2#3{{\it Nucl.~Phys.} {\bf{B#1}} (19#2) #3}
\def\PLB#1#2#3{{\it Phys.~Lett.} {\bf{B#1}} (19#2) #3}
\def\PRD#1#2#3{{\it Phys.~Rev.} {\bf{D#1}} (19#2) #3}
\def\PRL#1#2#3{{\it Phys.~Rev.~Lett.} {\bf{#1}} (19#2) #3}

\def\simlt{\stackrel{<}{{}_\sim}}
\def\simgt{\stackrel{>}{{}_\sim}}

\begin{document}
\begin{titlepage}
\begin{flushright}
IEM-FT-187/99 \\
{\sf hep-ph/9901312}\\
\end{flushright}
\vskip 0.25in
\vspace{1cm}
\begin{center}{\Large\bf
Finite Temperature Field Theory and Phase Transitions$^*$}
\vskip .2in
{\bf Mariano Quir\'os}
\vskip.1truecm
Instituto de Estructura de la Materia (CSIC), Serrano 123 \\
E--28006 Madrid, Spain
\end{center}
\vskip .2in
\vspace{.2cm}
\begin{abstract}
We review different aspects of field theory at zero and finite 
temperature, related to the theory of phase transitions. 
We discuss different renormalization conditions
for the effective potential at zero temperature, emphasizing in
particular the $\overline{\rm MS}$ renormalization scheme.
Finite temperature field theory is discussed in the real and
imaginary time formalisms, showing their equivalence in simple
examples. Bubble nucleation by thermal tunneling, and the subsequent 
development of the phase transition is described in some detail. 
Some attention is also devoted to the breakdown of the perturbative 
expansion and the infrared problem in the finite temperature field 
theory. Finally the application to baryogenesis at the electroweak 
phase transition is done in the Standard Model and in the Minimal 
Supersymmetric Standard Model. In all cases we have translated the 
condition of not washing out any previously generated baryon asymmetry 
by upper bounds on the Higgs mass. 
\end{abstract}

\vspace{1cm}
\begin{flushleft}
IEM-FT-187/99 \\
January 1999\\
\end{flushleft}
\vskip1.truecm
\hrule width 4.5cm
\vskip.1truecm
$^*$Based on lectures given at the {\it Summer School in High
Energy Physics and Cosmology}, ICTP, Trieste (Italy) 29 June--17 July 1998.

\end{titlepage}

\newpage

\section{The effective potential at zero temperature}

The effective potential for quantum field theories was
originally introduced by Euler, Heisenberg and Schwinger~\cite{EHS}, 
and applied to studies of spontaneous symmetry
breaking by Goldstone, Salam, Weinberg and Jona-Lasinio~\cite{GSWJL}. 
Calculations of the effective potential were
initially performed at one-loop by Coleman and E. Weinberg~\cite{CW} 
and at higher-loop by Jackiw~\cite{J} and Iliopoulos,
Itzykson and Martin~\cite{IIM}. More recently the effective
potential has been the subject of a vivid investigation,
especially related to its invariance under the renormalization
group. I will try to review, in this section, the main ideas and
update the latest developments on the effective potential. 

\subsection{Generating functionals}

To fix the ideas, let us consider the theory described by a
scalar field $\phi$ with a lagrangian density ${\cal
L}\{\phi(x)\}$ and an action
\begin{equation}
S[\phi]=\int d^4 x{\cal L}\{\phi(x)\}
\label{actsca}
\end{equation}
The generating functional (vacuum-to-vacuum amplitude) is given
by the path-integral representation,
\begin{equation}
Z[j]=\langle 0_{\rm out}\mid 0_{\rm in} \rangle_j \equiv \int
d\phi \exp\{i(S[\phi]+\phi j)\}
\label{zfunct}
\end{equation}
where we are using the notation
\begin{equation}
\phi j\equiv \int d^4x \phi(x) j(x)
\label{product}
\end{equation}
Using (\ref{zfunct}) one can obtain the connected generating
functional $W[j]$ defined as,
\begin{equation}
Z[j] \equiv \exp\{iW[j]\}
\label{wfunct}
\end{equation}
and the effective action $\Gamma[\overline{\phi}]$ as the
Legendre transform of (\ref{wfunct})
\begin{equation}
\Gamma[\overline{\phi}]=W[j]-\int d^4 x \frac{\delta
W[j]}{\delta j(x)} j(x)
\label{effaction}
\end{equation}
where
\begin{equation}
\overline{\phi}(x)=\frac{\delta W[j]}{\delta j(x)} 
\label{phibar}
\end{equation}
In particular, from (\ref{effaction}) and (\ref{phibar}), the
following equality can be easily proven,
\begin{equation}
\frac{\delta \Gamma[\overline{\phi}]}{\delta \overline{\phi}}= 
\frac{\delta W[j]}{\delta j}\frac{\delta j}{\delta
\overline{\phi}}- j-\overline{\phi}\frac{\delta j}{\delta
\overline{\phi}}= -j
\label{j}
\end{equation}
where we have made use of the notation (\ref{product}).
Eq.~(\ref{j}) implies in particular that,
\begin{equation}
\left.\frac{\delta \Gamma[\overline{\phi}]}{\delta \overline{\phi}}
\right|_{j=0}=0 
\label{vacuum}
\end{equation}
which defines de vacuum of the theory in the absence of external
sources.

We can now expand $Z[j]$ ($W[j]$) in a power series of $j$, to
obtain its representation in terms of Green functions $G_{(n)}$
(connected Green functions $G^{\ c}_{(n)}$) as,
\begin{equation}
Z[j]= \sum^{\infty}_{n=0} \frac{i^n}{n!} \int d^4x_1 \ldots
d^4x_n j(x_1)\ldots j(x_n) G_{(n)}(x_1, \ldots,x_n)
\label{green}
\end{equation}
and
\begin{equation}
iW[j]= \sum^{\infty}_{n=0} \frac{i^n}{n!} \int d^4x_1 \ldots
d^4x_n j(x_1)\ldots j(x_n) G^{\ c}_{(n)}(x_1, \ldots,x_n)
\label{greenc}
\end{equation}
Similarly the effective action can be expanded in powers of
$\overline{\phi}$ as
\begin{equation}
\Gamma[\overline{\phi}]= \sum^{\infty}_{n=0} \frac{1}{n!} \int d^4x_1 \ldots
d^4x_n \overline{\phi}(x_1)\ldots \overline{\phi}(x_n) 
\Gamma^{(n)}(x_1, \ldots,x_n)
\label{1pi}
\end{equation}
where $\Gamma^{(n)}$ are the one-particle irreducible (1PI)
Green functions.

We can Fourier transform $\Gamma^{(n)}$ and $\overline{\phi}$ as,
\begin{equation}
\Gamma^{(n)}(x)=\int \prod_{i=1}^n \left[\frac{d^4
p_i}{(2\pi)^4} \exp\{ip_i x_i\} \right] (2\pi)^4
\delta^{(4)}(p_1+\cdots+p_n) \Gamma^{(n)}(p)
\label{1pitrans}
\end{equation}
\begin{equation}
\tilde{\phi}(p)=\int d^4 x e^{-ipx}\overline{\phi}(x)
\label{phitrans}
\end{equation}
and obtain for (\ref{effaction}) the expression,
\begin{equation}
\Gamma[\overline{\phi}]=\sum_{n=0}^{\infty} \int 
\prod_{i=1}^n \left[\frac{d^4
p_i}{(2\pi)^4} \tilde{\phi}(-p_i) \right] 
(2\pi)^4
\delta^{(4)}(p_1+\cdots+p_n) \Gamma^{(n)}(p_1,\ldots,p_n)
\label{effactp}
\end{equation}

In a translationally invariant theory,
\begin{equation}
\overline{\phi}(x)=\phi_c
\label{phiconst}
\end{equation}
the field $\overline{\phi}$ is constant. Removing an overall
factor of space-time volume, we define the effective potential 
$V_{\rm eff}(\phi_c)$ as,
\begin{equation}
\Gamma[\phi_c]=-\int d^4 x V_{\rm eff}(\phi_c)
\label{effpotdef}
\end{equation}
Using now the definition of Dirac $\delta$-function,
\begin{equation}
\delta^{(4)}(p)=\int \frac{d^4 x}{(2\pi)^4} e^{-ipx}
\label{deltadef}
\end{equation}
and (\ref{phiconst}) in (\ref{phitrans}) we obtain,
\begin{equation}
\tilde{\phi}_c(p)=(2\pi)^4 \phi_c \delta^{(4)}(p).
\label{phiconstp}
\end{equation}
Replacing (\ref{phiconstp}) in (\ref{effactp}) we can write the
effective action for constant field configurations as,
\begin{equation}
\Gamma(\phi_c)=\sum_{n=0}^{\infty}\frac{1}{n!}\phi_c^n (2\pi)^4
\delta^{(4)}(0) \Gamma^{(n)}(p_i=0)=
\sum_{n=0}^{\infty}\frac{1}{n!}\phi_c^n  \Gamma^{(n)}(p_i=0)
\int d^4 x
\label{effactp0}
\end{equation}
and comparing it with (\ref{effpotdef}) we obtain the final expression,
\begin{equation}
V_{\rm eff}(\phi_c)=-\sum_{n=0}^{\infty}\frac{1}{n!}\phi_c^n 
\Gamma^{(n)}(p_i=0) 
\label{effpotent}
\end{equation} 
which will be used for explicit calculations of the effective potential.

Let us finally mention that there is an alternative way of
expanding the effective action: it can also be expanded in
powers of momentum, about the point where all external momenta
vanish. In configuration space that expansion reads as:
\begin{equation}
\label{effpotmom}
\Gamma[\overline{\phi}]=\int d^4 x \left[-V_{\rm eff}
(\overline{\phi})+\frac{1}{2}
(\partial_{\mu}\overline{\phi}(x))^2 Z(\overline{\phi})+\cdots \right]
\end{equation}

\subsection{The one-loop effective potential}

We are now ready to compute the effective potential. In
particular the zero-loop contribution is simply the
classical (tree-level)  potential.
The one-loop contribution is readily computed using the
previous techniques and can be written in closed form for any
field theory containing spinless particles, spin-$\frac{1}{2}$
fermions and gauge bosons. Here we will follow closely the
calculation of Ref.~\cite{CW}.

\subsubsection{Scalar fields}

We consider the simplest model of one self-interacting real
scalar field, described by the lagrangian
\begin{equation}
{\cal L}=\frac{1}{2}\partial^{\mu}\phi \partial_{\mu}\phi-V_0(\phi)
\label{scallag}
\end{equation}
with a tree-level potential
\begin{equation}
V_0=\frac{1}{2}m^2\phi^2+\frac{\lambda}{4!}\phi^4
\label{scalpot}
\end{equation}

\begin{figure}[htb]
\epsfxsize=10truecm
\centerline{\epsfbox{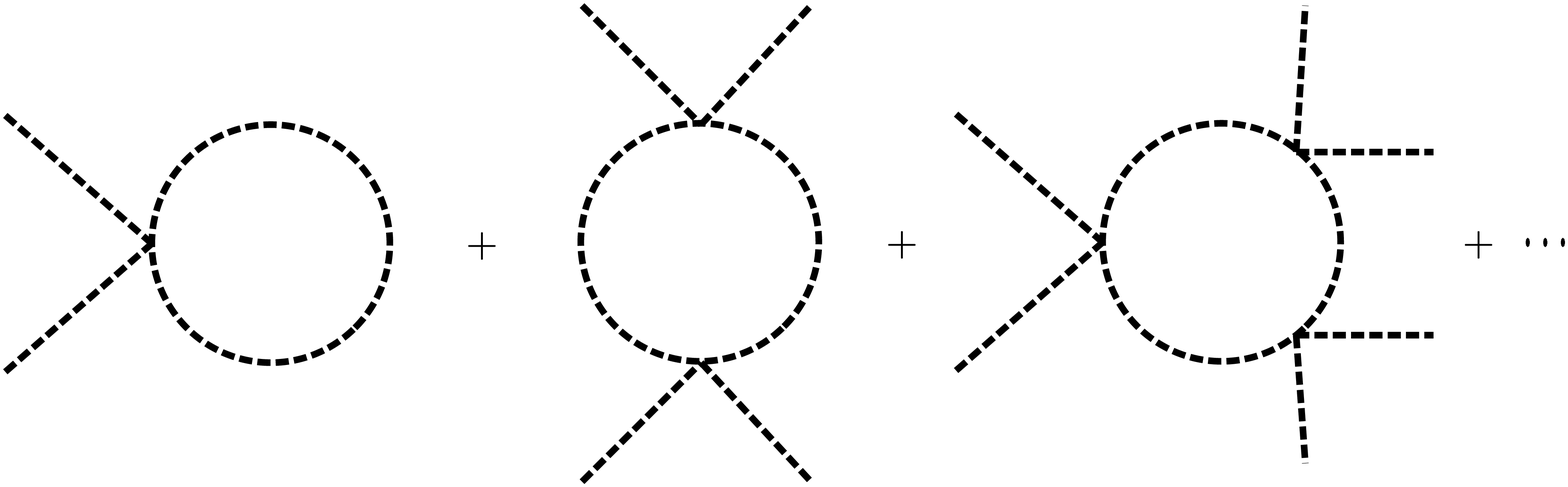}}
\caption[0]{1PI diagrams contributing to the one-loop effective
potential of (\ref{scallag}).}
\label{1loopsc}
\end{figure}

The one-loop correction to the tree-level potential should be
computed as the sum of all 1PI diagrams with a single loop and
zero external momenta. Diagrammatically they are displayed in
Fig.~\ref{1loopsc}, where each vertex has 2 external legs.

The $n$-th diagram has $n$ propagators, $n$ vertices and $2n$
external legs. The $n$ propagators will contribute a factor of
$i^n(p^2-m^2+i\epsilon)^{-n}$ \footnote{We are using the Bjorken
and Drell's~\cite{BjD} notation and conventions.}. 
The external lines contribute a
factor of $\phi_c^{2n}$ and each vertex a factor of
$-i\lambda/2$, where the factor $1/2$ comes from the fact that
interchanging the 2 external lines of the vertex does not change
the diagram. There is a global symmetry factor $\frac{1}{2n}$,
where $\frac{1}{n}$ comes from the symmetry of the diagram under
the discrete group of rotations ${\bf Z}_n$ and $\frac{1}{2}$
from the symmetry of the diagram under reflection. Finally there
is an integration over the loop momentum and an extra global
factor of $i$ from the definition of the generating functional.

Using the previous rules the one-loop effective potential can be
computed as, 
$$V_{\rm eff}(\phi_c)=V_0(\phi_c)+V_1(\phi_c),$$ with
\begin{eqnarray}
V_1(\phi_c)&=&{\displaystyle
i\sum_{n=1}^{\infty}\int\frac{d^4p}{(2\pi)^4}
\frac{1}{2n} \left[\frac{\lambda
\phi_c^2/2}{p^2-m^2+i\epsilon}\right]^n } \nonumber\\
&=&{\displaystyle
-\frac{i}{2}\int\frac{d^4p}{(2\pi)^4} \log\left[1-\frac{\lambda
\phi_c^2/2}{p^2-m^2+i\epsilon} \right] }
\label{1loopscmin}
\end{eqnarray}

After a Wick rotation 
\begin{equation}
\label{Wick}
p^0=ip^0_E,\ p_E=(-ip^0,\vec{p}\ ),\
p^2=(p^0)^2-\vec{p}^{\ 2}=-p_E^2, 
\end{equation}
Eq.~(\ref{1loopscmin}) can be cast as,
\begin{equation}
V_1(\phi_c)=\frac{1}{2}\int\frac{d^4p_E}{(2\pi)^4}\log\left[1+\frac{\lambda
\phi_c^2/2}{p_E^2+m^2}\right]
\label{1loopsceuc}
\end{equation}
Finally, using the {\it shifted} mass
\begin{equation}
m^2(\phi_c)=m^2+\frac{1}{2}\lambda \phi_c^2=\frac{d^2
V_0(\phi_c)}{d\phi_c^2} 
\label{shiftedm}
\end{equation}
and dropping the subindex $E$ from the euclidean momenta, we can
write the final expression of the one-loop contribution to the
effective potential as,
\begin{equation}
\label{1loopsca}
V_1(\phi_c)=\frac{1}{2}\int\frac{d^4p}{(2\pi)^4}\log\left[
p^2+m^2(\phi_c)\right]
\end{equation}
where a field independent term has been neglected.

The result of Eq.~(\ref{1loopsca}) can be trivially generalized
to the case of $N_s$ {\bf complex} scalar fields described by
the lagrangian,
\begin{equation}
\label{scalagc}
{\cal
L}=\partial^{\mu}\phi^a\partial_{\mu}\phi^{\dagger}_a-
V_0(\phi^a,\phi_a^{\dagger}).
\end{equation}
The one-loop contribution to the effective potential in the
theory described by the lagrangian (\ref{scalagc}) is given by
\begin{equation}
\label{1loopscc}
V_1=\frac{1}{2}Tr\int\frac{d^4p}{(2\pi)^4}\log\left[
p^2+M_s^2(\phi^a,\phi_b^{\dagger})\right]
\end{equation}
where
\begin{equation}
\label{msqr}
(M_s^2)^a_b\equiv V^a_b=\frac{\partial^2V}{\partial
\phi^{\dagger}_a \partial \phi^b}
\end{equation}
and $Tr\ M_s^2=2\ V^a_a$, where the factor of 2 comes from the
fact that each complex field contains two degrees of freedom.
Similarly $Tr$ {\bf 1}=2 $N_s$.

\subsubsection{Fermion fields}

We consider now a theory with fermion fields described by the lagrangian,
\begin{equation}
\label{fermlag}
{\cal L}=i\overline{\psi}_a \gamma \cdot \partial
\psi^a-\overline{\psi}_a (M_{f})^a_b \psi^b
\end{equation}
where the mass matrix $(M_{f})^a_b(\phi_c^i)$ is a function of the
scalar fields linear in $\phi_c^i$:
$(M_{f})^a_b=\Gamma^a_{bi}\phi_c^i$.

\begin{figure}[htb]
\epsfxsize=15truecm
\centerline{\epsfbox{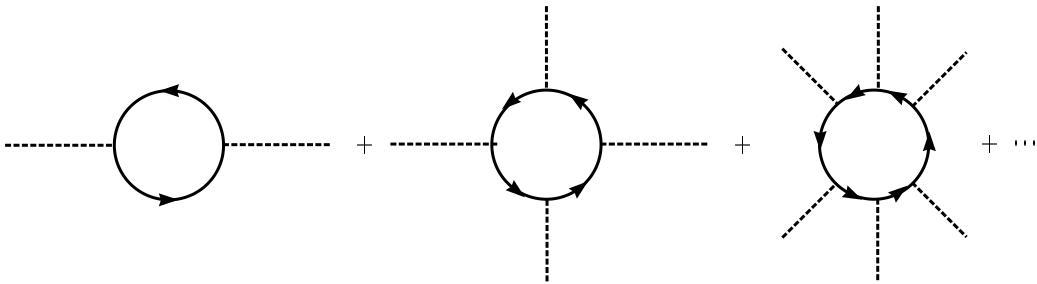}}
\caption[0]{1PI diagrams contributing to the one-loop effective
potential of (\ref{fermlag}).}
\label{1loopf}
\end{figure}

The diagrams contributing to the one-loop effective potential
are depicted in Fig.~\ref{1loopf}.

Diagrams with an odd number of vertices are zero because of the 
$\gamma$-matrices property:
$tr(\gamma^{\mu_1}\cdots\gamma^{\mu_{2n+1}})=0$.
The diagram with $2n$ vertices has $2n$ fermionic propagators.
The propagators yield a factor 
\begin{center}
$Tr_s[i^{2n} (\gamma \cdot p)^{2n}
(p^2+i\epsilon)^{-2n}]$ 
\end{center}
where $Tr_s$ refers to spinor indices. The vertices contribute as 
\begin{center}
$Tr[-i^{2n}M_{f}(\phi_c)^{2n}]$
\end{center}
where $Tr$ runs over the different fermionic fields. There is
also a combinatorial factor $\frac{1}{2n}$ 
(from the cyclic and anticyclic symmetry of diagrams)
and an overall $-1$
coming from the fermions loop. One finally obtains the total factor
$$-\frac{1}{2n}\frac{Tr(M_f^{2n})}{p^{2n}}\cdot Tr_s {\bf 1}. $$
The factor $Tr_s {\bf 1}$ just counts the number of degrees of
freedom of the fermions. It is equal to 4 if Dirac fermions are
used, and 2 if Weyl fermions (and $\sigma$-matrices) are present.
So we will write,
\begin{equation}
\label{trs}
Tr_s {\bf 1}=2 \lambda
\end{equation}
where $\lambda=1$ ($\lambda=2$) for Weyl (Dirac) fermions.
On the other hand we have grouped terms pairwise in the matrix
product and used,
$$ \tilde{p}^2=p^2 $$
where $\tilde{p}$ stands either for $p\cdot \gamma$ or $p \cdot
\sigma$, depending on the kind of fermions we are using.

Collecting everything together we can write the one-loop
contribution to the effective potential from fermion fields as,
\begin{equation}
\label{onefer}
V_1(\phi_c)=-2\lambda i Tr \sum_{n=1}^{\infty} \int
\frac{d^4p}{(2\pi)^4} \frac{1}{2n} \left[\frac{M_f^2}{p^2}\right]^n=
2\lambda \frac{i}{2}Tr \int
\frac{d^4p}{(2\pi)^4}\log\left[1-\frac{M_f^2}{p^2}\right] 
\end{equation}

As in the case of the scalar theory, after making a Wick
rotation to the Euclidean momenta space, and neglecting an
irrelevant field independent term, we can cast (\ref{onefer}) as
\begin{equation}
\label{oneferfin}
V_1=-2\lambda\frac{1}{2}Tr\int\frac{d^4p}{(2\pi)^4}\log\left[
p^2+M_f^2(\phi_c)\right]
\end{equation}

\subsubsection{Gauge bosons}
 
Consider now a theory described by the lagrangian,
\begin{equation}
\label{gaugelag}
{\cal L}= -\frac{1}{4} Tr(F_{\mu \nu}F^{\mu \nu})+\frac{1}{2} 
Tr(D_{\mu}\phi_a)^{\dagger}D^{\mu} \phi^a + \cdots
\end{equation}
In the Landau gauge, which does not require ghost-compensating
terms, the free gauge-boson propagator is
\begin{equation}
\label{gaugeprop}
\Pi^{\mu}_{\ \nu}=-\frac{i}{p^2+i\epsilon} \Delta^{\mu}_{\ \nu}
\end{equation}
with
\begin{equation}
\label{transverse}
\Delta^{\mu}_{\ \nu}=g^{\mu}_{\ \nu}-\frac{p^{\mu}p_{\nu}}{p^2}
\end{equation}
satisfying the property $p_{\mu}\Delta^{\mu}_{\ \nu}=0$ and
$\Delta^n=\Delta$, $n=1,2,\ldots$.

\begin{figure}[htb]
\epsfxsize=10truecm
\centerline{\epsfbox{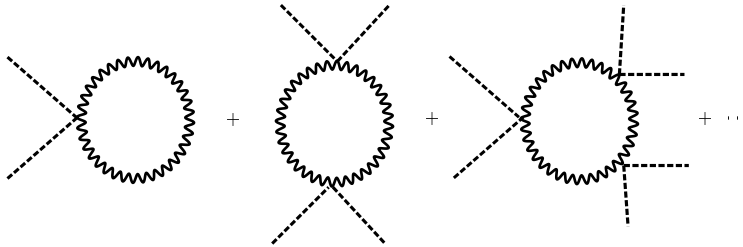}}
\caption[0]{1PI diagrams contributing to the one-loop effective
potential of (\ref{gaugelag}).}
\label{1loopgb}
\end{figure}

The only vertex which contributes to one-loop is
\begin{equation}
\label{massaa}
{\cal L}=\frac{1}{2} (M_{gb})^2_{\alpha \beta}A^{\alpha}_{\mu}A^{\mu
\beta}+\cdots
\end{equation}
where
\begin{equation}
\label{massgauge}
(M_{gb})^2_{\alpha \beta}(\phi_c)=g_{\alpha}g_{\beta} Tr\left[
\left(T^i_{\alpha \ell}\phi_i \right)^{\dagger} T^{\ell}_{\beta j}
\phi^j \right]
\end{equation}
In this way the diagrams contributing to the one-loop effective
potential are depicted in Fig.~\ref{1loopgb}.

A few comments about Eq.~(\ref{massgauge}):  {\bf (i)}
$g_{\alpha}$ is the gauge coupling constant associated to the
gauge field $A_{\mu}^{\alpha}$; if the gauge group is simple,
{\it e.g.} $SU(5)$, $SO(10)$, $E_6, \ldots$, then all gauge
couplings are equal; otherwise there is a distinct gauge
coupling per group factor. {\bf (ii)} $T_{\alpha}$ are the
generators of the Lie algebra of the gauge group in the
representation of the $\phi$-fields and the trace in
(\ref{massgauge}) is over indices of that representation. 

Taking into account the combinatorial factors, the graph with
$n$ propagators and $n$ vertices yields a total factor
$$
\frac{1}{2n} \frac{Tr((M_{gb})^2)^n}{p^{2n}} Tr(\Delta)
$$
where
\begin{equation}
\label{delta3}
Tr(\Delta)=3
\end{equation}
which is the number of degrees of freedom of a massive gauge boson.
Collecting together all factors, and making the Wick rotation to
the euclidean momenta space, we can cast the effective potential
from gauge bosons as,
\begin{equation}
\label{onegaugefin}
V_1=Tr(\Delta)\frac{1}{2}Tr\int\frac{d^4p}{(2\pi)^4}\log\left[
p^2+(M_{gb})^2(\phi_c)\right]
\end{equation}

\subsection{The higher-loop effective potential}

Calculating the effective potential by summing infinite series of
Feynman graphs at zero external momentum is an extremely onerous task
beyond the one-loop approximation. However, as has been shown in 
Ref.~\cite{J}, this task is trivial for the case of one-loop, and affordable
for the case of higher-loop. Here we will just summarize the result of
Ref.~\cite{J} \footnote{The interested reader can find in~\cite{J} all
calculational details.}.

We will start considering the theory described by a real scalar field,
with a lagrangian ${\cal L}$ given in (\ref{scallag}-\ref{scalpot}), and
an action as in (\ref{actsca}). We will define another lagrangian
$\hat{{\cal L}}$ by the following procedure:
\begin{equation}
\label{shiftedlag}
\int d^4x \hat{{\cal L}}\{\phi_c;\phi(x)\} \equiv
S[\phi_c+\phi]-S[\phi_c]-\phi \frac{\delta S[\phi_c]}{\delta \phi_c}
\end{equation}
where we have used in the last term the notation (\ref{product}). 
In (\ref{shiftedlag}), $\phi_c$ is an $x$-independent shifting field.
The second term in (\ref{shiftedlag}) makes the vacuum energy equal to
zero, and the third term is there to cancel the tadpole part of the shifted
action. 

If we denote by ${\cal D}\{\phi_c;x-y\}$ the propagator of the 
shifted theory,
\begin{equation}
\label{prop}\left.
i{\cal D}^{-1}\{\phi_c;x-y\}=\frac{\delta^2S[\phi]}{\delta \phi(x)
\delta \phi(y)}\right|_{\phi=\phi_c}
\end{equation}
and
$$ i{\cal D}^{-1}\{\phi_c;p\}$$
its Fourier transform, the effective potential is found to be given 
by~\cite{J}:
\begin{eqnarray}
V_{\rm eff}(\phi_c)&=&{\displaystyle
V_0(\phi_c)-\frac{i}{2}\int \frac{d^4p}{(2\pi)^4}
\log \det i{\cal D}^{-1}\{\phi_c;p\} } \nonumber\\
&+& {\displaystyle
i\left \langle\exp\left[i\int d^4 x \hat{\cal L}_I\{\phi_c;\phi(x)\}
\right] \right\rangle } 
\label{effpotall}
\end{eqnarray}

The first term in (\ref{effpotall}) is just the classical tree-level 
potential. The second term is the one-loop potential, where the determinant
operates on any possible internal indices defining the propagator. The third
term summarizes the following operation: 
Compute all 1PI vacuum diagrams, with
conventional Feynman rules, using the propagator of the shifted theory
${\cal D}\{\phi_c;p\}$ and the interaction provided by the interaction
lagrangian $\hat{\cal L}_I\{\phi_c;\phi(x)\}$,
and delete the overall factor of space-time volume
$\int d^4 x$ from the effective action (\ref{effpotdef}). It can be shown
that the last term in (\ref{effpotall}) starts at two-loop. Every term in
(\ref{effpotall}) resums an infinite number of Feynman diagrams of the
unshifted theory.

In the simple example of the lagrangian (\ref{scallag}-\ref{scalpot}) it
can be easily seen that the shifted potential is given by
\begin{equation}
\label{shiftpot}
\hat{V}\{\phi_c;\phi\}=\frac{1}{2}m^2(\phi_c)\phi^2+\frac{\lambda}{3!}
\phi_c \phi^3+\frac{\lambda}{4!}
\phi^4
\end{equation}
where the shifted mass is defined in (\ref{shiftedm}). The shifted 
propagator is found to be
\begin{equation}
\label{propeff}
i{\cal D}^{-1}\{\phi_c;p\}=p^2-m^2(\phi_c)
\end{equation}
and the second term of (\ref{effpotall}) easily computed to be
\begin{equation}
\label{1loopfunct}
V_1(\phi_c)=-\frac{i}{2}\int \frac{d^4 p}{(2\pi)^4} 
\log[p^2-m^2(\phi_c)]
\end{equation}
which is easily seen to coincide with (\ref{1loopscmin}), up to 
field independent terms, so that after the Wick rotation we recover
the result (\ref{1loopsca}).

The two-loop effective potential is harder than the one-loop term, but
affordable. The result can be found in Ref.~\cite{J}.
Diagrammatically, the one- and two-loop effective potentials are given
by the ``figure-eight'' plus the ``sunset'' diagrams,
where it is understood that we are using the
Feynman rules of the shifted theory, as stated above. 

Of course, the previous rules apply also to theories containing fermions and
gauge bosons. The Feynman rules of the shifted theory applied to all
1PI diagrams provide the total effective potential according to
(\ref{effpotall}). In particular it is trivial to obtain the one-loop
effective potential for fermions and gauge bosons, as given by 
Eqs.~(\ref{oneferfin}) and (\ref{onegaugefin}), respectively. Notice that the
masses $M^2_f(\phi_c)$, in (\ref{oneferfin}), and $M^2_{gb}(\phi_c)$, in
(\ref{onegaugefin}) are the masses in the corresponding shifted theories.
Diagrammatically (\ref{oneferfin}) and (\ref{onegaugefin}) can be
represented in the shifted theory as vacuum diagrams with one fermion
and one gauge boson loop, respectively.
 
\subsection{Renormalizations conditions}

The final expression of the effective potential we have deduced
in the previous section, Eqs.~(\ref{1loopsca}), (\ref{oneferfin})
and (\ref{onegaugefin}), is ultraviolet-divergent. To make sense
out of it we have to follow the renormalization procedure of
quantum field theories. First of all, to give a sense to the
ultraviolet behaviour of the theory we have to make it finite:
{\it i.e.} we have to {\bf regularize} the theory. Second of
all, all infinities have to be absorbed by appropriate {\bf
counterterms}, which were not explicitly written in our previous
expressions. The way these infinities are absorbed by the
counterterms depend on the definition of the {\bf renormalized
parameters}, {\it i.e.} on the choice of the {\bf
renormalization conditions}. Finally the theory, written as a
function of the renormalized parameters, is finite.

In this way, the first step towards renormalizing the theory is
choosing the regularization scheme. We will first present the
straightforward regularization using a cut-off of momenta.

\subsubsection{Cut-off regularization}

We will illustrate this scheme with the simplest theory: a
massless real scalar field, with a lagrangian
\begin{equation}
\label{renlag}
{\cal L}=\frac{1}{2}(1+\delta
Z)(\partial_{\mu}\phi)^2-\frac{1}{2}\delta m^2 \phi^2-
\frac{\lambda+\delta \lambda}{4!} \phi^4
\end{equation}
where $\delta Z$, $\delta m^2$ and $\delta \lambda$ are the
usual wave-function, mass and coupling constant renormalization
counterterms. They have to be defined self-consistently order by
order in the loop expansion. Here we will compute everything to
one-loop order. 

The conventional definition of the renormalized mass of the
scalar field is the negative inverse propagator at zero
momentum. In view of (\ref{effpotent}) we can write it as:
\begin{equation}
\label{renmass}\left.
m^2_R=-\Gamma^{(2)}(p=0)=\frac{d^2 V}{d\phi_c^2}\right|_{\phi_c=0}
\end{equation}
We can also define the renormalized coupling as the four-point
function at zero external momentum,
\begin{equation}
\label{rencoup}\left.
\lambda_R=-\Gamma^{(4)}(p=0)=\frac{d^4 V}{d\phi_c^4}\right|_{\phi_c=0}
\end{equation}
and the standard condition for the field renormalization is,
\begin{equation}
\label{renz}
Z(0)=1
\end{equation}

Now we will compute the effective potential (\ref{1loopsca})
cutting off the integral at $p^2=\Lambda^2$. First of all we can
integrate over angular variables. For that we can use,
\begin{equation}
\label{angint}
\int d^np f(\rho)=\frac{\pi^{n/2}}{\Gamma(\frac{n}{2})} \int
f(\rho)\rho^{n/2-1}\,d\rho 
\end{equation}
where $\rho=|p|^2$,
and we can cast (\ref{1loopsca}) as
\begin{equation}
\label{intermedio}
 V_1(\phi_c)=\frac{1}{32\pi^2}\int_0^{\Lambda^2}\rho
\log[\rho+m^2(\phi_c)] d\rho.
\end{equation}
This indefinite integral can be solved with the help of~\cite{GR}
$$
\int x\log(a+x)\,dx=\frac{1}{2}(x^2-a^2)\log(a+x)-\frac{1}{2}
\left(\frac{x^2}{2}-ax\right)
$$

Neglecting now in (\ref{intermedio}) field independent terms, and
terms which vanish in the limit $\Lambda \rightarrow \infty$, we
finally obtain,
\begin{equation}
\label{1loopcutoff}
V_1(\phi_c)=\frac{1}{32\pi^2}m^2(\phi_c)\Lambda^2+\frac{1}{64\pi^2}
m^4(\phi_c)\left[\log\frac{m^2(\phi_c)}{\Lambda^2}-\frac{1}{2}\right]
\end{equation}

Using now (\ref{1loopcutoff}) we can write the one-loop effective
potential of the theory (\ref{renlag}) as,
\begin{equation}
\label{potcutoff}
V=\frac{1}{2}\delta m^2 \phi_c^2+\frac{\lambda+\delta
\lambda}{4!} \phi_c^4+\frac{\lambda \phi_c^2}{64 \pi^2}
\Lambda^2 + \frac{\lambda^2\phi_c^4}{256
\pi^2}\left(\log\frac{\lambda
\phi_c^2}{2\Lambda^2}-\frac{1}{2}\right) 
\end{equation}

We will impose now a variant of the renormalization conditions
(\ref{renmass}), (\ref{rencoup}) and (\ref{renz}). For the
renormalized mass we can impose it to vanish, {\it i.e.},
\begin{equation}
\label{renmass2}\left.
\frac{d^2 V}{d\phi_c^2}\right|_{\phi_c=0}=0
\end{equation}
For the renormalized gauge coupling $\lambda$, we cannot use
Eq.~(\ref{rencoup}) at a value of the field equal to zero. There
is nothing wrong with using a different renormalization
prescription and using a different subtraction point. We can use,
\begin{equation}
\label{rencoup2}\left.
\lambda=\frac{d^4 V}{d\phi_c^4}\right|_{\phi_c=\mu}
\end{equation}
where $\mu$ is some mass scale. Different choices of the scale
lead to different definitions of the coupling constant, {\it
i.e.} to different parametrizations of the same theory, but in
principle any value of $\mu$ is as good as any other.

Imposing now the conditions (\ref{renmass2}) and
(\ref{rencoup2}) to (\ref{potcutoff}) we can write the
counterterms as,
\begin{equation}
\label{dm2}
\delta m^2=-\frac{\lambda}{32\pi^2}\Lambda^2
\end{equation}
and
\begin{equation}
\label{dl}
\delta \lambda=-\frac{11 \lambda^2}{32 \pi^2}-\frac{3
\lambda^2}{32 \pi^2}\log \frac{\lambda \mu^2}{2 \Lambda^2}
\end{equation}

Using now (\ref{dm2}) and (\ref{dl}) in (\ref{potcutoff}) we can
write the one-loop effective potential in the previous
renormalization scheme as,
\begin{equation}
\label{potcutfin}
V_{\rm
eff}=\frac{\lambda}{4!}\phi_c^4+\frac{\lambda^2\phi_c^4}{256\pi^2}
\log\left(\frac{\phi_c^2}{\mu^2}-\frac{25}{6} \right)
\end{equation}

A similar renormalization scheme can be defined also for
theories with fermions and/or gauge bosons. However for
gauge theories the regularization provided by the
cut-off explicitly break gauge invariance so that the
dimensional regularization is better suited for them. In the
next section we will review the calculation of the effective
potential in the dimensional regularization and define the
so-called $\overline{\rm MS}$ scheme.

\subsubsection{Dimensional regularization}

This regularization scheme was introduced by t'Hooft and Veltman~\cite{DR}. 
It consists in making an analytic continuation of
Feynman integrals to the complex plane in the number of
space-time dimensions $n$. The integrals have singularities
which arise as poles in $1/(n-4)$ and have to be subtracted out.
The particular prescription for subtraction is called a {\bf
renormalization scheme}. In working with the effective potential
it is customary to use the so-called $\overline{\rm MS}$
renormalization scheme~\cite{MS}.

We will compute now the one-loop effective potential
(\ref{1loopsca}) using dimensional regularization, {\it i.e.}
\begin{equation}
\label{1loopdr}
V_1(\phi_c)=\frac{1}{2} (\mu^2)^{2-\frac{n}{2}}
\int\frac{d^n p}{(2\pi)^n}\log\left[
p^2+m^2(\phi_c)\right]
\end{equation}
where $\mu$ is a scale with mass dimension which needs to be
introduced to balance the dimension of the integration measure.
It is simpler to compute the tadpole
\begin{equation}
\label{derivative}
V'=\frac{1}{2} (\mu^2)^{2-\frac{n}{2}}
\int \frac{d^n p}{(2\pi)^n}\frac{1}{p^2+m^2(\phi_c)}
\end{equation}
where the meaning of $V'$ is the derivative with respect to
$m^2(\phi_c)$, using the basic formula of dimensional regularization,
\begin{equation}
\label{integral}
\int
d^np\frac{(p^2)^{\alpha}}{(p^2+M^2)^{\beta}}=\pi^{\frac{n}{2}}
(M^2)^{\frac{n}{2}+\alpha-\beta}
\frac{\Gamma(\alpha+\frac{n}{2})
\Gamma(\beta-\alpha-\frac{n}{2})}{\Gamma(\frac{n}{2})\Gamma(\beta)}
\end{equation}
and integrating the resulting integral with respect to
$m^2(\phi_c)$. One can then write the regularized potential
(\ref{1loopdr}) as,
\begin{equation}
\label{1loopreg}
V_1(\phi_c)=-\frac{1}{32 \pi^2}
\frac{1}{\frac{n}{2}(\frac{n}{2}-1)} \left(\frac{m^2(\phi_c)}{4
\pi \mu^2} \right)^{\frac{n}{2}-2} \Gamma\left(2-\frac{n}{2}\right)
m^4(\phi_c)
\end{equation}

We can expand (\ref{1loopreg}) in powers of $2-n/2$ and use the
expansion 
\begin{equation}
\label{gamma}
\Gamma(z)=\frac{1}{z}-\gamma_E+{\cal O}(z)
\end{equation}
where $\gamma_E=0.5772\ldots$ is the Euler-Masccheroni constant~\cite{GR}. 
We obtain for (\ref{1loopreg})
\begin{eqnarray}
V_1(\phi_c)&=&{\displaystyle
\frac{m^4(\phi_c)}{64 \pi^2}\left\{
-\left[\frac{1}{2-\frac{n}{2}} -\gamma_E+\log
4\pi\right]\right. }\nonumber\\
&+&{\displaystyle \left.
\log\frac{m^2(\phi_c)}{\mu^2} -\frac{3}{2}+{\cal
O}(\frac{n}{2}-2)\right\} } 
\label{1loopreg2}
\end{eqnarray}

Now the $\overline{\rm MS}$ renormalization scheme consists in
subtracting the term proportional to
\begin{equation}
\label{msbar}
C_{\rm UV}\equiv\left[\frac{1}{2-\frac{n}{2}} -\gamma_E+\log 4\pi\right]
\end{equation}
in the regularized potential (\ref{1loopreg2}). Therefore the
divergent piece,
$$
-\frac{m^4(\phi_c)}{64 \pi^2}\left\{
\left[\frac{1}{2-\frac{n}{2}} -\gamma_E+\log 4\pi\right]\right\}
$$
has to be absorbed by the counterterms. Therefore the final
expression for the one-loop potential, written in terms of the
renormalized parameters, is
\begin{equation}
\label{1loopms}
V_1(\phi_c)=\frac{1}{64 \pi^2}m^4(\phi_c)\left\{
\log\frac{m^2(\phi_c)}{\mu^2} -\frac{3}{2}\right\} 
\end{equation}

For instance, in the theory described by lagrangian
(\ref{renlag}), the counterterms are given by,
\begin{eqnarray}
\delta m^2&=& 0 \\
\delta \lambda &=& \frac{3 \lambda^2}{32 \pi^2}
\left[\frac{1}{2-\frac{n}{2}} -\gamma_E+\log 4\pi\right]
\nonumber
\end{eqnarray}
and the effective potential is,
\begin{equation}
\label{potmsfin}
V_{\rm
eff}=\frac{\lambda}{4!}\phi_c^4+\frac{\lambda^2\phi_c^4}{256\pi^2}
\log\left(\frac{\lambda \phi_c^2}{2 \mu^2}-\frac{3}{2} \right)
\end{equation}

The scale $\mu$ along this section is related to the
renormalization group behaviour of the renormalized couplings
and masses. 

For a theory with fermion fields, one needs a trace operation
in dimensional regularization, as $Tr{\bf 1}=f(n)$. For instance,
for an even dimension one could choose,
$f(n)=2^{n/2}$ for Dirac fermions, and
$f(n)=2^{n/2-1}$ for Weyl fermions. However the difference
$f(n)-f(4)$ is only relevant for divergent graphs and can
therefore be absorbed by a renormalization-group transformation.
It is usually convenient to choose $f(n)=f(4)=2\lambda$ for all
values of $n$~\cite{COLL}.
The effective potential
(\ref{oneferfin}) can be computed as in (\ref{1loopdr}), leading
to,
\begin{eqnarray}
V_1(\phi_c)&=&{\displaystyle
-\lambda \frac{M_f^4(\phi_c)}{32 \pi^2}\left\{
-\left[\frac{1}{2-\frac{n}{2}} -\gamma_E+\log 4\pi\right]\right. }
\nonumber\\
&+& {\displaystyle \left.
\log\frac{M_f^2(\phi_c)}{\mu^2} -\frac{3}{2}+{\cal
O}(\frac{n}{2}-2)\right\}  } 
\label{1loopregfer}
\end{eqnarray}
In the $\overline{\rm MS}$ renormalization scheme, after
subtracting the term proportional to (\ref{msbar}) we obtain,
\begin{equation}
\label{1loopmsfer}
V_1(\phi_c)=-\lambda\frac{1}{32 \pi^2}M_f^4(\phi_c)\left\{
\log\frac{M_f^2(\phi_c)}{\mu^2} -\frac{3}{2}\right\} 
\end{equation}

Similarly, in a theory with gauge bosons as in (\ref{gaugelag}),
the effective potential (\ref{onegaugefin}) is computed as,
\begin{eqnarray}
V_1(\phi_c)&=& {\displaystyle
Tr(\Delta)\frac{M_{gb}^4(\phi_c)}{64 \pi^2}\left\{
-\left[\frac{1}{2-\frac{n}{2}} -\gamma_E+\log 4\pi\right]\right. }
\nonumber\\
&+& {\displaystyle \left.
\log\frac{M_{gb}^2(\phi_c)}{\mu^2} -\frac{3}{2}+{\cal
O}(\frac{n}{2}-2)\right\} } 
\label{1loopreggb}
\end{eqnarray}
where
\begin{equation}
\label{deltan}
Tr(\Delta)=n-1
\end{equation}

In the $\overline{\rm MS}$ renormalization scheme, subtracting
as usual the term proportional to (\ref{msbar}) one obtains the
effective potential,
\begin{equation}
\label{1loopmsgb}
V_1(\phi_c)=3\frac{1}{64 \pi^2}M_{gb}^4(\phi_c)\left\{
\log\frac{M_{gb}^2(\phi_c)}{\mu^2} -\frac{5}{6}\right\} 
\end{equation}

A variant of the $\overline{\rm MS}$ renormalization scheme is
the $\overline{\rm DR}$ renormalization scheme~\cite{DRED}, where the
dimensional regularization is applied only to the scalar part of
the integrals, while all fermion and tensor indices are
considered in four dimensions. In this case $Tr(\Delta)$ is
taken equal to 3, as in (\ref{delta3}), and subtracting from
(\ref{1loopreggb}) the term proportional to (\ref{msbar}) one
obtains, 
\begin{equation}
\label{1loopdrgb}
V_1(\phi_c)=3\frac{1}{64 \pi^2}M_{gb}^4(\phi_c)\left\{
\log\frac{M_{gb}^2(\phi_c)}{\mu^2} -\frac{3}{2}\right\} 
\end{equation}

\subsection{One-loop effective potential for the Standard Model}

In this subsection we will apply the above ideas to compute the
one loop effective potential for the Standard Model of
electroweak interactions. The spin-zero fields of the Standard
Model are described by the $SU(2)$ doublet,
\begin{equation}
\label{doublet}
\Phi=\left(
\begin{array}{c}
\chi_1+i\chi_2 \\
{\displaystyle \frac{\phi_c+h+i\chi_3}{\sqrt{2}}  }
\end{array}
\right)
\end{equation}
where $\phi_c$ is the real constant background, $h$ the Higgs
field, and $\chi_a$ ($a$=1,2,3) are the three Goldstone bosons.
The tree level potential reads, in terms of the background
field, as
\begin{equation}
\label{treesm}
V_0(\phi_c)=-\frac{m^2}{2}\phi_c^2+\frac{\lambda}{4}\phi_c^4 
\end{equation}
with positive $\lambda$ and $m^2$, and the tree level minimum
corresponding to 
$$
v^2=\frac{m^2}{\lambda}.
$$
The spin-zero field dependent masses are
\begin{eqnarray}
\label{masses0sm}
m^2_h(\phi_c) & = & 3\lambda\phi_c^2-m^2 \nonumber \\
m^2_{\chi}(\phi_c) & = & \lambda\phi_c^2-m^2 
\end{eqnarray}
so that $m_h^2(v)=2\lambda v^2=2 m^2$ and $m_{\chi}^2(v)=0$. The
gauge bosons contributing to the one-loop effective potential
are $W^{\pm}$ and $Z$, with tree level field dependent masses,
\begin{eqnarray}
\label{masses1sm}
m^2_W(\phi_c) & = & \frac{g^2}{4}\phi_c^2 \\
m^2_Z(\phi_c) & = & \frac{g^2+g'^2}{4} \phi_c^2 \nonumber
\end{eqnarray}
Finally, the only fermion which can give a significant
contribution to the one loop effective potential is the top
quark, with a field-dependent mass
\begin{equation}
\label{mass1/2sm}
m^2_t(\phi_c)=\frac{h_t^2}{2}\phi_c^2
\end{equation}
where $h_t$ is the top quark Yukawa coupling.

The one-loop effective potential $V_1(\phi_c)$ can be computed
using Eqs.~(\ref{1loopsca}), (\ref{oneferfin}) and
(\ref{onegaugefin}). As we have said in the previous subsection,
these integrals are ultraviolet divergent. They have to be
regularized and the divergent contributions cancelled by the
counterterms 
\begin{equation}
\label{counter}
V_1^{\rm c.t.}=\delta\Omega+\frac{\delta
m^2}{2}\phi_c^2+\frac{\delta\lambda}{4} \phi_c^4 
\end{equation}
where we have introduced a counterterm $\delta\Omega$ for the
vacuum energy or cosmological constant (see next section).

The final expression for the effective potential is finite and
depends on the used regularization and, correspondingly, on the
renormalization conditions. Next we will describe the two most
commonly used renormalization conditions for the Standard Model.

\subsubsection{$\overline{\rm MS}$ renormalization}

In this case we can use Eqs.~(\ref{1loopreg2}),
(\ref{1loopregfer}) and (\ref{1loopreggb}) for the contribution
to $V_1(\phi_c)$ of the scalars, fermions and gauge bosons,
respectively. In the $\overline{\rm MS}$ renormalization scheme
we subtract the terms proportional to $C_{\rm UV}$, see 
Eq.~(\ref{msbar}), which are cancelled by the counterterms in
(\ref{counter}). One easily arrives to the finite effective
potential provided by
\begin{equation}
\label{msbsm}
V(\phi_c)=V_0(\phi_c)+\frac{1}{64\pi^2}\sum_{i=W,Z,h,\chi,t} n_i
m_i^4(\phi_c)\left[\log\frac{m_i^2(\phi_c)}{\mu^2}-C_i\right] 
\end{equation}
where $C_i$ are constants given by,
\begin{eqnarray}
\label{csm}
C_W=C_Z & = & \frac{5}{6} \\
C_h=C_{\chi}=C_t & = & \frac{3}{2} \nonumber
\end{eqnarray}
and $n_i$ are the degrees of freedom
\begin{equation}
\label{nsm}
n_W=6,\ n_Z=3,\ n_h=1,\ n_{\chi}=3,\ n_t=-12
\end{equation}

The counterterms which cancel the infinities are provided by,
\begin{eqnarray}
\label{countermsb}
\delta\Omega & = & \frac{m^4}{64\pi^2}\left(n_h+n_{\chi}\right)
C_{\rm UV} \nonumber \\
\delta m^2 & = & -\frac{3\lambda m^2}{16\pi^2}
\left(n_h+\frac{1}{3}n_{\chi}\right) C_{\rm UV} \\
\delta\lambda & = &
\frac{3}{16\pi^2}\left[\frac{2g^4+(g^2+g'^2)^2}{16}
-h_t^4+\left(3n_h+\frac{1}{3}n_{\chi}\right)
\lambda^2 \right] C_{\rm UV} \nonumber
\end{eqnarray}
where $C_{\rm UV}$ is defined in (\ref{msbar}). We have explicitly
written in (\ref{countermsb}) the contribution to the
counterterms from the Higgs sector, $n_h$ and $n_{\chi}$. The
latter give rise entirely to the mass counterterms $\delta m^2$
and $\delta\Omega$. For Higgs masses lighter than $W$ masses,
the Higgs sector can be ignored in the one loop radiative
corrections (as it is usually done) and the massive counterterms
are not generated.

\subsubsection{Cut-off regularization}

A very useful scheme~\cite{AH}
is obtained by regularizing the theory with
a cut-off and imposing that the minimum, 
at $v=246.22\ GeV$, and the Higgs mass does not
change with respect to their tree level values, {\it i.e.},
\begin{eqnarray}
\label{ahregcond}
\left.\frac{d(V_1+V_1^{c.t.})}{d\phi_c}\right|_{\phi_c=v} & = & 0 \\
\left.\frac{d^2(V_1+V_1^{c.t.})}{d\phi_c^2}\right|_{\phi_c=v} & = & 0
\nonumber 
\end{eqnarray}

Now we can use (\ref{1loopcutoff}) to write
\begin{equation}
\label{1loopcosm}
V_1(\phi_c)=\frac{1}{32\pi^2}\sum_{i=W,Z,t,h,\chi}n_i
\left[m_i^2(\phi_c)\Lambda^2+\frac{m_i^4(\phi_c)}{2}
\left(\log\frac{m_i^2(\phi_c)}{\Lambda^2}-\frac{1}{2}\right)\right]
\end{equation}

Imposing now the conditions (\ref{ahregcond}) the infinities in
(\ref{1loopcosm}) cancel against those in $V_1^{c.t.}$, and the
resulting $\phi_c$-dependent potential is finite, and given by,
\begin{equation}
\label{potsm}
V(\phi_c)=V_0(\phi_c)+\frac{1}{64\pi^2}\sum_{i}\left\{m_i^4(\phi_c)\left(
\log\frac{m_i^2(\phi_c)}{m_i^2(v)}-\frac{3}{2}\right)+2 m_i^2(v)
m_i^2(\phi_c) \right\}
\end{equation}
The counterterms $\delta\Omega$, $\delta m^2$ and $\delta\lambda$ in
(\ref{counter}) turn out to be given by 
\begin{eqnarray}
\label{countersm}
\delta\lambda & = &
-\frac{1}{16\pi^2}\sum_i n_i\left(\frac{m_i^2(v)-b_i}{v^2} \right)^2
\left(\log\frac{m_i^2(v)}{\Lambda^2}+1\right) \\
\delta m^2 & = & -\frac{1}{16\pi^2}\sum_i n_i
\frac{m_i^2-b_i}{v^2} \left[\Lambda^2-m_i^2(v)+
b_i\left(\log\frac{m_i^2(v)}{\Lambda^2}+1\right)\right]
\nonumber \\
\delta \Omega & = & \frac{m^2}{32\pi^2}\sum_{i=h,\chi}n_i 
\left[\Lambda^2-m_i^2(v)+\frac{1}{2}
b_i\left(\log\frac{m_i^2(v)}{\Lambda^2}+1\right)\right]
\nonumber
\end{eqnarray}
where $b_W=b_Z=b_t=0$ and $b_h=b_{\chi}=-m^2$.

We can see again in (\ref{countersm}) that ignoring the
contribution to the one loop effective potential from the Higgs
sector results in the non appearance of a cosmological constant.
However, unlike the $\overline{\rm MS}$ scheme, $\delta m^2$ is
also generated by the contribution of the gauge boson and top
quark loops.
Of course the one loop counterterms we are computing along this
section are only useful for two loop calculations.

\subsection{Improved effective potential and renormalization
group} 

As we have seen in the previous section, the calculation of the
effective action involves a mass $\mu$ which is not physical in
the sense that all the theory should be independent of the
chosen value of $\mu$. In fact a change in $\mu$ should be
accompanied by a change in the renormalized parameters
(couplings and masses) such that all the theory remains
unchanged. This statement for the effective action can be
expressed as an equation~\cite{CW}
\begin{equation}
\label{gammainv}
\left[\mu\frac{\partial}{\partial
\mu}+\beta_i\frac{\partial}{\partial \lambda_i}-\gamma \phi_c
\frac{\delta}{\delta \phi_c} \right]\Gamma[\phi_c]=0
\end{equation}
for an appropriate choice of the coefficients $\beta_i$ and
$\gamma$, where $\lambda_i$ denotes collectively all couplings
and masses of the theory. In the last term of (\ref{gammainv})
we have made use of the notation (\ref{product}).

We define the effective potential $\hat{V}$ as in Eq.~(\ref{effpotent}),
\begin{equation}
\hat{V}\equiv \hat{V}(\mu,\lambda_i,\phi_c)=
\hat{V}(\mu,\lambda_i,0)
-\sum_{n=1}^{\infty}\frac{1}{n!}\phi_c^n 
\Gamma^{(n)}(p_i=0) 
\label{effpotent2}
\end{equation} 
The role of the vacuum energy $\hat{\Omega}$,
$$
\hat{\Omega}=\hat{V}(\mu,\lambda_i,0)
$$
has been recently stressed in Ref.~\cite{IEP}. Using now the
renormalization group equation (RGE) satisfied by the effective
action (\ref{gammainv}), we obtain the RGE satisfied by
$\hat{V}$ as
\begin{equation}
\label{rgepot}
\left[\mu\frac{\partial}{\partial
\mu}+\beta_i\frac{\partial}{\partial \lambda_i}-\gamma \phi_c
\frac{\partial}{\partial \phi_c} \right]\hat{V}=
\left[\mu\frac{\partial}{\partial
\mu}+\beta_i\frac{\partial}{\partial \lambda_i} \right]\hat{\Omega}
\end{equation}

If we make a $\phi_c$-independent shift to $\hat{V}$ such that,
\begin{eqnarray}
\label{shift}
V=\hat{V}+\Delta \hat{\Omega}(\mu,\lambda_i) \nonumber \\
\Omega \equiv \hat{\Omega}+\Delta\hat{\Omega}
\end{eqnarray}
with the condition,
\begin{equation}
\label{rgevac}
\left[\mu\frac{\partial}{\partial
\mu}+\beta_i\frac{\partial}{\partial \lambda_i} \right]\Omega=0
\end{equation}
then the potential $V$ satisfies the well known RGE,
\begin{equation}
\label{rgepotfin}
\left[\mu\frac{\partial}{\partial
\mu}+\beta_i\frac{\partial}{\partial \lambda_i}-\gamma \phi_c
\frac{\partial}{\partial \phi_c} \right]V=0
\end{equation}

The formal solutions to Eqs.~(\ref{rgevac}) and
(\ref{rgepotfin}) can be written as,
\begin{eqnarray}
\label{formsol}
V\equiv V(\mu,\lambda_i,\phi_c)=V(\mu(t),\lambda_i(t),\phi(t))
\nonumber \\
\Omega \equiv \Omega(\mu, \lambda_i)=\Omega(\mu(t),\lambda_i(t))
\end{eqnarray}
where
\begin{eqnarray}
\label{renparam}
\mu(t)& = & \mu \exp(t) \nonumber \\
\phi(t) & = & \phi_c \xi(t) \\
\xi(t) & = & \exp\left\{ -\int_0^t \gamma(\lambda_i(t'))dt'
\right\} \nonumber \\
\beta_i(\lambda(t))& = & \frac{d \lambda_i(t)}{dt} \nonumber
\end{eqnarray}
with the boundary conditions,
\begin{eqnarray}
\label{boundary}
\mu(0) & = & \mu \nonumber \\
\phi(0) & = & \phi_c \nonumber \\
\xi(0) & = & 1 \nonumber \\
\lambda_i(0) & = & \lambda_i
\end{eqnarray}

In fact Eqs.~(\ref{rgevac}) and (\ref{rgepotfin}) can be simply
written as,
\begin{eqnarray}
\label{rgeinv}
\frac{d}{dt}\Omega & = & 0 \\
\frac{d}{dt}V & = & 0 \nonumber
\end{eqnarray}
which state that $\Omega$ and $V$ are scale-independent. Of
course the same happens to all derivatives of $V$,
\begin{equation}
\label{derivn}
V^{(n)}(\mu,\lambda_i,\phi_c) \equiv \frac{\partial^n
V(\mu,\lambda_i,\phi_c)}{\partial \phi_c^n}
\end{equation}
which by virtue of (\ref{renparam}) satisfies 
\begin{equation}
\label{derrel}
V^{(n)}=\xi(t)^n \frac{\partial^n}{\partial \phi(t)^n}
V(\mu(t),\lambda_i(t),\phi(t)) 
\end{equation}

The RGE satisfied by $V^{(n)}$ can be obtained from
(\ref{rgepotfin}) and the property,
$$
\frac{\partial^n}{\partial \phi_c^n}\left[-\gamma
\phi_c\frac{\partial}{\partial
\phi_c}\right]=\left[-\gamma\phi_c\frac{\partial}{\partial
\phi_c}\right] \frac{\partial^n}{\partial
\phi_c^n}-n\gamma\frac{\partial^n}{\partial \phi_c^n}
$$
It is given by,
\begin{equation}
\label{rgeder}
\left[\mu\frac{\partial}{\partial
\mu}+\beta_i\frac{\partial}{\partial \lambda_i}-\gamma \phi_c
\frac{\partial}{\partial \phi_c} \right]V^{(n)}=n\gamma V^{(n)} 
\end{equation}
which implies that $V^{(n)}$ is scale independent.

In particular the scale independence of $V^{(n)}$,
$n=0,1,\ldots$, means that we can fix the scale $t$ at any
value, even $\phi_c$-dependent. Suppose we fix $t$ by the
arbitrary conditions,
\begin{eqnarray}
\label{tfixing}
\mu(t)&=&f(\phi_c) \nonumber \\
t=t(\phi_c)&=&\log\{f(\phi_c)/\mu\} \\
\phi(t)&=&\xi(t(\phi_c))\phi_c \nonumber
\end{eqnarray}

Using (\ref{tfixing}) we can write the effective potential and
its derivatives (\ref{derivn}) as $\phi_c$-functions,
\begin{equation}
\label{potphic}
V(\phi_c) \equiv V\left[f(\phi_c),\lambda_i(t(\phi_c)),
\phi(t(\phi_c))\right] 
\end{equation}
and
\begin{equation}
\label{derphic}\left.
V^{(n)}(\phi_c)\equiv \xi(t(\phi_c))^n
\frac{\partial^n}{\partial \phi(t)^n}
V(\mu(t),\lambda_i(t),\phi(t)) \right|_{t=t(\phi_c)} 
\end{equation}

Using Eq.~(\ref{rgeder}) one can easily prove that~\cite{CEQR},
\begin{equation}
\label{derequiv}
V^{(n)}(\phi_c)=\frac{d^nV(\phi_c)}{d \phi_c^n}
\end{equation}

Fixing the scale is a matter of convention. Fixing the scale, as
we have just described, as a function of $\phi_c$ ({\it i.e.}
giving different scales for different values of the field) is usually
done to optimize the validity of the perturbative expansion,
{\it i.e.} minimizing the value of radiative corrections to the
effective potential around the minimum of the field. A very
interesting result obtained in Ref.~\cite{IEP} is: {\it The RGE
improved effective potential exact up to (next-to-leading)$^L$
log order
\footnote{The convention is (next-to-leading)$^0\equiv
$leading, {\it i.e.} $L=0$. For $L=1$ the potential is exact to
next-to-leading log.}
is obtained using the L-loop effective potential and
the (L+1)-loop RGE $\beta$-functions.}

\section{Field Theory at Finite Temperature}

The formalism used in conventional quantum field theory is 
suitable to
describe observables ({\it e.g.} cross-sections) measured in
empty space-time, as particle interactions in an accelerator.
However, in the early stages of the universe, at high
temperature, the environment had a non-negligible matter and
radiation density, making the hypotheses of
conventional field theories impracticable. For that reason, under those
circumstances, the methods of conventional field theories are no
longer in use, and should be replaced by others, closer to
thermodynamics, where the background state is a thermal bath. 
This field has been called field theory at finite temperature
and it is extremely useful to study all phenomena which happened
in the early universe: phase transitions, inflationary
cosmology, ... Excellent articles~\cite{DJ,W}, 
review articles~\cite{B,LW,HPA} and
textbooks~\cite{Kapusta} exist which discuss different aspects of
these issues. In this section we will review the main methods
which will be useful for the theory of phase transitions at
finite temperature.

\subsection{Grand-canonical ensemble}

In this section we shall give some definitions borrowed from
thermodynamics and statistical mechanics. The {\bf microcanonical
ensemble} is used to describe an isolated system with fixed energy
$E$, particle number $N$ and volume $V$. The {\bf canonical ensemble}
describes a system in contact with a heat reservoir at temperature $T$:
the energy can be exchanged between them and $T$, $N$ and $V$ are fixed.
Finally, in the {\bf grand canonical ensemble} the system can exchange
energy and particles with the reservoir: $T$, $V$ and the chemical potentials
are fixed. 

Consider now a dynamical
system characterized by a hamiltonian~\footnote{All operators
will be considered in the Heisenberg picture.} $H$ and a set of
conserved (mutually commuting) charges $Q_A$. The equilibrium
state of the system at rest in the large volume $V$ is described
by the {\bf grand-canonical density operator}
\begin{equation}
\label{grandcan}
\rho=\exp(-\Phi)\exp\left\{-\sum_A \alpha_A Q_A - \beta H \right\}
\end{equation}
where $\Phi \equiv \log Tr \exp\left\{-\sum_A \alpha_A Q_A - \beta H
\right\}$ is called the Massieu function (Legendre transform of the
entropy), $\alpha_A$ and $\beta$ are Lagrange multipliers
given by $\beta = T^{-1}$, $\alpha_A =  - \beta \mu_A$,
$T$ is the temperature and $\mu_A$ are the chemical potentials.

Using (\ref{grandcan}) one defines the {\bf grand canonical
average} of an arbitrary operator ${\cal O}$, as
\begin{equation}
\label{average}
\langle {\cal O} \rangle \equiv Tr({\cal O} \rho)
\end{equation}
satisfying the property $\langle {\bf 1}\rangle = 1$.

In the following of this section we will always consider the
case of zero chemical potential. It will be re-introduced when
necessary. 

\subsection{Generating functionals}

We will start considering the case
of a real scalar field $\phi(x)$, carrying no charges
($\mu_A=0$), with hamiltonian $H$, {\it i.e.}
\begin{equation}
\label{field}
\phi(x)=e^{itH}\phi(0,\vec{x})e^{-itH}
\end{equation}
where the time $x^0=t$ is analytically continued to the complex
plane. 

We define the thermal Green function as the grand canonical
average of the ordered product of the $n$ field operators
\begin{equation}
\label{thgreen}
G^{(C)}(x_1,\ldots,x_n) \equiv \langle T_C
\phi(x_1),\ldots,\phi(x_n) \rangle
\end{equation}
where the $T_C$ ordering means that fields should be ordered
along the path $C$ in the complex $t$-plane. For instance the
product of two fields is defined as,
\begin{equation}
\label{ordproduct}
T_C\phi(x)\phi(y)=\theta_C(x^0-y^0)\phi(x)\phi(y)+
\theta_C(y^0-x^0)\phi(y)\phi(x)
\end{equation}

If we parameterize
$C$ as $t=z(\tau)$, where $\tau$ is a real parameter, $T_C$
ordering means standard ordering along $\tau$. Therefore the
step and delta functions can be given as
$\theta_C(t) =  \theta(\tau)$, 
$\delta_C(t) =  \left( \partial z/ \partial
\tau\right)^{-1} \delta(\tau)$.

The rules of the functional formalism can be applied as usual,
with the prescription
$\delta j(y)/ \delta
j(x)=\delta_C(x^0-y^0)\delta^{(3)}(\vec{x}-\vec{y})$, 
and the generating functional $Z^{\beta}[j]$ for the full Green
functions,
\begin{equation}
\label{zgreen}
Z^{\beta}[j]= \sum^{\infty}_{n=0} \frac{i^n}{n!} \int_C d^4x_1 \ldots
d^4x_n j(x_1)\ldots j(x_n) G^{(C)}(x_1, \ldots,x_n)
\end{equation}
can also be written as,
\begin{equation}
\label{thgenfunct}
Z^{\beta}[j]=\left\langle T_C \exp\left\{i\int_C d^4 x j(x) \phi(x)
\right\} \right\rangle
\end{equation}
which is normalized to $Z^{\beta}[0]=\langle {\bf 1} \rangle = 1$,
as in (\ref{average}), and
where the integral along $t$ is supposed to follow the path C
in the complex plane.

Similarly, the generating functional for connected Green
functions $W^{\beta}[j]$ is defined as 
$Z^{\beta}[j] \equiv \exp\{iW^{\beta}[j]\}$,
and the generating functional for 1PI Green functions
$\Gamma^{\beta}[\overline{\phi}]$, by the
Legendre transformation,
\begin{equation}
\Gamma^{\beta}[\overline{\phi}]=W^{\beta}[j]-\int_C d^4 x \frac{\delta
W^{\beta}[j]}{\delta j(x)} j(x)
\label{theffaction}
\end{equation}
where the current $j(x)$ is eliminated in favor of the classical
field $\overline{\phi}(x)$ as
$\overline{\phi}(x)=\delta W^{\beta}[j]/\delta j(x)$. 
It follows that
$\delta \Gamma^{\beta}[\overline{\phi}] /\delta
\overline{\phi}(x)=-j(x)$, 
and $\overline{\phi}(x)=\langle \phi(x) \rangle$
is the grand canonical average of the field $\phi(x)$.

Symmetry violation is signaled by 
\begin{equation}\left.
\frac{\delta \Gamma^{\beta}[\overline{\phi}]}{\delta \overline{\phi}}
\right|_{j=0}=0 
\label{thvacuum}
\end{equation}
for a value of the field different from zero.

As in field theory at zero temperature, in a
translationally invariant theory $\overline{\phi}(x)=\phi_c$ is
a constant. In this case, by removing the overall factor of
space-time volume arising in each term of
$\Gamma^{\beta}[\phi_c]$, we can define the effective potential
at finite temperature as,
\begin{equation}
\Gamma^{\beta}[\phi_c]=-\int d^4 x V^{\beta}_{\rm eff}(\phi_c)
\label{theffpotdef}
\end{equation}
and symmetry breaking occurs when
\begin{equation}
\label{nosymm}
\frac{\partial V^{\beta}_{\rm eff}(\phi_c)}{\partial \phi_c}=0 
\end{equation}
for $\phi_c \ne 0$.

\subsection{Green functions}

\subsubsection{Scalar fields}

Not all the contours are allowed if we require Green functions
to be analytic with respect to $t$. Using (\ref{ordproduct}) we
can write the two-point Green function as,
\begin{equation}
\label{twogreen}
G^{(C)}(x-y)=\theta_C(x^0-y^0)G_{+}(x-y)+\theta_C(y^0-x^0)G_{-}(x-y)
\end{equation}
where
\begin{equation}
\label{redgreen}
G_{+}(x-y)=\langle \phi(x) \phi(y) \rangle, \quad
G_{-}(x-y)=G_{+}(y-x) 
\end{equation}

Now, take the complete set of states $|n\rangle$ with
eigenvalues $E_n$:
$H |n \rangle = E_n |n\rangle$.
One can readily compute (\ref{redgreen}) at the point
$\vec{x}=\vec{y}=0$ as
\begin{equation}
\label{greencont}
G_{+}(x^0-y^0)=e^{-\Phi}\sum_{m,n} \left|\langle m| \phi(0) 
|n\rangle
\right|^2 e^{-iE_n(x^0-y^0)}e^{iE_m(x^0-y^0+i\beta)} 
\end{equation}
so that the convergence of the sum implies that
$
-\beta \le Im(x^0-y^0) \le 0
$
which requires $\theta_C(x^0-y^0)=0$ for $Im(x^0-y^0)>0$.
From (\ref{redgreen}) it follows that the similar property for
the convergence of $G_{-}(x^0-y^0)$ is that
$
0 \le Im(x^0-y^0) \le \beta
$,
which requires $\theta_C(y^0-x^0)=0$ for $Im(x^0-y^0)<0$,
and the final condition for the convergence of the complete
Green function on the strip
\begin{equation}
\label{strip}
-\beta \le Im(x^0-y^0) \le \beta
\end{equation}
is that we define the function $\theta_C(t)$ such that
$
\theta_C(t)=0\ \ {\rm for}\ \ Im(t)> 0.
$
The latter condition implies
that $C$ must be such that {\it a point moving along it has a
monotonically decreasing or constant imaginary part}.

A very important periodicity relation affecting Green functions
can be easily deduced from the very definition of $G_{+}(x)$ and
$G_{-}(x)$, Eq.~(\ref{redgreen}). By using the definition of the
grand canonical average and the cyclic permutation property of
the trace of a product of operators, it can be easily deduced,
\begin{equation}
\label{kms}
G_{+}(t-i\beta,\vec{x})=G_{-}(t,\vec{x})
\end{equation}
which is known as the Kubo-Martin-Schwinger relation~\cite{KMS}.

We can now compute the two-point Green function (\ref{twogreen})
for a free scalar field,
\begin{equation}
\label{scfielddec}
\phi(x)=\int \frac{d^3 p}{(2\pi)^{3/2} (2\omega_p)^{1/2}}
\left[a(p)e^{-ipx}+a^{\dagger}(p) e^{ipx}\right]
\end{equation}
where $\omega_p=\sqrt{\vec{p}^{\ 2}+m^2}$,
which satisfies the equation 
\begin{equation}
\label{greeneq}
\left[ \partial^{\mu}\partial_{\mu} + m^2 \right]G^{(C)}(x-y)
=-i\delta_C(x-y) \equiv
-i\delta_C(x^0-y^0)\delta^{(3)}(\vec{x}-\vec{y}) 
\end{equation}

Using the time derivative of (\ref{scfielddec}),
and the equal time commutation relation,
\begin{equation}
\label{etime}
\left[\phi(t,\vec{x}),\dot{\phi}(t,\vec{y})\right]=i\delta^{(3)}
(\vec{x}-\vec{y})
\end{equation}
one easily obtains the commutation relation for creation and
annihilation operators,
\begin{equation}
\label{commut}
\left[a(p),a^{\dagger}(k)\right]=\delta^{(3)}(\vec{p}-\vec{k}) 
\end{equation}
and defining the Hamiltonian of the field as,
\begin{equation}
\label{hamilton}
H=\int\frac{d^3 p}{(2\pi)^3} \omega_p a^{\dagger}(p) a(p)
\end{equation}
one can obtain, using (\ref{commut}) the thermodynamical averages,
\begin{eqnarray}
\label{averages}
\langle a^{\dagger}(p) a(k) \rangle = n_B(\omega_p)
\delta^{(3)}(\vec{p}-\vec{k}) \\
\langle a(p) a^{\dagger}(k) \rangle =[1+ n_B(\omega_p)]
\delta^{(3)}(\vec{p}-\vec{k}) \nonumber
\end{eqnarray}
where $n_B(\omega)$ is the Bose distribution function,
\begin{equation}
\label{bose}
n_B(\omega)=\frac{1}{e^{\beta\omega}-1}
\end{equation}

We will give here a simplified derivation of expression
(\ref{averages}). Consider the simpler example of a quantum
mechanical state occupied by bosons of the {\bf same} energy
$\omega$. There may be any number of bosons in that state and no
interaction between the particles: we will denote that state by
$|n\rangle$. The set $\{|n\rangle\}$ is complete.
Creation and annihilation operators are denoted by
$a^{\dagger}$ and $a$, respectively. They act on the states
$|n\rangle$ as,
$
a^{\dagger}|n\rangle=\sqrt{n+1}|n+1\rangle
$
and
$
a|n\rangle=\sqrt{n}|n-1\rangle,
$
and satisfy the commutation relation,
$\left[a,a^{\dagger}\right]=1$.
The hamiltonian and number operators are defined as
$
H=\omega N
$
and 
$
N=a^{\dagger}a
$, 
with eigenvalues $\omega n$ and $n$, respectively. 

It is very easy to compute now $\langle a^{\dagger}a \rangle$ and $\langle
a a^{\dagger} \rangle$ as in (\ref{averages}) using the
completeness of $\{|n\rangle\}$. In particular,
$$
Tr (e^{-\beta H})=\sum_{n=0}^{\infty} \langle n| e^{-\beta H}
|n\rangle =\sum_{n=0}^{\infty} e^{-\beta\omega
n}=\frac{1}{1-e^{-\beta \omega}}
$$
and
$$
Tr (e^{-\beta H} a^{\dagger}
a)=\sum_{n=0}^{\infty}ne^{-\beta\omega
n}=\frac{e^{-\beta\omega}}{(1-e^{-\beta\omega})^2} 
$$
from where
$\langle a^{\dagger} a \rangle = n_B(\omega)$,
and
$\langle a a^{\dagger} \rangle =1+ n_B(\omega)$,
as we wanted to prove.

Using now (\ref{averages}) we can cast the two-point Green function as,
\begin{equation}
\label{twogreensc}
G^{(C)}(x-y)=\int \frac{d^4 p}{(2\pi)^4} \rho(p) e^{-ip(x-y)}
\left[\theta_C(x^0-y^0)+n_B(p^0)\right] 
\end{equation}
where the function $\rho(p)$ is defined by
$\rho(p)=2\pi [\theta(p^0)-\theta(-p^0)]\delta(p^2-m^2)$.
Now the particular value of the Green function
(\ref{twogreensc}) depends on the chosen contour $C$. We will
show later on two particular contours giving rise to the
so-called imaginary and real time formalisms. Before coming to
them we will describe how the previous formulae apply to the
case of fermion fields.

\subsubsection{Fermion fields}

We will replace here (\ref{twogreen}) and (\ref{redgreen}) by,
\begin{equation}
\label{twogreenf}
S^{(C)}_{\alpha \beta}(x-y)   
\equiv \langle T_C\psi_{\alpha}(x) \overline{\psi}_{\beta}(y)
\rangle 
=  \theta_C(x^0-y^0)S^+_{\alpha \beta} - \theta_C(y^0-x^0)
S^-_{\alpha \beta} \nonumber
\end{equation}
where $\alpha$ and $\beta$ are spinor indices, and
\begin{equation}
\label{redgreenf}
S^+_{\alpha\beta}(x-y)=\langle
\psi_{\alpha}(x)\overline{\psi}_{\beta}(y) \rangle
\end{equation}
are the reduced Green function, which satisfy the
Kubo-Martin-Schwinger relation,
\begin{equation}
\label{kmsf}
S^+_{\alpha\beta}(t-i\beta,\vec{x})=-S^-_{\alpha\beta}(t,\vec{x})
\end{equation}

The calculation of the two-Green function for a free fermion
field, satisfying the equation
\begin{equation}
\label{greeneqf}
\left( i\gamma \cdot \partial -m\right)_{\alpha
\sigma}S^{(C)}_{\sigma \beta}(x-y)=i \delta_C(x-y)
\delta_{\alpha \beta}
\end{equation}
follows lines similar to Eqs.~(\ref{scfielddec}) to (\ref{twogreensc}).
In particular, one can define a Green function $S^{(C)}$ as
\begin{equation}
\label{defscalar}
S^{(C)}_{\alpha\beta}(x-y) \equiv (i\gamma \cdot
\partial+m)_{\alpha \beta} S^{(C)}(x-y)
\end{equation}
where $S^{(C)}(x-y)$ satisfies the Klein-Gordon propagator equation
(\ref{greeneq}). One can obtain for $S^{(C)}$ the expression, 
\begin{equation}
\label{twogreenfer}
S^{(C)}(x-y)=\int \frac{d^4 p}{(2\pi)^4} \rho(p) e^{-ip(x-y)}
\left[\theta_C(x^0-y^0)-n_F(p^0)\right] 
\end{equation}
where $n_F(\omega)$ is the Fermi distribution function
\begin{equation}
\label{fermi}
n_F(\omega)=\frac{1}{e^{\beta\omega}+1}.
\end{equation}

Eq.~(\ref{fermi}) can be derived similarly to 
(\ref{bose}) as the mean number of
fermions for a Fermi gas. This time the Pauli exclusion
principle forbids more than one fermion occupying a single
state, so that only the states $|0\rangle$ and $|1\rangle$
exist. They are acted on by creation and annihilation operators
$b^{\dagger}$ and $b$, respectively as:
$b^{\dagger}|0\rangle=|1\rangle,$
$b^{\dagger}|1\rangle=0,$
$b|0\rangle=0,$
$b|1\rangle=|0\rangle,$
and satisfy anticommutation rules,
$\left\{b,b^{\dagger}\right\}=1$.
Defining the hamiltonian and number operators as $H=\omega N$
and $N=b^{\dagger}b$, we can compute now the statistical
averages of $\langle b^{\dagger}b \rangle$ and $\langle b
b^{\dagger} \rangle$ using the completeness of $\{|n\rangle\}$.
$$
Tr (e^{-\beta H})=\sum_{n=0}^{1} \langle n| e^{-\beta H}
|n\rangle =\sum_{n=0}^{1} e^{-\beta\omega
n}=1+e^{-\beta \omega}
$$
and
$$
Tr (e^{-\beta H} b^{\dagger}
b)=\sum_{n=0}^{1}ne^{-\beta\omega
n}=e^{-\beta\omega} 
$$
from where
$\langle b^{\dagger} b \rangle = n_F(\omega)$,
and
$\langle b b^{\dagger} \rangle =1-n_F(\omega)$,
as we wanted to prove.

\subsection{Imaginary time formalism}

The calculation of the propagators in the previous sections
depends on the chosen path $C$ going from an initial arbitrary
time $t$ to $t-i\beta$, provided by the Kubo-Martin-Schwinger
periodicity properties (\ref{kms}) and (\ref{kmsf}) of Green
functions. The simplest path is to take a straight line along the
imaginary axis $t=-i\tau$. It is called Matsubara contour, since
Matsubara~\cite{M} was the first to set up a perturbation theory
based upon this contour. In that case
$\delta_C(t)=i\delta(\tau)$.

The two-point Green functions for scalar (\ref{twogreensc}) and
fermion (\ref{twogreenfer}) fields can be written as,
\begin{equation}
\label{twogreenim}
G(\tau,\vec{x})=\int \frac{d^4 p}{(2\pi)^4} \rho(p)
e^{i\vec{p}\vec{x}} e^{-\tau p^0}
\left[\theta(\tau)+\eta n(p^0)\right] 
\end{equation}
where the symbol $\eta$ stands as:
$\eta_B=1$ ($\eta_F=-1$) for bosons (for fermions). 
Analogously,
$n(p^0)$ stands either for $n_B(p^0)$, as given by (\ref{bose})
for bosons, or $n_F(p^0)$, as given by (\ref{fermi}) for
fermions. It can be defined as a function of $\eta$ as,
\begin{equation}
\label{bosefermi}
n(\omega)=\frac{1}{e^{\beta \omega}-\eta}
\end{equation}

The Green function (\ref{twogreenim}) can be
decomposed as in (\ref{twogreen})
\begin{equation}
\label{twodecim}
G(\tau,\vec{x})=G_{+}(\tau,\vec{x})\theta(\tau)+
G_{-}(\tau,\vec{x})\theta(-\tau)
\end{equation}

Using now the Kubo-Martin-Schwinger relations, Eqs.~(\ref{kms})
and (\ref{kmsf}), we can write
$G(\tau+\beta) =  \eta G(\tau) \ {\rm for}\ -\beta\le\tau\le 0$,
$G(\tau-\beta)  =  \eta G(\tau) \ {\rm for}\ \ 0\le\tau\le \beta$,
which means that the propagator for bosons (fermions) is
periodic (antiperiodic) in the {\it time} variable $\tau$, with
period $\beta$.

It follows that the Fourier transform of (\ref{twogreenim})
\begin{equation}
\label{fourierprop}
\widetilde{G}(\omega_n,\vec{p})=
\int_{\alpha-\beta}^{\alpha} d\tau \int d^3 x
e^{i\omega_n \tau-i\vec{x}\vec{p}} G(\tau,\vec{x})
\end{equation}
(where $0 \le \alpha \le \beta$) is independent of $\alpha$ and
the discrete frequencies satisfy the relation
$\eta e^{i\omega_n \beta}=1$,
{\it i.e.}
$\omega_n=2n\pi\beta^{-1}$
for bosons, and
$\omega_n=(2n+1)\pi\beta^{-1}$
for fermions.

Inserting now (\ref{twogreenim}) into (\ref{fourierprop}) we can
obtain the propagator in momentum space $\widetilde{G}$
\begin{equation}
\label{promomfinal}
\widetilde{G}(\omega_n,\vec{p})=\frac{1}{\vec{p}^{\ 2}+m^2+\omega_n^2}.
\end{equation}

We can now define the euclidean propagator,
$\Delta(-i\tau,\vec{x})$, by
\begin{equation}
\label{propeuc}
G(\tau,\vec{x})=i\Delta(-i\tau,\vec{x})
\end{equation}
where $G(\tau,\vec{x})$ is the propagator defined in
(\ref{twogreenim}). Therefore, using (\ref{promomfinal}), we can
write the inverse Fourier transformation,
\begin{equation}
\label{inveuc}
\Delta(x)=\frac{1}{\beta}\sum_{n=-\infty}^{\infty}\int\frac{d^3
p}{(2\pi)^3} e^{-i\omega_n\tau+i\vec{p}\vec{x}}
\frac{-i}{\vec{p}^{\ 2}+m^2+\omega_n^2}
\end{equation}
where the Matsubara frequencies $\omega_n$ are either
for bosons or for fermions. 

From (\ref{inveuc}) one can deduce the Feynman rules for the
different fields in the imaginary time formalism. We can
summarize them in the following way:
\begin{eqnarray}
\label{feynmanim}
{\rm Boson\ propagator}&:& \frac{i}{p^2-m^2};\
p^{\mu}=[2ni\pi\beta^{-1},\vec{p}\ ]\nonumber \\
{\rm Fermion\ propagator}&:& \frac{i}{\gamma \cdot p-m};\ p^{\mu}=
[(2n+1)i\pi\beta^{-1},\vec{p}\ ] \nonumber \\
{\rm Loop \ integral}&:&
\frac{i}{\beta}\sum_{n=-\infty}^{\infty}\int\frac{d^3
p}{(2\pi)^3} \\
{\rm Vertex\ function}& : & -i\beta (2\pi)^3 \delta_{\sum\omega_i}
\delta^{(3)}(\sum_i \vec{p}_i)
\nonumber
\end{eqnarray}

There is a standard trick to perform infinite summations as in
(\ref{feynmanim}). For the case of bosons we can have frequency
sums as,
\begin{equation}
\label{insum}
\frac{1}{\beta}\sum_{n=-\infty}^{\infty}f(p^0=i\omega_n)
\end{equation}
with $\omega_n=2n\pi\beta^{-1}$. Since the function
$
{\displaystyle \frac{1}{2}\beta\coth(\frac{1}{2}\beta z)}
$
has poles at $z=i\omega_n$ and is analytic and bounded
everywhere else, we can write (\ref{insum}) as,
$$
\frac{1}{2\pi i \beta}\int_{\gamma} dz
f(z)\frac{\beta}{2}\coth(\frac{1}{2}\beta z)
$$
where the contour $\gamma$ encircles anticlockwise all the previous
poles of the imaginary axis. We are assuming that $f(z)$ does
not have singularities along the imaginary axis (otherwise the
previous expression is obviously not correct). The contour
$\gamma$ can be deformed to a new contour consisting in two
straight lines: the first 
one starting at $-i\infty+\epsilon$ and going to
$i\infty+\epsilon$, and the second one starting at
$i\infty-\epsilon$ and ending at $-i\infty-\epsilon$.
Rearranging the exponentials in the hyperbolic cotangent one can
write the previous expression as,
$$
\frac{1}{2\pi i}\int_{-i\infty}^{i\infty}
dz\frac{1}{2}[f(z)+f(-z)] +\frac{1}{2\pi
i}\int_{-i\infty+\epsilon}^{i\infty+\epsilon}dz[f(z)+f(-z)]\frac{1}{e^{\beta
z}-1} 
$$
and the contour of the second integral can be deformed to a
contour $C$ which encircles clockwise all singularities of the
functions $f(z)$ and $f(-z)$ in the right half plane. Therefore
we can write (\ref{insum}) as
\begin{equation}
\label{insumbos}
\frac{1}{\beta}\sum_{n=-\infty}^{\infty}f(p^0=i\omega_n)=
\int_{-i\infty}^{i\infty} \frac{dz}{4\pi
i}[f(z)+f(-z)]+\int_{C}\frac{dz}{2\pi i} n_B(z)[f(z)+f(-z)]
\end{equation}
where $n_B(z)$ is the Bose distribution function (\ref{bose}).

Eq.~(\ref{insumbos}) can be generalized for both bosons and
fermions as,
\begin{equation}
\label{sumbosfer}
\frac{1}{\beta}\sum_{n=-\infty}^{\infty}f(p^0=i\omega_n)=
\int_{-i\infty}^{i\infty} \frac{dz}{4\pi
i}[f(z)+f(-z)]+\eta \int_{C}\frac{dz}{2\pi i} n(z)[f(z)+f(-z)]
\end{equation}
where the distribution functions $n(z)$ are defined in 
(\ref{bosefermi}). Eq.~(\ref{sumbosfer}) 
shows that the frequency sum naturally
separates into a $T$ independent piece, which should coincide
with the similar quantity computed in the field theory at zero
temperature, and a $T$ dependent piece which vanishes in the
limit $T\rightarrow 0$, {\it i.e.} $\beta \rightarrow \infty$.

\subsection{Real time formalism}

\begin{figure}[htb]
\epsfxsize=10truecm
\centerline{\epsfbox{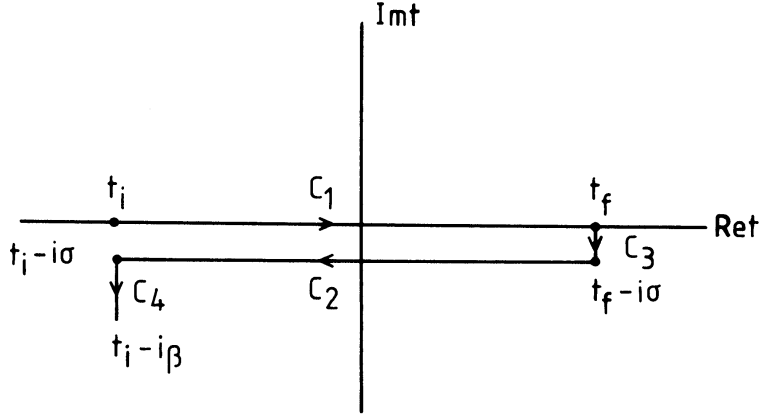}}
\caption[0]{Contour used in the real time formalism.}
\label{realcontour}
\end{figure}

The obvious disadvantage of the imaginary time formalism is to
compute Green functions along imaginary time, so that going to
the real time has to be done through a process of analytic
continuation. However, a direct evaluation of Green function in
the real time is possible by a judicious choice of the contour
$C$ in (\ref{thgreen}). The family of such real time contours is
depicted in Fig.~\ref{realcontour}
where the contour $C$ is
${\displaystyle C=C_1\bigcup C_2 \bigcup C_3 \bigcup C_4}$
where $C_1$ goes from the initial time $t_i$ to the final time
$t_f$, $C_3$ from $t_f$ to $t_f-i\sigma$, with
$0\le\sigma\le\beta$, $C_2$ from $t_f-i\sigma$ to $t_i-i\sigma$,
and $C_4$ from $t_i-i\sigma$ to $t_i-i\beta$. Different choices
of $\sigma$ lead to an equivalence class of quantum field
theories at finite temperature~\cite{MATSU}. For instance the
choice $\sigma=0$ leads to the Keldysh perturbation expansion~\cite{KEL}, 
while the choice 
$\sigma=\beta/2$
is the preferred one to compute Green functions.

Computing the Green function for scalar (\ref{twogreensc}) and
fermion (\ref{twogreenfer}) fields taking path $C$ 
is a matter of calculation, as we did
for the imaginary time formalism in
(\ref{twogreenim})-(\ref{promomfinal}). One can prove that the
contribution from the contours $C_3$ and $C_4$ can be neglected~\cite{LW,EA}. 
Therefore, for the propagator between $x^0$ and
$y^0$ there are four possibilities depending on whether they are
on $C_1$ or $C_2$. Correspondingly, there are four propagators
which are labeled by (11), (12), (21) and (22).

Making the choice $\sigma=\beta/2$, the propagators for scalar
fields (\ref{twogreensc}) can be written, in momentum space, as 
\begin{equation}
\label{scalreal}
G(p)\equiv \left(
\begin{array}{cc} G^{(11)}(p) & G^{(12)}(p) \\
                  G^{(21)}(p) & G^{(22)}(p)
\end{array} \right)=M_B(\beta,p)\left( 
\begin{array}{cc} \Delta(p) & 0 \\
                  0 & \Delta^*(p)
\end{array} \right)M_B(\beta,p)
\end{equation}
where $\Delta(p)$ is the boson propagator at zero temperature,
and the matrix $M_B(\beta,p)$ is given by,
\begin{equation}
\label{matrixrt}
M_B(\beta,p)=\left(
\begin{array}{cc} 
\cosh \theta(p) & \sinh \theta(p) \\
\sinh \theta(p) & \cosh \theta(p)
\end{array}
\right)
\end{equation}
where
\begin{eqnarray}
\label{sinhcosh}
\sinh \theta(p) & = & e^{-\beta \omega_p/2}\left(1-e^{-\beta \omega_p}
\right)^{-1/2} \nonumber \\
\cosh \theta(p) & = & \left(1-e^{-\beta \omega_p}\right)^{-1/2}
\end{eqnarray}

Using now (\ref{scalreal}), (\ref{matrixrt}), (\ref{sinhcosh}), 
one can easily write the expression for the
four bosonic propagators, as
\begin{eqnarray}
\label{realpropbos}
G^{(11)}(p)& =&\Delta(p)+2\pi n_B(\omega_p)\delta(p^2-m^2) \nonumber \\
G^{(22)}(p) & = & G^{(11)*} \\
G^{(12)}&=& 2\pi e^{\beta \omega_p/2}n_B(\omega_p) \delta(p^2-m^2) \nonumber
\\
G^{(21)}&=& G^{(12)} \nonumber
\end{eqnarray}

Similarly, the propagators for fermion fields can be written as  
\begin{eqnarray}
\label{ferreal}
S(p)_{\alpha \beta} & \equiv & \left(
\begin{array}{cc} S^{(11)}_{\alpha \beta} (p) & S^{(12)}_{\alpha
\beta} (p) \\
S^{(21)}_{\alpha \beta}(p) & S^{(22)}_{\alpha \beta}(p)
\end{array} \right) \\
&=& M_F(\beta,p)\left( 
\begin{array}{cc}(\gamma\cdot p+m)_{\alpha \beta} \Delta(p) & 0 \\
 0 & (\gamma\cdot p+m)_{\alpha \beta}\Delta^*(p)
\end{array} \right) M_F(\beta,p) \nonumber
\end{eqnarray}
where the matrix $M_F(\beta,p)$ is,
\begin{equation}
\label{matrixrtf}
M_F(\beta,p)=\left(
\begin{array}{cc} 
\cos \theta(p) & \sin \theta(p) \\
\sin \theta(p) & \cos \theta(p)
\end{array}
\right)
\end{equation}
with
\begin{eqnarray}
\label{sincos}
\sin \theta(p) & = & e^{-\beta \omega_p/2}\left(1+e^{-\beta \omega_p}
\right)^{-1/2} \nonumber \\
\cos \theta(p) & = &[\theta(p^0)-\theta(-p^0)] 
\left(1+e^{-\beta \omega_p}\right)^{-1/2}
\end{eqnarray}

In the same way, using now (\ref{ferreal}), (\ref{matrixrtf}),
and  (\ref{sincos}) one can easily write the expression for the
four fermionic propagators, as
\begin{eqnarray}
\label{realpropfer}
S^{(11)}(p)& =&(\gamma \cdot p + m) \left( \Delta(p)-
2\pi n_F(\omega_p)\delta(p^2-m^2)\right) \nonumber \\
S^{(22)}(p) & = & S^{(11)*} \\
S^{(12)}&=&-2 \pi (\gamma \cdot p + m) [\theta(p^0)-\theta(-p^0)]
e^{\beta \omega_p/2} n_F(\omega_p) \delta(p^2-m^2) \nonumber
\\
S^{(21)}&=&-S^{(12)} \nonumber
\end{eqnarray}

As one can see from (\ref{realpropbos}) and (\ref{realpropfer}),
the main feature of the real time formalism is that the
propagators come in two terms: one which is the same as in the zero
temperature field theory, and a second one where all the
temperature dependence is contained. This is welcome. However
the propagators (12), (21) and (22) are unphysical since one of
their time arguments has an imaginary component. They are
required for the consistency of the theory. The only physical
propagator is the (11) component in (\ref{realpropbos}) and
(\ref{realpropfer}). 

Now the Feynman rules in the real time formalism are very similar
to those in the zero temperature field theory. 
In fact all diagrams have
the same topology as in the zero temperature field theory and
the same symmetry factors. However, associated to every field
there are two possible vertices, 1 and 2, and four possible
propagators, (11), (12), (21) and (22) connecting them. 
All of them have to be considered for the consistency of the
theory. In the Feynman rules, type 2 vertices are hermitian
conjugate with respect to type 1 vertices. The golden rule is
that: {\it physical legs must always be attached to type 1
vertices}. Apart from the previous prescription, one must sum
over all the configurations of type 1 and type 2 vertices, and
use the propagator $G^{(ab)}$ or $S^{(ab)}$ to connect vertex
$a$ with vertex $b$.

There is now a general agreement in the sense that the imaginary
time formalism and the real time formalism should give the same
physical answer~\cite{RVI}. Using one or the other is sometimes
a matter of taste, though in some cases the choice is dictated
by calculational simplicity depending on the physical 
problem one is dealing with.

\section{The effective potential at finite temperature}

In this section we will construct the (one-loop) effective
potential at finite temperature, using all the tools provided in
the previous sections. As we will see, in particular, the
effective potential at finite temperature contains the effective
potential at zero temperature computed in section 1. The
usefulness of this construction is addressed to the theory of
phase transitions at finite temperature. The latter being
essential for the understanding of phenomena as: inflation,
baryon asymmetry generation, quark-gluon plasma transition in
QCD,... We will compare different methods leading to the same
result, including the use of both the imaginary and the real
time formalisms. This exercise can be useful mainly to face more
complicated problems than those which will be developed in this
course. 

\subsection{Scalar fields}

We will consider here the simplest model of one self-interacting
scalar fields described by the lagrangian (\ref{scallag}) and
(\ref{scalpot}). We have to compute the diagrams contained in
Fig.~\ref{1loopsc} using the Feynman rules described in
(\ref{feynmanim}), for the imaginary time formalism, or in
(\ref{realpropbos}) for the real time formalism. We will write
the result as,
\begin{equation}
\label{potdecomp}
V^{\beta}_{\rm eff}(\phi_c)=V_0(\phi_c)+V_1^{\beta}(\phi_c)
\end{equation}
where $V_0(\phi_c)$ is the tree level potential. 

\subsubsection{Imaginary time formalism}
 
We will compute the diagrams in Fig.~\ref{1loopsc}. Using the
Feynman rules in Eq.~(\ref{feynmanim}), Eq.~(\ref{1loopsca})
translates into,
\begin{equation}
\label{1loopscat}
V^{\beta}_1(\phi_c)=\frac{1}{2\beta} \sum_{n=-\infty}^{\infty}
\int \frac{d^3 p}{(2\pi)^3}\log(\omega_n^2+\omega^2)
\end{equation}
where $\omega_n$ are the bosonic Matsubara frequencies and 
\begin{equation}
\label{omegashifted}
\omega^2=\vec{p}^{\ 2}+m^2(\phi_c)
\end{equation}
$m^2$ being defined in (\ref{shiftedm}).

The sum over $n$ in (\ref{1loopscat}) diverges, but the infinite
part does not depend on $\phi_c$. The finite part, which
contains the $\phi_c$ dependence, can be computed by the
following method~\cite{DJ}. Define,
\begin{equation}
\label{ve}
v(\omega)=\sum_{n=-\infty}^{\infty} \log(\omega_n^2+\omega^2)
\end{equation}
then,
\begin{equation}
\label{dve}
\frac{\partial v}{\partial\omega}=
\sum_{n=-\infty}^{\infty}\frac{2\omega}{\omega_n^2+
\omega^2}
\end{equation}
Using the identity,
\begin{eqnarray}
\label{fy}
f(y)=\sum_{n=1}^{\infty} \frac{y}{y^2+n^2} & = &
-\frac{1}{2y}+\frac{1}{2} \pi \coth \pi y  \\
& = & -\frac{1}{2y}+\frac{\pi}{2}+\pi \frac{e^{-2\pi
y}}{1-e^{-2\pi y}} \nonumber
\end{eqnarray}
with $y=\beta \omega/2\pi$ we obtain,
\begin{equation}
\label{dvefin}
\frac{\partial
v}{\partial\omega}=2\beta\left[\frac{1}{2}+\frac{e^{-\beta
\omega}}{1-e^{-\beta\omega}} \right]
\end{equation}
and 
\begin{equation}
\label{vefin}
v(\omega)=2\beta\left[\frac{w}{2}+\frac{1}{\beta}\log\left(
1-e^{-\beta\omega}\right) \right]+\omega-{\rm independent\ \ terms}
\end{equation}

Substituting finally (\ref{vefin}) into (\ref{1loopscat}) one gets,
\begin{equation}
\label{v1sct}
V_1^{\beta}(\phi_c)=\int \frac{d^3
p}{(2\pi)^3}\left[\frac{\omega}{2}+\frac{1}{\beta}
\log\left(1-e^{-\beta\omega} \right) \right]
\end{equation}

One can easily prove that the first integral in (\ref{v1sct}) is
the one-loop effective potential at zero temperature. For that
we have to prove the identity,
\begin{equation}
\label{identity}
-\frac{i}{2}\int_{-\infty}^{\infty} \frac{dx}{2\pi}
\log(-x^2+\omega^2-i\epsilon) =\frac{\omega}{2}+{\rm constant}
\end{equation}
{\it i.e.}
\begin{equation}
\label{deridentity}
\omega\int_{-\infty}^{\infty}\frac{dx}{2\pi
i}\frac{1}{-x^2+\omega^2-i\epsilon} = \frac{1}{2}
\end{equation}
Integral (\ref{deridentity}) can be performed closing the
integration interval $(-\infty,\infty)$ 
in the complex $x$  plane along a contour going anticlockwise and
picking the pole of the integrand at
$x=-\sqrt{\omega^2-i\epsilon}$ with a residue $1/2\omega$. Using
the residues theorem Eq.~(\ref{deridentity}) can be easily
checked. Now we can use identity (\ref{identity}) to write the
temperature independent part of (\ref{v1sct}) as
\begin{equation}
\label{bind}
\frac{1}{2}\int\frac{d^3p}{(2\pi)^3}\omega
=-\frac{i}{2}\int\frac{d^4p}{(2\pi)^4}
\log(-p_o^2+\omega^2-i\epsilon)
\end{equation}
and, after making the Wick rotation $p^0=ip_E$ in (\ref{bind})
we obtain,
\begin{equation}
\label{bindfin}
\frac{1}{2}\int\frac{d^3p}{(2\pi)^3}\omega
=\frac{1}{2}\int\frac{d^4p}{(2\pi)^4}
\log[p^2+m^2(\phi_c)]
\end{equation}
which is the same result we obtained in the zero temperature
field theory, see Eq.~(\ref{1loopsca}).

Now the temperature dependent part in (\ref{v1sct}) can be
easily written as,
\begin{equation}
\label{bdepfin}
\frac{1}{\beta}\int\frac{d^3 p}{(2\pi)^3}
\log\left(1-e^{-\beta\omega} \right)=\frac{1}{2\pi^2\beta^4}
J_B[m^2(\phi_c)\beta^2] 
\end{equation}
where the thermal bosonic function $J_B$ is defined as,
\begin{equation}
\label{jb}
J_B[m^2\beta^2]=\int_0^{\infty} dx\
x^2\log\left[1-e^{-\sqrt{x^2+\beta^2 m^2}}\right]
\end{equation}

The integral (\ref{jb}) and therefore the thermal bosonic
effective potential admits a high-temperature expansion which
will be very useful for practical applications. It is given by
\begin{eqnarray}
\label{jbexp}
J_B(m^2/T^2) & = & -\frac{\pi^4}{45}+
\frac{\pi^2}{12}\frac{m^2}{T^2}-\frac{\pi}{6}
\left(\frac{m^2}{T^2}\right)^{3/2}-\frac{1}{32}
\frac{m^4}{T^4}\log\frac{m^2}{a_b T^2}  \\
& &
-2\pi^{7/2}\sum_{\ell=1}^{\infty}(-1)^{\ell}\frac{\zeta(2\ell+1)}
{(\ell+1)!}\Gamma\left(\ell+\frac{1}{2}\right)
\left(\frac{m^2}{4\pi^2 T^2}
\right)^{\ell+2} \nonumber
\end{eqnarray}
where $a_b=16\pi^2\exp(3/2-2\gamma_E)$ 
($\log a_b=5.4076$)
and $\zeta$ is the Riemann $\zeta$-function.

There is a very simple way of computing the effective potential:
it consists in {\it computing its derivative 
{\bf in the shifted theory} and then
integrating}! In fact the derivative of the effective potential
$$
\frac{dV_1^{\beta}}{d\phi_c}
$$
is described diagrammatically by the tadpole diagram. 
In fact using the Feynman rules in (\ref{feynmanim}) one can 
easily write for the tadpole the expression,
\begin{equation}
\label{tadsc}
\frac{dV_1^{\beta}}{d\phi_c}=\frac{\lambda\phi_c}{2}\frac{1}{\beta}
\sum_{n=-\infty}^{\infty} \int\frac{d^3 p}{(2\pi)^3}
\frac{1}{\omega_n^2+\omega^2} 
\end{equation}
or, using the expression (\ref{shiftedm}) for $m^2(\phi_c)$,
\begin{equation}
\label{tadsc2}
\frac{dV_1^{\beta}}{dm^2(\phi_c)}=\frac{1}{2\beta}
\sum_{n=-\infty}^{\infty} \int\frac{d^3 p}{(2\pi)^3}
\frac{1}{\omega_n^2+\omega^2} 
\end{equation}

Now we can perform the infinite sum in (\ref{tadsc2}) using the
result in Eq.~(\ref{insumbos}) with a function $f$ defined as,
\begin{equation}
\label{fofz}
f(z)=\frac{1}{\omega^2-z^2}
\end{equation}
and obtain for the tadpole (\ref{tadsc2}) the result
\begin{equation}
\label{tadscfin}
\frac{dV_1^{\beta}}{dm^2(\phi_c)}=\int \frac{d^3 p}{(2\pi)^3}
\left\{
\frac{1}{2} \int_{-i\infty}^{i\infty}\frac{dz}{2\pi
i}\frac{1}{\omega^2-z^2} +\int_C\frac{dz}{2\pi
i}\frac{1}{e^{\beta z}-1}\frac{1}{\omega^2-z^2}
\right\}
\end{equation}
The first term in (\ref{tadscfin}) gives the $\beta$-independent 
part of the tadpole contribution as,
\begin{equation}
\label{bindsc}
\frac{1}{2} \int_{-i\infty}^{i\infty}\frac{dz}{2\pi
i}\frac{1}{\omega^2-z^2} 
\end{equation}
We can now close the integration contour of (\ref{bindsc})
anticlockwise and pick the pole of (\ref{fofz}) at $z=-\omega$
with a residue $1/2\omega$. The result of (\ref{bindsc}) is
\begin{equation}
\label{bindsc2}
\frac{1}{4\omega}
\end{equation}
The second term in (\ref{tadscfin}) gives the $\beta$-dependent
part of the tadpole contribution. Here the integration contour
encircles the pole at $z=\omega$ with a residue
\begin{equation}
\label{bdepsc}
-\frac{1}{2\omega}\frac{1}{e^{\beta\omega}-1}
\end{equation}
Adding (\ref{bindsc2}) and (\ref{bdepsc}) we obtain for the
tadpole the final expression,
\begin{equation}
\label{tadpolescfin}
\frac{dV_1(\phi_c)}{dm^2(\phi_c)}=\frac{1}{2}\int\frac{d^3
p}{(2\pi)^3}
\left[\frac{1}{2\omega}+\frac{1}{\omega}\frac{1}{e^{\beta\omega}-1}
\right]
\end{equation}
Now, integration of (\ref{tadpolescfin}) with respect to
$m^2(\phi_c)$ leads to the expression (\ref{v1sct}) for the
thermal effective potential and, therefore, to the final
expression given by (\ref{bindfin}) and (\ref{bdepfin}).

\subsubsection{Real time formalism}

As we will see in this section, the final result for the
effective potential (\ref{v1sct}) can be also obtained using the
real time formalism. Let us compute the tadpole diagram. 
Since physical legs must be attached to type
1 vertices, the vertex in the tadpole must be
considered of type 1, and the propagator circulating around the
loop has to be considered as a (11) propagator.
Application of the Feynman rules
(\ref{realpropbos}) to the tadpole diagram leads to the expression 
\footnote{We are replacing in (\ref{realpropbos}) the value of
$\omega_p$ by the corresponding value 
$\omega$ given by (\ref{omegashifted}) in the shifted theory.}
\begin{equation}
\label{tadscr}
\frac{dV_1^{\beta}}{d\phi_c}=\frac{\lambda\phi_c}{2}\int\frac{d^4
p}{(2\pi)^4}\left[\frac{i}{p^2-m^2(\phi_c)+i\epsilon}+2\pi
n_B(\omega)\delta(p^2-m^2(\phi_c)) \right]
\end{equation}
or, using as before the expression (\ref{shiftedm}) for $m^2(\phi_c)$,
\begin{equation}
\label{tadscr2}
\frac{dV_1^{\beta}}{dm^2(\phi_c)}=\frac{1}{2}\int\frac{d^4
p}{(2\pi)^4}\left[\frac{-i}{-p^2+m^2(\phi_c)-i\epsilon}+2\pi
n_B(\omega)\delta(p^2-m^2(\phi_c)) \right]
\end{equation}

Now the $\beta$-independent part of (\ref{tadscr2}), after
integration on $m^2(\phi_c)$ contributes to the effective
potential as
\begin{equation}
\label{bindpot}
-\frac{i}{2}\int\frac{d^4 p}{(2\pi)^4}\log(-p^2+m^2(\phi_c)-i\epsilon)
\end{equation}
Finally using Eq.~(\ref{identity}) to perform the $p^0$
integral, we can cast Eq.~(\ref{bindpot}) as
\begin{equation}
\label{bindpotfin}
\int\frac{d^3 p}{(2\pi)^3}\frac{\omega}{2}
\end{equation}
which coincides with the first term in (\ref{v1sct}).

Integration over $p^0$ in the $\beta$-dependent part of
(\ref{tadscr2}) can be easily performed with the help of the
identity 
\begin{equation}
\delta(p^2-m^2)=\frac{1}{2\omega_p}\left[\delta(p^0+\omega_p)
+\delta(p^0-\omega_p)\right]
\label{deltap2}
\end{equation}
leading to,
\begin{equation}
\label{bdeptadsc}
\int\frac{d^3 p}{(2\pi)^3}\frac{1}{2\omega}n_B(\omega)
\end{equation}
which, upon integration over $m^2(\phi_c)$ leads to the second term
of Eq.~(\ref{v1sct}).

We have checked that trivially the real time and imaginary time
formalisms lead to the same expression of the thermal effective
potential, in the one loop approximation.

\subsection{Fermion fields}

We will consider here a theory with fermion fields described by
the lagrangian (\ref{fermlag}). As in the scalar case, we have
to compute the diagrams contained in Fig.~\ref{1loopf}, using the
Feynman rules either for the imaginary or for the real time
formalism, and decompose the thermal effective potential as in
(\ref{potdecomp}). 

\subsubsection{Imaginary time formalism}

The calculation of the diagrams in Fig.~\ref{1loopf}, using the
Feynman rules (\ref{feynmanim}), yields,
\begin{equation}
\label{1loopfert}
V^{\beta}_1(\phi_c)=-\frac{2\lambda}{2\beta} \sum_{n=-\infty}^{\infty}
\int \frac{d^3 p}{(2\pi)^3}\log(\omega_n^2+\omega^2)
\end{equation}
where $\omega_n$ are the fermionic Matsubara frequencies and 
\begin{equation}
\label{omegafermion}
\omega^2=\vec{p}^{\ 2}+M_f^2.
\end{equation}

The sum over $n$ is done with the help of the same trick
employed in (\ref{ve})-(\ref{vefin}). Let $f(y)$ be given by
(\ref{fy}), then,
\begin{eqnarray}
\label{parsums}
\sum_{m=2,4,\ldots}\frac{y}{y^2+m^2} & = &
\sum_{n=1}^{\infty}\frac{y}{y^2+4n^2} = \frac{1}{2}
f\left(\frac{y}{2}\right) \nonumber \\
\sum_{m=1,3,\ldots}\frac{y}{y^2+m^2} & = & f(y)-\frac{1}{2}
f\left(\frac{y}{2}\right)
\end{eqnarray}
and using (\ref{fy}) we get,
\begin{equation}
\label{fyodd}
\sum_{m=1,3,\ldots}\frac{y}{y^2+m^2}=\frac{\pi}{4}-\frac{\pi}{2}
\frac{1}{e^{\pi y}+1}
\end{equation}

The function $v(\omega)$ in this case can be written as,
\begin{equation}
\label{vefer}
v(\omega)=2\sum_{n=1,3,\ldots}\log\left[\frac{\pi^2n^2}{\beta^2}
+\omega^2\right]
\end{equation}
and its derivative,
\begin{equation}
\label{dervefer}
\frac{\partial
v}{\partial\omega}=\frac{4\beta}{\pi}\sum_{1,3,\ldots}
\frac{y}{y^2+n^2} 
\end{equation}
where $y=\beta\omega/\pi$. Then using (\ref{fyodd}) we get
\begin{equation}
\label{dvefinfer}
\frac{\partial
v}{\partial\omega}=2\beta\left[\frac{1}{2}-\frac{1}{1+
e^{\beta\omega}}\right] 
\end{equation}
and, after integration with respect to $\omega$, 
\begin{equation}
\label{vefinfer}
v(\omega)=2\beta\left[\frac{w}{2}+\frac{1}{\beta}\log\left(
1+e^{-\beta\omega}\right) \right]+\omega-{\rm independent\ \ terms}
\end{equation}

Replacing finally (\ref{vefinfer}) into (\ref{1loopfert}) one
gets, 
\begin{equation}
\label{v1fert}
V_1^{\beta}(\phi_c)=-2\lambda\int \frac{d^3
p}{(2\pi)^3}\left[\frac{\omega}{2}+\frac{1}{\beta}
\log\left(1+e^{-\beta\omega} \right) \right]
\end{equation}
The first integral in (\ref{v1fert}) can be proven, as in
(\ref{identity})-(\ref{bindfin}), to lead to the one-loop
effective potential at zero temperature (\ref{oneferfin}).
The second integral, which contains all the temperature
dependent part, can be written as,
\begin{equation}
\label{bdepfer}
-2\lambda\frac{1}{\beta}\int\frac{d^3 p}{(2\pi)^3}
\log\left(1+e^{-\beta\omega} \right)=
-2\lambda\frac{1}{2\pi^2\beta^4}
J_F[M_f^2(\phi_c)\beta^2] 
\end{equation}
where the thermal fermionic function $J_F$ is defined as,
\begin{equation}
\label{jf}
J_F[m^2\beta^2]=\int_0^{\infty} dx\
x^2\log\left[1+e^{-\sqrt{x^2+\beta^2 m^2}}\right]
\end{equation}

As in the scalar field,
the integral (\ref{jf}) and therefore the thermal fermionic
effective potential admits a high-temperature expansion which
will be very useful for practical applications. It is given by
\begin{eqnarray}
\label{jfexp}
J_F(m^2/T^2) & = & \frac{7\pi^4}{360}-
\frac{\pi^2}{24}\frac{m^2}{T^2}-\frac{1}{32}
\frac{m^4}{T^4}\log\frac{m^2}{a_f T^2} \\
& &
-\frac{\pi^{7/2}}{4}\sum_{\ell=1}^{\infty}(-1)^{\ell}
\frac{\zeta(2\ell+1)}{(\ell+1)!}
\left(1-2^{-2\ell-1}\right)
\Gamma\left(\ell+\frac{1}{2}\right)\left(\frac{m^2}{\pi^2 T^2}
\right)^{\ell+2} \nonumber 
\end{eqnarray}
where $a_f=\pi^2\exp(3/2-2\gamma_E)$ 
($\log a_f=2.6351$) and $\zeta$ is the
Riemann $\zeta$-function.

As we did in the case of the scalar field, there is a very
simple way of obtaining the effective potential, computing the
tadpole in the shifted theory,
and integrating over $\phi_c$. Using for the fermion propagator
(\ref{feynmanim})
$$
i\frac{\gamma \cdot p +M_f}{p^2-M^2_f}
$$
and the trace formula,
$$
Tr\left(\gamma \cdot p+M_f\right)=2\lambda M_f
$$
we can write for the tadpole the expression,
\begin{equation}
\label{tadfer}
\frac{dV_1^{\beta}}{d\phi_c}=-2\lambda\Gamma M_f\frac{1}{\beta}
\sum_{n=-\infty}^{\infty} \int\frac{d^3 p}{(2\pi)^3}
\frac{1}{\omega_n^2+\omega^2} 
\end{equation}
or, using the expression $M_f(\phi_c)=\Gamma \phi_c$, where
$\Gamma$ is the Yukawa coupling,
\begin{equation}
\label{tadfer2}
\frac{dV_1^{\beta}}{dM_f^2(\phi_c)}=-2\lambda\frac{1}{2\beta}
\sum_{n=-\infty}^{\infty} \int\frac{d^3 p}{(2\pi)^3}
\frac{1}{\omega_n^2+\omega^2} 
\end{equation}

Now the infinite sum in (\ref{tadfer2}) can be done with the
help of (\ref{sumbosfer}), with $f(z)$ given by (\ref{fofz}), as
\begin{equation}
\label{tadferfin}
\frac{dV_1^{\beta}}{dM_f^2(\phi_c)}=-2\lambda
\int\frac{d^3 p}{(2\pi)^3}
\left\{
\frac{1}{2} \int_{-i\infty}^{i\infty}\frac{dz}{2\pi
i}\frac{1}{\omega^2-z^2} -\int_C\frac{dz}{2\pi
i}\frac{1}{e^{\beta z}+1}\frac{1}{\omega^2-z^2}\right\}
\end{equation}

The first term of (\ref{tadferfin}) reproduces the zero
temperature result (\ref{oneferfin}), after $M_f^2$
integration, by closing the integration contour of
(\ref{bindsc}) anticlockwise and picking the pole at $z=-\omega$
with a residue $1/2\omega$. The second term in (\ref{tadferfin})
gives the $\beta$-dependent part of the tadpole contribution.
Here the integration contour $C$ encircles the pole at
$z=\omega$ with a residue
\begin{equation}
\label{bdepferfin}
(-2\lambda)\frac{1}{2\omega}\frac{1}{e^{\beta\omega}+1}
\end{equation}
Adding all of them together, we obtain for the tadpole the final
expression
\begin{equation}
\label{tadpoleferfin}
\frac{dV_1(\phi_c)}{dM_f^2(\phi_c)}=-\lambda\int\frac{d^3
p}{(2\pi)^3}
\left[\frac{1}{2\omega}-\frac{1}{\omega}\frac{1}{e^{\beta\omega}+1}
\right]
\end{equation}
and, upon integration with respect to $M_f^2$ we obtain the
result previously presented in Eq.~(\ref{v1fert}).

\subsubsection{Real time formalism}

As for the case of scalar fields, the thermal effective
potential for fermions (\ref{v1fert}) can also be very easily
obtained using the real time formalism. We compute again the
tadpole diagram, where the vertex between
the two fermions and the scalar is of type 1 and the fermion
propagator circulating along the loop is a (11) propagator.
Application of the Feynman rules (\ref{realpropfer}) leads to
the expression
\begin{equation}
\label{tadferr}
\frac{dV_1^{\beta}}{d\phi_c}=-\Gamma Tr
\int\frac{d^4
p}{(2\pi)^4}(\gamma \cdot p+M_f)
\left[\frac{i}{p^2-M_f^2+i\epsilon}-2\pi
n_F(\omega)\delta(p^2-M_f^2) \right]
\end{equation}
or, using as before the expression for $M_f^2$,
\begin{equation}
\label{tadferr2}
\frac{dV_1^{\beta}}{dM_f^2(\phi_c)}=-\frac{Tr {\bf 1}}{2}\int\frac{d^4
p}{(2\pi)^4}\left[\frac{-i}{-p^2+M_f^2-i\epsilon}-2\pi
n_F(\omega)\delta(p^2-M_f^2) \right]
\end{equation}

Now the $\beta$-independent part of (\ref{tadferr2}), after
integration on $M^2_f$, contributes to the effective potential,
\begin{equation}
\label{bindpotfer}
-Tr{\bf 1}\int\frac{d^3 p}{(2\pi)^3}\frac{\omega}{2}
\end{equation}
which coincides with the first term in (\ref{v1fert}).

Integration over $p^0$ in the $\beta$-dependent part of
(\ref{tadferr2}) can be easily performed with the help of the
identity (\ref{deltap2}) leading to,
\begin{equation}
\label{bdeptadfer}-Tr{\bf 1}
\int\frac{d^3 p}{(2\pi)^3}\frac{1}{2\omega}[-n_F(\omega)]
\end{equation}
which, upon integration over $M^2_f$ leads to the second term of
Eq.~(\ref{v1fert}).

\subsection{Gauge bosons}

The thermal effective potential for gauge bosons in a theory
described by the lagrangian (\ref{gaugelag}) is computed in the
same way as for previous fields. The simplest thing is to
compute the tadpole diagram using the
shifted mass for the gauge boson.
In the Landau gauge, the gauge boson propagator reads as,
\begin{equation}
\label{gbprop}
\Pi^{\mu}_{\
\nu}(p)^{(\alpha\beta)}=\Delta^{\mu}_{\ \nu}G^{(\alpha\beta)}(p)
\end{equation}
where $\Delta$ is the projector defined in (\ref{transverse})
with a trace equal to 3 (see Eq.~(\ref{delta3})). Therefore the
final expression for the thermal effective potential is computed
as,
\begin{equation}
\label{gb1loopt}
V_1^{\beta}(\phi_c)=
Tr(\Delta)\left\{
\frac{1}{2}\int\frac{d^4p}{(2\pi)^4}
\log[p^2+M^2_{gb}(\phi_c)]+
\frac{1}{2\pi^2\beta^4}
J_B[M^2_{gb}(\phi_c)\beta^2] \right\}
\end{equation}
where the thermal bosonic function $J_B$ in (\ref{jb}).
The first term of (\ref{gb1loopt}) agrees with the zero
temperature effective potential computed in (\ref{onegaugefin}),
and the second one just counts that of a scalar field theory a
number of times equal to the number of degrees of freedom (3) of
the gauge boson.

\subsection{The Standard Model case}

The Standard Model of electroweak interactions was previously
defined through Eqs.~(\ref{doublet})-(\ref{mass1/2sm}), and the
corresponding one loop effective potential at zero temperature
computed through Eqs.~(\ref{counter})-(\ref{countersm}) using
various renormalization schemes and the contribution of gauge
and Higgs bosons and the top quark fermion to radiative
corrections. Here we will compute
the corresponding one loop effective potential at finite
temperature. We will use the renormalization scheme of 
Eq.~(\ref{ahregcond}), so that the renormalized effective potential
at zero temperature is given by Eq.~(\ref{potsm}), and consider
only the contribution of $W$ and $Z$ bosons, and the top quark
to radiative corrections. This is expected to be a good enough
approximation for Higgs masses lighter than the $W$ mass.

Using Eqs.~(\ref{bdepfer}) and (\ref{gb1loopt}) one can
easily see that
the finite-temperature part of the one-loop effective potential
can be written as,
\begin{equation}
\label{deltatsm}
\Delta V^{(1)}(\phi_c,T)=\frac{T^4}{2\pi^2}\left[\sum_{i=W,Z}n_i
J_B[m_i^2(\phi_c)/T^2]+n_t J_F[m_t^2(\phi_c)/T^2] \right]
\end{equation}
where the function $J_B$ and $J_F$ where defined in 
Eqs.~(\ref{jb}) and (\ref{jf}), respectively.

Using now the high temperature expansions (\ref{jbexp}) and
(\ref{jfexp}), and the one loop effective potential at zero
temperature, Eq.~(\ref{potsm}), one can write the  total
potential as,
\begin{equation}
V(\phi_c,T)=D(T^2-T_o^2)\phi_c^2-ET\phi_c^3+\frac{\lambda(T)}{4}\phi_c^4
\label{potsmtemp}
\end{equation}
where the coefficients are given by
\begin{equation}
\label{valord}
D=\frac{2m_W^2+m_Z^2+2m_t^2}{8v^2}
\end{equation}
\begin{equation}
\label{valore}
E=\frac{2m_W^3+m_Z^3}{4\pi v^3}
\end{equation}
\begin{equation}
\label{valort02}
T_o^2=\frac{m_h^2-8Bv^2}{4D}
\end{equation}
\begin{equation}
\label{valorb}
B=\frac{3}{64\pi^2 v^4}\left(2m_W^4+m_Z^4-4m_t^4\right)
\end{equation}
\begin{equation}
\label{valorlamb}
\lambda(T)=\lambda-\frac{3}{16\pi^2
v^4}\left(2m_W^4\log\frac{m_W^2}{A_B
T^2}+m_Z^4\log\frac{m_Z^2}{A_B T^2}-4m_t^4\log\frac{m_t^2}{A_F
T^2} \right)
\end{equation}
where $\log A_B=\log a_b-3/2$ and $\log A_F=\log a_F-3/2$, 
and $a_B$, $a_F$ are
given in (\ref{jbexp}) and (\ref{jfexp}). All the masses which
appear in the definition of the coefficients, 
Eqs.~(\ref{valord}) to (\ref{valorlamb}), are the physical masses,
{\it i.e.} the masses at the zero temperature minimum. The
peculiar form of the potential, as given by Eq.~(\ref{potsmtemp}), 
will be useful to study the associated phase
transition, as we will see in subsequent sections.

\section{Cosmological phase transitions}

All cosmological applications of field theories are based on the
theory of phase transitions at finite temperature, that we will
briefly describe throughout this section. The main point here is
that at finite temperature, the equilibrium value of the scalar
field $\phi$, $\langle\phi(T)\rangle$, 
does not correspond to the minimum of
the effective potential $V_{\rm eff}^{T=0}(\phi)$, but to the minimum
of the finite temperature effective potential $V_{\rm
eff}^{\beta}(\phi)$, as given by (\ref{potdecomp}). Thus, even
if the minimum of $V_{\rm eff}^{T=0}(\phi)$ occurs at $\langle
\phi \rangle = \sigma\ne 0$, very often, for sufficiently large
temperatures, the minimum of $V_{\rm
eff}^{\beta}(\phi)$ occurs at $\langle \phi(T) \rangle=0$: 
this phenomenon is known as {\bf symmetry restoration} at high
temperature, and gives rise to the phase transition from
$\phi(T)=0$ to $\phi=\sigma$. It was discovered by Kirzhnits~\cite{K} 
in the context of the electroweak theory (symmetry
breaking between weak and electromagnetic interactions occurs
when the universe cools down to a critical temperature $T_c\sim
10^2\ GeV$) and subsequently confirmed and developed by other
authors~\cite{KL,DJ,W,LINDE}.

The cosmological scenario can be drawn as follows: In the theory
of the hot big bang, the universe is initially at very high
temperature and, depending on the function $V_{\rm
eff}^{\beta}(\phi)$, it can be in the {\bf symmetric phase}
$\langle\phi(T)\rangle=0$, {\it i.e.} 
$\phi=0$ can be the stable absolute
minimum. At some critical temperature $T_c$ the minimum at
$\phi=0$ becomes metastable and the phase transition may
proceed. The phase transition may be {\bf first} or {\bf second}
order. First-order phase transitions have supercooled (out of
equilibrium) symmetric states when the temperature decreases and
are of use for baryogenesis purposes. Second-order phase
transitions are used in the so-called new
inflationary models~\cite{INFLATION}.
We will illustrate these
kinds of phase transitions with very simple examples.

\subsection{First and second order phase transitions}

We will illustrate the difference between {\bf first} and {\bf
second} order phase transitions by considering first 
the simple example of a potential~\footnote{The $\phi$
independent terms in (\ref{potsecond}), {\it i.e.} $V(0,T)$, are
not explicitly considered.}
described by the function,
\begin{equation}
\label{potsecond}
V(\phi,T)=D(T^2-T_o^2)\phi^2+\frac{\lambda(T)}{4}\phi^4
\end{equation}
where $D$ and $T_o^2$ are constant terms and $\lambda$ is a
slowly varying function of $T$~\footnote{The $T$ dependence of
$\lambda$ will often be neglected in this section.}. A quick glance at
(\ref{jbexp}) and (\ref{jfexp}) shows that the potential
(\ref{potsecond}) can be part of the 
one-loop finite temperature
effective potential in field theories.

At zero temperature, the potential
has a negative mass-squared term, which indicates that the state
$\phi=0$ is unstable, and the energetically favored state
corresponds to the minimum at $\phi(0)=\pm
\sqrt{\frac{2D}{\lambda}} T_o$,
where the symmetry $\phi \leftrightarrow -\phi$ of the original
theory is spontaneously broken.

The curvature of the finite temperature potential (\ref{potsecond})
is now $T$-dependent,
\begin{equation}
\label{curvature}
m^2(\phi,T)=3\lambda\phi^2+2D(T^2-T_o^2)
\end{equation}
and its stationary points, {\it i.e.} solutions to
$dV(\phi,T)/d\phi=0$, given by,
\begin{eqnarray}
\label{station}
\phi(T)& = & 0 \nonumber \\
{\rm and}& &   \\
\phi(T)& = & \sqrt{\frac{2D(T_o^2-T^2)}{\lambda(T)}} \nonumber
\end{eqnarray}
Therefore the critical temperature is given by $T_o$.
At $T>T_o$, $m^2(0,T)>0$ and the origin $\phi=0$ is a minimum.
At the same time only the solution $\phi=0$ in (\ref{station})
does exist. At $T=T_o$, $m^2(0,T_o)=0$ and both solutions in
(\ref{station}) collapse at $\phi=0$. The potential
(\ref{potsecond}) becomes,
\begin{equation}
\label{potseccrit}
V(\phi,T_o)=\frac{\lambda(T_o)}{4}\phi^4
\end{equation}
At $T<T_o$, $m^2(0,T)<0$ and the origin becomes a maximum.
Simultaneously, the solution $\phi(T)\ne 0$ does appear in
(\ref{station}). This phase transition is called of {\bf second
order}, because there is no barrier between the symmetric and
broken phases. Actually, when the broken phase is formed, the
origin (symmetric phase) becomes a maximum. 
The phase transition may be achieved by a thermal fluctuation
for a field located at the origin.

However, in many interesting theories there is a {\bf barrier}
between the symmetric and broken phases. This is characteristic
of {\bf first order} phase transitions. A typical example is
provided by the potential~\footnote{See, 
{\it e.g.} the one-loop effective potential for the
Standard Model, Eq.~(\ref{potsm}).},
\begin{equation}
\label{potfirst}
V(\phi,T)=D(T^2-T_o^2)\phi^2-ET\phi^3+\frac{\lambda(T)}{4}\phi^4
\end{equation}
where, as before, $D$, $T_0$ and $E$ are $T$ independent
coefficients, and $\lambda$ is a slowly varying 
$T$-dependent function.
Notice that the difference between (\ref{potfirst}) and
(\ref{potsecond}) is the cubic term with coefficient $E$. This
term can be provided by the contribution to the effective
potential of bosonic fields (\ref{jbexp}). The behaviour of
(\ref{potfirst}) for the different temperatures is
reviewed in Refs.~\cite{AH,DLHLL}.
At $T>T_1$ the only minimum is at $\phi=0$. At $T=T_1$ 
\begin{equation}
\label{t1}
T_1^2=\frac{8\lambda(T_1) DT_o^2}{8\lambda(T_1) D-9E^2}
\end{equation}
a local minimum at $\phi(T) \ne 0$ appears as an inflection
point.  
The value of the field $\phi$ at $T=T_1$ is,
\begin{equation}
\label{phi1}
\langle \phi(T_1)\rangle=\frac{3ET_1}{2\lambda(T_1)}
\end{equation}
A barrier between
the latter and the minimum at $\phi=0$ starts to develop at lower
temperatures. Then the point (\ref{phi1}) splits into a maximum
\begin{equation}
\label{maximo}
\phi_M(T)=\frac{3ET}{2\lambda(T)}-\frac{1}{2\lambda(T)}
\sqrt{9E^2T^2- 8\lambda(T)D(T^2-T_o^2)} 
\end{equation}
and a local minimum
\begin{equation}
\label{minimo}
\phi_m(T)=\frac{3ET}{2\lambda(T)}+\frac{1}{2\lambda(T)}
\sqrt{9E^2T^2- 8\lambda(T)D(T^2-T_o^2)} 
\end{equation}

At a given temperature $T=T_c$ 
\begin{equation}
\label{tc}
T_c^2=\frac{\lambda(T_c)D T_o^2}{\lambda(T_c)D-E^2}
\end{equation}
the origin and the minimum (\ref{minimo}) become degenerate. 
From (\ref{maximo}) and (\ref{minimo}) we find that
\begin{equation}
\label{maximotc}
\phi_M(T_c)=\frac{ET_c}{\lambda(T_c)}
\end{equation}
and
\begin{equation}
\label{minimotc}
\phi_m(T_c)=\frac{2ET_c}{\lambda(T_c)}
\end{equation}

For $T<T_c$ the minimum at $\phi=0$ becomes metastable and the
minimum at $\phi_m(T)\ne 0$ becomes the global one. At $T=T_o$
the barrier disappears, the origin becomes a maximum
\begin{equation}
\label{maximot0}
\phi_M(T_o)=0
\end{equation}
and the second minimum becomes equal to
\begin{equation}
\label{minimot0}
\phi_m(T_o)=\frac{3ET_o}{\lambda(T_o)}
\end{equation}
The phase transition starts at $T=T_c$ by tunneling. However,
if the barrier is high enough the tunneling effect is very
small and the phase transition does effectively start at a
temperature $T_c>T_t>T_o$. In some models $T_o$ can be equal
to zero. The details of the phase transition depend therefore on
the process of tunneling from the false to the global minimum.
These details will be studied in the rest of this section.

\subsection{Thermal tunneling}

The transition from the false to the true vacuum proceeds
via {\bf thermal tunneling} at finite temperature.
It can be understood
in terms of formation of bubbles of the broken phase in the sea
of the symmetric phase. Once this has happened, the bubble
spreads throughout the universe converting false vacuum into
true one.

The tunneling rate~\cite{C,CC,CDL} is computed by
using the rules of
field theory at finite temperature~\cite{LTUN}. In the previous
section we defined the critical temperature $T_c$ as the
temperature at which the two minima of the potential $V(\phi,T)$
have the same depth. However, tunneling with
formation of bubbles of the field $\phi$ corresponding to the
second minimum  starts somewhat later, and goes sufficiently
fast to fill the universe with bubbles of the new phase only at
some lower temperature $T_t$ when the corresponding euclidean
action $S_E=S_3/T$
suppressing the tunneling becomes ${\cal O}(130-140)$~\cite{LSTV,DHS,AH}, 
as we will see in the next section. 

We will use as prototype the potential of Eq.~(\ref{potfirst})
which can trigger, as we showed in this section, a first order
phase transition. In this case the false minimum is $\phi=0$,
and the value of the potential at 
the origin is zero, $V(0,T)=0$. 
The tunneling probability per unit time per unit volume is
given by~\cite{LTUN}
\begin{equation}
\label{probt}
\frac{\Gamma}{\nu}\sim A(T)e^{-S_3/T}
\end{equation}
In (\ref{probt}) the prefactor 
$A(T)$ is roughly of
${\cal O}(T^4)$ while $S_3$ is the three-dimensional euclidean
action defined as 
\begin{equation}
\label{eucactt}
S_3=\int d^3x\left[\frac{1}{2}\left(\vec{\nabla}\phi\right)^2+
V(\phi,T)\right] .
\end{equation}

At very high temperature the {\bf bounce} solution has $O(3)$
symmetry~\cite{LTUN} and the euclidean action is then simplified
to,
\begin{equation}
\label{eucactsymt}
S_3=4\pi\int_0^{\infty}r^2
dr\left[\frac{1}{2}\left(\frac{d\phi}{dr}\right)^2
+V(\phi(r),T)\right] 
\end{equation}
where $r^2=\vec{x}^{\ 2}$, and the euclidean equation of motion
yields,
\begin{equation}
\label{euceqsymt}
\frac{d^2\phi}{dr^2}+\frac{2}{r}\frac{d\phi}{dr}=V^{\prime}(\phi,T)
\end{equation}
with the boundary conditions
\begin{eqnarray}
\label{bcinfsymt}
\lim_{r\rightarrow\infty}\phi(r)& = & 0 \\
\label{bczerosymt}
\left.\frac{d\phi}{dr}\right|_{r=0} & = & 0 
\end{eqnarray}

From here on we will follow the discussion in Ref.~\cite{AH}.
Let us take $\phi=0$ outside a bubble. Then (\ref{eucactsymt}),
which is also the surplus free energy of a true vacuum bubble, 
can be written as
\begin{equation}
\label{surplus}
S_3=4\pi\int_0^{R}r^2
dr\left[\frac{1}{2}\left(\frac{d\phi}{dr}\right)^2
+V(\phi(r),T)\right] 
\end{equation}
where $R$ is the bubble radius. There are two contributions to
(\ref{surplus}): a surface term $F_S$, coming from the
derivative term in (\ref{surplus}), and a volume term $F_V$,
coming from the second term in (\ref{surplus}). They scale like,
\begin{equation}
\label{surplusapp}
S_3\sim 2\pi R^2\left(\frac{\delta\phi}{\delta R}\right)^2
\delta R + \frac{4\pi R^3 \langle V \rangle}{3} 
\end{equation}
where $\delta R$ is the thickness of the bubble wall, $\delta
\phi=\phi_m$ and $\langle V\rangle$ is the average of the
potential inside the bubble. 

For temperatures just below $T_c$, the height of the barrier
$V(\phi_M,T)$ is large compared to the depth of the potential at
the minimum, $-V(\phi_m,T)$. In that case, the solution of
minimal action corresponds to minimizing the contribution to
$F_V$ coming from the region $\phi=\phi_M$. This amounts to a very
small bubble wall $\delta R/R \ll 1$ and 
so a very quick change of the field
from $\phi=0$ outside the bubble to $\phi=\phi_m$ inside the
bubble. Therefore, the first formed bubbles after $T_c$ are {\bf
thin wall} bubbles.

Subsequently, when the temperature drops towards $T_o$ the
height of the barrier $V(\phi_M,T)$ becomes small as compared
with the depth of the potential at the minimum $-V(\phi_m,T)$. 
In that case the contribution to $F_V$ from the region
$\phi=\phi_M$ is negligible, and the minimal action corresponds
to minimizing the surface term $F_S$. This amounts to a
configuration where $\delta R$ is as large as possible, {\it i.e.}
$\delta R/R = {\cal O}(1)$: {\bf thick wall} bubbles. So whether
the phase transition proceeds through thin or thick wall bubbles
depends on how large the bubble nucleation rate (\ref{probt}) is,
or how small $S_3$ is, before thick bubbles are energetically
favoured. 

For the potential (\ref{potfirst}) an
analytic formula has been obtained in Ref.~\cite{DLHLL} without
assuming the thin wall approximation. It is given by,
\begin{equation}
\label{s3ex}
\frac{S_3}{T}=\frac{13.72}{E^2}\left[D\left(1-\frac{T_o^2}{T^2}\right)
\right]^{3/2}f\left[\frac{\lambda(T)D}{E^2}
\left(1-\frac{T_o^2}{T^2}\right)\right]
\end{equation}
\begin{equation}
\label{flinde}
f(x)=1+\frac{x}{4}\left[1+\frac{2.4}{1-x}+\frac{0.26}{(1-x)^2}\right]
\end{equation}

The case of two fields is extremely more complicated. In particular the
two-Higgs situation in the supersymmetric standard theory has been 
recently solved in Ref.~\cite{Marcos}.
The connection between zero
temperature and finite temperature tunneling is manifest.
In particular at temperatures much less than the
inverse radius the ($T=0$) $O(4)$ solution has the least action. This
can happen for theories with a supercooled symmetric phase: for
instance in the presence of a barrier that does not disappear
when the temperature drops to zero. 
At temperatures much larger than the
inverse radius, the $O(3)$ solution has the least action.

\subsection{Bubble nucleation}

In the previous subsection we have established the free energy
and the critical radius of a bubble large enough to grow after
formation. The subsequent progress of the phase transition
depends on the ratio of the rate of production of bubbles of
true vacuum, as given by (\ref{probt}), over
the expansion rate of the universe. For example if the former
remains always smaller than the latter, then the state will be
trapped in the supercooled false vacuum. Otherwise the
phase transition will start at some temperature $T_t$
by {\bf bubble nucleation}. 
The probability of bubble formation per unit time per unit
volume is given by (\ref{probt})
where $B(T)=S_3(T)/T$, $A(T)=\omega T^4$,  where 
the parameter $\omega$ will be taken
of ${\cal O}(1)$. 

Since the progress of the phase transition should depend on the
expansion rate of the universe, we have to describe the universe
at temperatures close to the electroweak phase transition. A
homogeneous and isotropic (flat) universe is described by a
Robertson-Walker metric which, in comoving coordinates, is given
by
$ds^2=dt^2-a(t)^2\left(dr^2+r^2d\Omega^2\right)$,
where $a(t)$ is the scale factor of the universe. The universe
expansion is governed by the equation
\begin{equation}
\label{einstein}
\left(\frac{\dot{a}}{a}\right)^2=\frac{8\pi}{3M_{P\ell}^2}\rho 
\end{equation}
where $M_{P\ell}$ is the Planck mass,
and $\rho$ is the energy density. For temperatures $T\sim 10^2\
GeV$ the universe is radiation dominated, and its energy density
is given by,
\begin{equation}
\label{rhorad}
\rho=\frac{\pi^2}{30}g(T)T^4
\end{equation}
where
$g(T)=g_B(T)+\frac{7}{8}g_F(T)$,
and $g_B(T)$ ($g_F(T)$) is the effective number of bosonic
(fermionic) degrees of freedom
at the temperature $T$. For the standard model we have 
$g^{SM}=106.75$
which can be considered as temperature independent. 

The equation of motion (\ref{einstein})
can be solved, and assuming an adiabatic expansion of the universe,
$a(T_1)T_1=a(T_2)T_2$, one obtains the following relationship,
\begin{equation}
\label{tTrel}
t=\zeta \frac{M_{P\ell}}{T^2}
\end{equation}
where
$
{\displaystyle 
\zeta=\frac{1}{4\pi}\sqrt{\frac{45}{\pi g}}\sim 3\times 10^{-2} }
$.
Using (\ref{tTrel}) the horizon volume is given by
\begin{equation}
\label{horvol}
V_H(t)=8\zeta^3\frac{M_{P\ell}^3}{T^6}
\end{equation}

The onset of nucleation happens at a temperature $T_t$ such that
{\bf the probability for a single bubble to be nucleated within
one horizon volume is $\sim$ 1}, i.e.~\cite{Marcos}
\begin{equation}
\label{probabilidad1}
\int_{T_t}^{\infty}\frac{dT}{T}\left(\frac{2\zeta M_{\rm
Pl}}{T}\right)^4 \exp\{-S_3(T)/T\}={\cal O}(1)\ .
\end{equation}
which implies numerically,
\begin{equation}
\label{BTt}
B(T_t)\sim 137+\log\frac{10^2 E^2}{\lambda D}+4\log\frac{100\
GeV}{T_t} 
\end{equation}
where we have normalized $T_t\sim 100\ GeV$ and $E^2/(\lambda
D)\sim 10^{-2}$ which are typical values which will be obtained
in the standard model of electroweak interactions.

\section{Baryogenesis at phase transitions}

There are two essential problems to be understood related with
the baryon number of the universe:

{\bf i)} There is no evidence of antimatter in the universe. In
fact, there is no antimatter in the solar system, and only
$\overline{p}$ in cosmic rays. However antiprotons can be produced as
secondaries in collisions ($pp\rightarrow 3p+\overline{p}$) at a rate
similar to the observed one. Numerically,
$
{\displaystyle \frac{n_{\overline{p}}}{n_p}\sim 3\times 10^{-4} }
$,
and 
$
{\displaystyle \frac{n_{^4 He}}{n_{^4\overline{He}}}\sim 10^{-5}}
$.
We can conclude that $n_B\gg n_{\overline{B}}$, so $n_{\Delta B}
\equiv n_B-n_{\overline{B}} \sim n_B$.

{\bf ii)} The second problem is to understand the origin of 
\begin{equation}
\label{etaprim}
\eta\equiv \frac{n_B}{n_{\gamma}}\sim (0.3-1.0)\times 10^{-9}
\end{equation}
today. This parameter is essential for primordial
nucleosynthesis~\cite{KT}.
$\eta$ may not have changed since nucleosynthesis. At these
energy scales ($\sim$ 1 MeV) baryon number is conserved if there
are no processes which would have produced entropy to change the
photon number. We can easily estimate from $\eta$ the baryon to
entropy ratio by using

\begin{equation}
\label{esefot}
s=\frac{\pi^4}{45\zeta(3)}3.91\ n_\gamma=7.04\ n_\gamma
\end{equation}
and the range (\ref{etaprim}). 

In the standard cosmological model there is no
explanation for the smallness of the ratio (\ref{etaprim}) if we
start from $n_{\Delta B}=0$. An initial asymmetry has to be
imposed by hand as an initial condition (which violates any
naturalness principle) or has to be dynamically generated at phase
transitions, which is the way we will explore all along this
section. 

\subsection{Conditions for baryogenesis}

As we have seen in the previous subsection the universe was
initially baryon symmetric ($n_B\simeq n_{\overline{B}}$) although
the matter-antimatter asymmetry appears to be large today
($n_{\Delta B}\simeq n_B \gg n_{\overline{B}}$). In the standard
cosmological model there is no explanation for the value of
$\eta$ consistent with nucleosynthesis, Eq.~(\ref{etaprim}), and
it has to be imposed by hand as an initial condition. However,
it was suggested by Sakharov long ago~\cite{SAKHA} that a tiny
$n_{\Delta B}$ might have been produced in the early universe
leading, after $p\overline{p}$ annihilations, to (\ref{etaprim}). The
three ingredients necessary for baryogenesis are:

\subsubsection{$B$-nonconserving interactions}
\label{sakha1}

This condition is obvious since we want to start with a baryon
symmetric universe ($\Delta B=0$) and evolve it to a universe
where $\Delta B\neq 0$. B-nonconserving interactions might
mediate proton decay; in that case the phenomenological
constraints are provided by the proton lifetime measurements~\cite{PDG} 
$\tau_p\simgt 10^{32} yr$.

\subsubsection{C and CP violation}
\label{sakha2}

The action of $C$ (charge conjugation) and $CP$ (combined action
of charge conjugation and parity) interchanges particles with
antiparticles, changing therefore the sign of $B$. For instance
if we describe spin-$\frac{1}{2}$ fermions by two-component
fields of definite chirality (left-handed fields $\psi_L$ and
right-handed fields $\psi_R$) the action of $C$ and $CP$ over
them is given by
\begin{eqnarray}
\label{pccp}
P & : & \psi_L\longrightarrow \psi_R,\ \ \psi_R\longrightarrow
\psi_L \\
C & : & \psi_L\longrightarrow \psi_L^C\equiv \sigma_2\psi_R^*,
\ \ \psi_R \longrightarrow \psi_R^C\equiv-\sigma_2\psi_L^* \nonumber
\\
CP & : & \psi_L\longrightarrow\psi_R^C,\ \ \psi_R\longrightarrow
\psi_L^C \nonumber
\end{eqnarray}

If the universe is initially matter-antimatter symmetric, and
without a preferred direction of time as in the standard
cosmological model, it is represented by a $C$ and $CP$
invariant state, $|\phi_o\rangle$, with $B=0$. If $C$ and $CP$
were conserved, {\it i.e.}
$[C,H]=[CP,H]=0$,
$H$ being the hamiltonian, then the state of the universe at a
later time $t$,
$|\phi(t)\rangle=e^{iHt}|\phi_o\rangle$
would be $C$ and $CP$ invariant and, therefore, baryon number
conserving, $\Delta B=0$. The only way to generate a net $\Delta
B \neq 0$ is to have $C$ and $CP$ violating interactions.

\subsubsection*{Departure from thermal equilibrium}
\label{sakha3}
 If all particles in the universe remained in thermal
equilibrium, then no direction for time would be defined and
$CPT$ invariance would prevent the appearance of any baryon
excess, rendering $CP$ violating interactions irrelevant~\cite{KW}.

A particle species is in thermal equilibrium if all its reaction
rates, $\Gamma$, are much faster than the expansion rate of the
universe, $H$. On the other hand a departure from thermal
equilibrium is expected whenever a rate crucial for maintaining
it is less than the expansion rate ($\Gamma<H$). Deviation from
thermal equilibrium cannot occur in a homogeneous isotropic
universe containing only massless species: massive species are
needed in general for such deviations to occur.

\subsection{Baryogenesis at the electroweak phase transitions}

It has been recently realized~\cite{KRS,AnnRev} that the three
Sakharov's conditions for baryogenesis can be fulfilled at the
electroweak phase transition:

\begin{itemize}
\item
Baryonic charge non-conservation was discovered by 't Hooft~\cite{Hooft}. 
In fact baryon and lepton number are conserved
anomalous global symmetries in the Standard Model. They are
violated by non-perturbative effects.
\item
CP violation can be generated in the Standard Model from 
phases in the fermion mass matrix,
Cabibbo, Kobayashi, Maskawa (CKM) phases~\cite{CKM}. 
This effect is
much too small to explain the observed baryon to entropy ratio.
However, in extensions of the Standard Model as 
the minimal supersymmetric standard model
(MSSM), a sizeable CP violation can happen through an extended
Higgs sector.
\item
The out of equilibrium condition can be achieved, if the phase
transition is strong enough first order, in the bubble walls.
In that case the B-violating interactions are out of equilibrium
in the bubble walls and a net B-number can be generated during
the phase transition~\cite{AnnRev}.
\end{itemize}

\subsubsection{Baryon and lepton number violation in the electroweak
theory} 

Violation of baryon and lepton number in the electroweak theory
is a very striking phenomenon. Classically, baryonic and
leptonic currents are conserved in the electroweak theory.
However, that conservation is spoiled by quantum corrections
through the chiral anomaly associated with triangle fermionic
loop in external gauge fields. The calculation gives,
\begin{equation}
\label{anomalia}
\partial_\mu j^\mu_B=\partial_\mu j^\mu_L=N_f\left(
\frac{g^2}{32\pi^2}
W\widetilde{W}-\frac{g'^2}{32\pi^2}Y\widetilde{Y}\right) 
\end{equation}
where $N_f$ is the number of fermion generations, 
$W_{\mu\nu}$ and $Y_{\mu\nu}$ are
the gauge field strength tensors for $SU(2)$ and $U(1)_Y$,
respectively, and the tilde means the dual tensor.

A very important feature of (\ref{anomalia}) is that the
difference $B-L$ is strictly conserved, and so only the sum
$B+L$ is anomalous and can be violated. Another feature is that
fluctuations of the gauge field strengths can lead to
fluctuations of the corresponding value of $B+L$. The product of
gauge field strengths on the right hand side of 
Eq.~(\ref{anomalia}) can be written as four-divergences,
$W\widetilde{W}= \partial_\mu k_W^\mu$,
$Y\widetilde{Y} = \partial_\mu k_Y^\mu$,
where
\begin{eqnarray}
\label{vectors}
k_Y^\mu & = & \epsilon^{\mu\nu\alpha\beta}Y_{\nu\alpha}Y_\beta \\
k_W^\mu & = & \epsilon^{\mu\nu\alpha\beta}\left(W^a_{\nu\alpha}W^a_\beta-
\frac{g}{3}\epsilon_{abc}W^a_\nu W^b_\alpha W^c_\beta \right) \nonumber
\end{eqnarray}
and $W_\mu$, $Y_\mu$ are the gauge fields of $SU(2)$ and
$U(1)_Y$, respectively. In general total derivatives are
unobservable because they can be integrated by parts and drop
from the integrals. This is true for the terms in the
four-vectors (\ref{vectors}) proportional to the field
strengths $W_{\mu\nu}$ and $Y_{\mu\nu}$. 
This means that for the
abelian subgroup $U(1)_Y$ the current non conservation induced
by quantum effects becomes non observable. However this is not
mandatory for gauge fields, for which the integral can be
nonzero. Hence only for non-abelian groups can the current non
conservation induced by quantum effects become observable. In
particular one can write
$\Delta B=\Delta L=N_f\Delta N_{CS}$,
where $N_{CS}$ is the so-called Chern-Simons number
characterizing the topology of the gauge field configuration,
\begin{equation}
\label{NCS}
N_{CS}=\frac{g^2}{32\pi^2}\int d^3x\epsilon^{ijk}
\left(W^a_{ij}W^a_k -\frac{g}{3}\epsilon_{abc}W^a_iW^b_jW^c_k
\right) 
\end{equation}
Note that though $N_{CS}$ is not gauge invariant, its variation
$\Delta N_{CS}$ is. 

We want to compute now $\Delta B$ between an initial and a final
configuration of gauge fields. We are
considering vacuum field strength tensors $W_{\mu\nu}$ which
vanish. The corresponding potentials are not necessarily zero
but can be represented by purely gauge fields,
\begin{equation}
\label{gaugepot}
W_\mu=-\frac{i}{g}U(x)\partial_\mu U^{-1}(x)
\end{equation}
There are two classes of gauge transformations keeping
$W_{\mu\nu}=0$: 
\begin{itemize}
\item
Continuous transformations of the potentials yielding $\Delta
N_{CS}=0$. 
\item
If one tries to generate $\Delta N_{CS}\neq 0$ by a continuous
variation of the potentials, then one has to enter a region
where $W_{\mu\nu}\neq 0$. This means that vacuum states with
different topological charges are separated by potential
barriers. 
\end{itemize}

The probability of barrier penetration can be calculated using
the quasi-classical approximation~\cite{C}. In
euclidean space time, the trajectory in field space
configuration which connects two vacua differing by a unit of
topological charge is called instanton. The euclidean action
evaluated at this trajectory gives the probability for barrier
penetration as
$\Gamma\sim \exp\left(-\frac{4\pi}{\alpha_W}\right)\sim 10^{-162}$,
where $\alpha_W=g^2/4\pi$. This number is so
small that the calculation of the pre-exponential is unnecessary
and the probability for barrier penetration is essentially zero.

\subsubsection{Baryon violation at finite temperature: sphalerons}

However, in a system with non
zero temperature a particle may classically go over the barrier
with a probability determined by the Boltzmann exponent, as we
have seen.

What we have is a potential which depends on the gauge field
configuration $W_\mu$. This potential has an infinite number of
degenerate minima, labeled as $\Omega_n$. These minima are
characterized by different values of the Chern-Simons number.
The minimum $\Omega_0$ corresponds to the configuration
$W_\mu=0$ and we can take conventionally the value of the
potential at this point to be zero. Other minima have gauge
fields given by (\ref{gaugepot}). In the temporal gauge $W_0=0$,
the gauge transformation $U$ must be time independent (since we
are considering gauge configurations with $W_{\mu\nu}=0$), {\it
i.e.} $U=U(\vec{x})$, and so functions $U$ define maps,
$$
U: S^3\longrightarrow SU(2)
$$
All the minima with $W_{\mu\nu}=0$ have equally zero potential
energy, but those defined by a map $U(\vec{x})$ with nonzero
Chern-Simons number 
\begin{equation}
\label{ChSi}
n[U]=\frac{1}{24\pi^2}\int d^3x
\epsilon^{ijk}Tr(U\partial_iU^{-1} U\partial_jU^{-1} U\partial_k
U^{-1}) 
\end{equation}
correspond to degenerate minima in the
configuration space with non-zero baryon and lepton number.

Degenerate minima are separated by a potential barrier. The
field configuration at the top of the barrier is called {\bf
sphaleron}, which is a {\it static unstable} solution to the
classic equations of motion~\cite{Sph}. The sphaleron solution
has been explicitly computed in Ref.~\cite{Sph} for the case of
zero Weinberg angle, ({\it i.e.} neglecting terms ${\cal
O}(g')$), and for an arbitrary value of $\sin^2\theta_W$ in 
Ref.~\cite{Sph2}. 

An ansatz for the sphaleron solution for the case of zero
Weinberg angle was given (for the zero temperature potential) 
in Ref.~\cite{Sph}, for the Standard Model with a
single Higgs doublet, as,
\begin{equation}
\label{ansatzW}
W_i^a\sigma^a dx^i=-\frac{2i}{g}f(\xi)dU\ U^{-1}
\end{equation}
for the gauge field, and
\begin{equation}
\label{ansatzfi}
\Phi=\frac{v}{\sqrt{2}}h(\xi)U\left(
\begin{array}{c}
0 \\
1
\end{array}
\right)
\end{equation}
for the Higgs field, where the gauge transformation $U$ is taken
to be,
\begin{equation}
\label{ansatzgtr}
U=\frac{1}{r}\left(
\begin{array}{cc}
z & x+iy \\
-x+iy & z
\end{array}
\right)
\end{equation}
and we have introduced the dimensionless radial distance
$\xi=gvr$. 

Using the ansatz (\ref{ansatzW}), (\ref{ansatzfi}) and
(\ref{ansatzgtr}) the field equations reduce to,
\begin{eqnarray}
\label{eqspha}
\xi^2\frac{d^2
f}{d\xi^2} & = & 2f(1-f)(1-2f)-\frac{\xi^2}{4}h^2(1-f) \\
\frac{d}{d\xi}\left(\xi^2\frac{dh}{d\xi}\right) & = &
2h(1-f)^2+\frac{\lambda}{g^2} \xi^2(h^2-1)h \nonumber
\end{eqnarray}
with the boundary conditions, $f(0)=h(0)=0$ and
$f(\infty)=h(\infty)=1$. The energy functional becomes then,
\begin{eqnarray}
\label{energyspha}
E & = & \frac{4\pi v}{g}
\int^\infty_0\left\{4\left(\frac{df}{d\xi}\right)^2+
\frac{8}{\xi^2}[f(1-f)]^2
+\frac{1}{2}\xi^2\left(\frac{dh}{d\xi}\right)^2 \right.
\nonumber \\
& + & \left.
[h(1-f)]^2+\frac{1}{4}\left(\frac{\lambda}{g^2}\right)\xi^2(h^2-1)^2
\right\} d\xi  
\end{eqnarray}

The solution to Eqs.~(\ref{eqspha}) has to be found numerically.
The solutions depend on the gauge and quartic couplings, $g$ and
$\lambda$. Once replaced into the energy functional
(\ref{energyspha}) they give the sphaleron energy which is the
height of the barrier between different degenerate minima. It is
customary to write the solution as,
\begin{equation}
\label{energiaesfa}
E_{\rm sph}=\frac{2m_W}{\alpha_W}B(\lambda/g^2)
\end{equation}
where $B$ is the constant which requires numerical evaluation.
For the standard model with a single Higgs doublet this
parameter ranges from $B(0)=1.5$ to $B(\infty)=2.7$. A fit valid
for values of the Higgs mass
$
25\ GeV\leq m_h\leq 250\ GeV
$
can be written as,
\begin{equation}
\label{bfit}
B(x)=1.58+0.32x-0.05x^2
\end{equation}
where $x=m_h/m_W$.

The previous calculation of the sphaleron energy was performed
at zero temperature. The sphaleron at finite temperature was
computed in Ref.~\cite{Sph3,David}. Its energy follows the approximate
scaling law,
$E_{\rm sph}(T)=E_{\rm
sph}\langle\phi(T)\rangle/\langle\phi(0)\rangle $
which, using (\ref{energiaesfa}), can be written as,
\begin{equation}
\label{energiaesfaT}
E_{\rm sph}(T)=\frac{2m_W(T)}{\alpha_W}B(\lambda/g^2)
\end{equation}
where $m_W(T)=\frac{1}{2}g\langle\phi(T)\rangle$

\subsubsection{Baryon violation rate at $T>T_c$}

The calculation of the baryon violation rate at $T>T_c$, {\it
i.e.} in the symmetric phase, is very different from that in the
broken phase, that will be reviewed in the next section. In the
symmetric phase, at $\phi=0$, the perturbation theory is spoiled
by infrared divergences, and so we
cannot rely upon perturbative calculations to compute the baryon
violation rate in this phase. In fact, the infrared divergences
are cut off by the non-perturbative generation of a {\bf magnetic
mass}, 
$m_M\sim \alpha_W T$,
{\it i.e.} a {\bf magnetic screening length}
$\xi_M\sim(\alpha_WT)^{-1}$. The rate of baryon violation per
unit time and unit volume $\Gamma$ does not contain any
exponential Boltzmann factor~\footnote{It would disappear from
(\ref{probt}) in the limit $T\rightarrow\infty$.}. The
pre-exponential can be computed from dimensional grounds~\cite{ratesym} as
\begin{equation}
\label{anchurasim}
\Gamma=k (\alpha_WT)^4
\end{equation}
where the coefficient $k$ has been evaluated numerically in
Ref.~\cite{kappa} with the result~\footnote{In fact recent lattice 
computations~\cite{extra} seem to provide an extra factor in 
(\ref{anchurasim}) as $k' \alpha_W$ which is roughly speaking 
${\cal O}(1)$.} 
$
0.1\simlt k \simlt 1.0
$.

\subsubsection{Baryon violation rate at $T<T_c$}

After the phase transition, the calculation of baryon violation
rate can be done using the semiclassical approximations 
given by Eq.~(\ref{probt}). The rate per unit time
and unit volume for fluctuations between neighboring minima
contains a Boltzmann suppression factor $\exp\left(-E_{\rm sph}
(T)/T\right)$, where $E_{\rm sph}(T)$ is given by
(\ref{energiaesfaT}), and a pre-factor containing the
determinant of all zero and non-zero modes. 
The prefactor was computed in Ref.~\cite{CLMW} as
\begin{equation}
\Gamma\sim 2.8\times
10^5T^4\left(\frac{\alpha_W}{4\pi}\right)^4\kappa \frac{\zeta^7}{B^7}
e^{-\zeta} 
\label{anchuranos}
\end{equation}
where we have defined 
$\zeta(T)=E_{\rm sph}(T)/T$,
the coefficient $B$ is the function of $\lambda/g^2$ defined in
(\ref{bfit}) and $\kappa$  is the functional determinant
associated with fluctuations about the sphaleron. It has been
estimated~\cite{DHS} to be in the range,
$10^{-4}\simlt\kappa\simlt 10^{-1}$.

The equation describing the dilution $S$ of the baryon asymmetry in
the anomalous electroweak processes reads~\cite{VB}
\begin{equation}
\label{kinetic}
\frac{\partial S}{\partial t}=-V_B(t) S
\end{equation}
where $V_B(t)$ is the rate of the baryon number
non-conserving processes. Assuming $T$ is constant during the
phase transition the integration of (\ref{kinetic}) yields
$
S=e^{-X}
$
and
${\displaystyle X=\frac{13}{2}N_f\frac{\Gamma}{T^3}t }$.
Using now (\ref{anchuranos}) and (\ref{tTrel})
we can write the exponent $X$ as,
$X\sim 10^{10}\kappa \zeta^7 e^{-\zeta}$,
where we have taken the values of the parameters,
$B = 1.87 $,  $\alpha_W  =  0.0336 $, 
$N_f  =  3 $,
$T_c  \sim  10^2\ GeV$.
Imposing now the condition
$S\simgt 10^{-5}$,
or
$X\simlt 10$,
leads to the condition on $\zeta(T_c)$,
\begin{equation}
\label{condspha}
\zeta(T_c)\simgt 7\log\zeta(T_c)+9\log 10+\log\kappa
\end{equation}

Now, taking $\kappa$ at its upper bound, 
$\kappa=10^{-1}$, we obtain from (\ref{condspha}) the bound~\cite{boundsph} 
\begin{equation}
\label{boundsph}
\frac{E_{\rm sph}(T_c)}{T_c}\simgt 45,
\end{equation}
and using the lower bound, $\kappa=10^{-4}$
we obtain,
\begin{equation}
\label{boundsphmax}
\frac{E_{\rm sph}(T_c)}{T_c}\simgt 37,
\end{equation}
Eq.~(\ref{boundsph}) is the usual bound used to test
different theories while Eq.~(\ref{boundsphmax}) gives an idea on
how much can one move away from the bound (\ref{boundsph}), {\it
i.e.} the uncertainty on the bound (\ref{boundsph}).

The bounds (\ref{boundsph}) and (\ref{boundsphmax}) can be
translated into bounds on $\phi(T_c)/T_c$. Using the relation
(\ref{energiaesfaT})
we can write
\begin{equation}
\label{fiesfa}
\frac{\phi(T_c)}{T_c}=\frac{g}{4\pi B}\frac{E_{\rm
sph}(T_c)}{T_c}\sim\frac{1}{36}\frac{E_{\rm sph}(T_c)}{T_c} 
\end{equation}
where we have used the previous values 
of the parameters. The bound 
(\ref{boundsph})
translates into
\begin{equation}
\label{boundfi}
\frac{\phi(T_c)}{T_c}\simgt 1.3
\end{equation}
while the bound (\ref{boundsphmax}) translates into,
\begin{equation}
\label{boundfimax}
\frac{\phi(T_c)}{T_c}\simgt 1.0
\end{equation}

These bounds, Eqs.~(\ref{boundfi}) and (\ref{boundsph}), require
that the phase transition is strong enough first order. In fact
for a second order phase transition, $\phi(T_c)\simeq 0$ and any
previously generated baryon asymmetry would be washed out during
the phase transition. For the case of the Standard Model
the previous bounds translate into a bound on the
Higgs mass, as we will see.

\section{On the validity of the perturbative expansion}

The approach of Ref.~\cite{W} to the finite temperature
effective potential relied on the observation that {\bf
symmetry restoration implies that ordinary
perturbation theory must break down at high temperature}. In
fact, otherwise perturbation theory should hold and, since the
tree level potential is temperature independent, radiative
corrections (which are temperature dependent) should be unable
to restore the symmetry. We will see that the failure of
perturbative expansion is intimately linked to the appearance of
infrared divergences for the zero Matsubara modes of
bosonic degrees of freedom.
This just means that the usual
perturbative expansion in powers of the coupling constant fails
at temperatures beyond the critical temperature. It has to be
replaced by an improved perturbative expansion where an infinite
number of diagrams are resummed at each order in the new
expansion. We will review the actual situation in this section.

\subsection{The breakdown of perturbative expansion}

\begin{figure}[htb]
\epsfxsize=5truecm
\centerline{\epsfbox{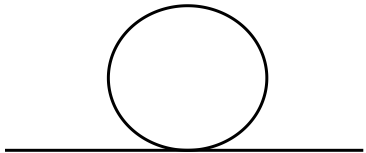}}
\caption[0]{One-loop contribution to the self-energy for the
scalar theory.}
\label{1daisy}
\end{figure}

\begin{figure}[htb]
\epsfxsize=5truecm
\centerline{\epsfbox{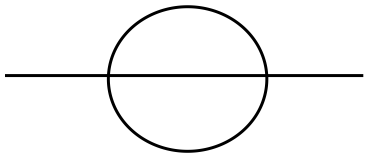}}
\caption[0]{Two-loop contribution to the self-energy for the
scalar theory.}
\label{2nondaisy}
\end{figure}

\begin{figure}[htb]
\epsfxsize=5truecm
\centerline{\epsfbox{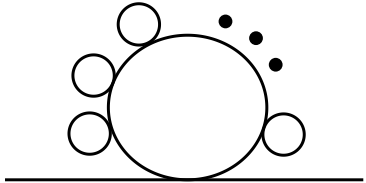}}
\caption[0]{Daisy (n+1)-loop contribution to the self-energy for the
scalar theory.}
\label{DaisyT}
\end{figure}

We will examine the simplest model of one self-interacting real
scalar field, described by a lagrangian with a squared-mass term,
$m^2$ and a  quartic coupling $\lambda$.
We will use now power counting arguments to
investigate the high temperature behaviour of higher loop
diagrams contributing to the effective potential~\cite{W,F,EQZ1}. 
After rescaling all loop momenta and energies
by $T$, a loop amplitude with superficial divergence $D$ takes
the form, $T^D f(m/T)$.
If there are no infrared divergences when $m/T\rightarrow 0$,
then the loop goes like $T^D$. For instance the diagram
contributing to the self-energy of Fig.~\ref{1daisy}
is quadratically divergent ($D=2$), and so behaves like
$\lambda T^2$.
For $D\le 0$, there are infrared divergences associated to the
zero modes of bosonic propagators in the imaginary time
formalism [$n=0$ in (\ref{feynmanim})] and the only $T$
dependence comes from the $T$ in front of the loop integral in
(\ref{feynmanim}). Then every logarithmically divergent or
convergent loop contributes a factor of $T$. For instance the
diagram contributing to the self-energy in Fig.~\ref{2nondaisy}
contains two logarithmically divergent loops and so behaves like,
$\lambda^2 T^2=\lambda(\lambda T^2)$.
It is clear that to a fixed order in the loop expansion the
largest graphs are those with the maximum number of
quadratically divergent loops. These diagrams are obtained from
the diagram in Fig.~\ref{1daisy} by adding $n$ quadratically
divergent loops on top of it, as shown in Fig.~\ref{DaisyT}. 

They behave as,
\begin{equation}
\label{ndaisyT}
\lambda^{n+1}\frac{T^{2n+1}}{M^{2n-1}}=\lambda^2\frac{T^3}{M}
\left(\frac{\lambda T^2}{M^2}\right)^{n-1}
\end{equation}
where $M$ is the mass scale of the theory, and has been
introduced to rescale the powers of the temperature~\footnote{In 
fact, the mass $M$ has a different meaning for the
improved and the unimproved theories, as we shall see. For the 
unimproved theory, $M$ is the mass in the shifted lagrangian,
$M^2=m^2(\phi)$, while in the improved theory, $M$ is given
by the thermal mass.}. As was
clear from Eq.~(\ref{ndaisyT}), adding a quadratically divergent
bubble to a propagator which is part of a logarithmically
divergent or finite loop amounts to multiplying the diagram by 
\begin{equation}
\label{primeralfa}
\alpha\equiv\lambda \frac{T^2}{M^2}
\end{equation}
This means that for the one-loop approximation to be valid it is
required that 
$$
\lambda \frac{T^2}{M^2}\ll 1
$$
along with the usual requirement for the ordinary perturbation
expansion 
$$
\lambda \ll 1
$$
However at the critical temperature we have that $T_c\sim
M/\sqrt{\lambda}$ [see Eqs.~(\ref{potsecond})-(\ref{curvature})]. 
Therefore {\bf at the
critical temperature the one-loop approximation is not valid}
and higher loop diagrams where multiple quadratically divergent
bubbles are inserted cannot be neglected. 
Daisy resummation~\cite{BOOKS}
consists precisely to resum all powers of $\alpha$ and
provides a theory where $m^2(\phi_c)\rightarrow 
m^2_{\rm eff}\equiv
m^2(\phi_c)+\Pi$, where $\Pi$ is the self-energy corresponding to the
one-loop resummed diagrams to leading order in powers of
the temperature $T$ (e.g. $\sim T^2$). This method was pursued
systematically by Parwani~\cite{P} and applied to the Standard
Model by Arnold and Espinosa~\cite{AE}.

What about the diagrams which are not considered in the {\bf
improved} expansion?
The two-loop diagram of
Fig.~\ref{2nondaisy} is suppressed with respect to the diagram
of Fig.~\ref{1daisy} by $\lambda$. 
\begin{figure}[htb]
\epsfxsize=5truecm
\centerline{\epsfbox{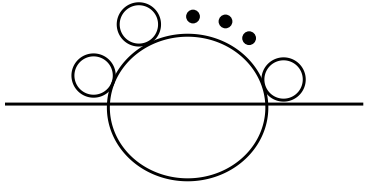}}
\caption[0]{Non-daisy (n+2)-loop contribution to 
the self-energy for the
scalar theory.}
\label{NondaisyT}
\end{figure}
On the other hand the
multiple loop diagram obtained from that of Fig.~\ref{2nondaisy}
by adding $n$ quadratically divergent loops on top of it,
see Fig.~\ref{NondaisyT},
 behaves as
\begin{equation}
\label{nnondaisyT}
\lambda^{n+2}\frac{T^{2n+2}}{M^{2n}}=\lambda^{n+1}\frac{T^{2n+1}}{M^{2n-1}}
\left(\lambda\frac{T}{M}\right)
\end{equation}
and it is suppressed with respect to the multiple loop diagram
of Eq.~(\ref{ndaisyT}) by $\lambda T/M$~\footnote{Non-daisy 
contributions to the self-energy are suppressed,
with respect to daisy contributions, by ${\cal O}(\beta$), where $\beta$
is defined in (\ref{impvalidity}). The corresponding contributions to the
vacuum diagrams ({\it i.e.} effective potential) are suppressed by
${\cal O}(\beta^2$)~\cite{EQZ1}.}. 
Therefore the validity
of the improved expansion is guaranteed provided that,
\begin{eqnarray}
\label{impvalidity}
\lambda\ll 1 \\
\beta\equiv\lambda\frac{T}{M}\ll 1 \nonumber
\end{eqnarray}

This simplified discussion can be done of course in more complicated
field theories, as e.g. the Standard Model, with similar arguments~\cite{AE}. 
We will now apply these results to two different models.

\subsection{The scalar theory}

In the case just considered of the scalar theory, the effective potential
can be written in the one-loop daisy resummed approximation as
\begin{equation}
V(\phi,T)=\frac{1}{2}m_{\rm eff}^2 \phi^2
-\frac{1}{12\pi}\left(m_{\rm eff}^2+\frac{1}{2}\lambda \phi^2\right)^{3/2}
+\frac{\lambda}{4!}\phi^4+\cdots
\label{potescalar}
\end{equation}
where $m_{\rm eff}^2\equiv m^2+\frac{\lambda}{24} T^2$.

The potential (\ref{potescalar}) apparently yields a first order
phase transition~\cite{CARRINGTON}. However,
the symmetry breaking minimum occurs when the three terms in the potential 
are similar. Then, at $T\sim T_c$, at the minimum
\begin{equation}
\phi\sim \sqrt{\lambda} T,\quad m_{\rm eff}^2 \sim \lambda^2 T^2
\end{equation}
and then,
\begin{equation}
\beta\sim \lambda \frac{T}{m_{\rm eff}}\sim \lambda \frac{T}{\lambda T}
={\cal O}(1)
\end{equation}
which shows that the result of perturbation theory is fake~\cite{AE}.

\subsection{The Standard Model}

In cases where there are gauge and/or Yukawa interactions, as the Standard
Model case, the situation is completely different. The effective potential
in the one-loop improved approximation, Eq.~(\ref{potsmtemp}) 
can be written as
\begin{equation}
V(\phi,T)=\frac{1}{2}m_{\rm eff}^2 \phi^2
-b g^3\phi^3 T
+\frac{\lambda}{4!}\phi^4+\cdots
\label{potsmodel}
\end{equation}
where 
$$
b g^3\phi^3 T\equiv \left[4 M_W^3(\phi)+2 M_Z^3(\phi)\right]\frac{T}{12\pi}
\equiv \left[\frac{1}{2}g^3+\left(g^2+g'^2\right)^{3/2}\right]\frac{T}{12\pi}
\phi^3
$$
and $m_{\rm eff}^2=m^2+a g^2 T^2$, $ag^2$ denoting the contribution of 
gauge and Yukawa couplings to the one loop self-energy. We have only 
considered the contribution of transverse gauge bosons to the phase
transition strength, and neglected that of the (screened) longitudinal
gauge bosons.

Now  again, symmetry breaking occurs when
all terms are similar,
\begin{equation}
\phi\sim \frac{g^3}{\lambda} T,\quad m_{\rm eff}^2\sim \frac{g^6}{\lambda}
T^2
\end{equation}
and the expansion parameter $\beta$,
\begin{equation}
\beta_{\rm SM}\sim g^2 \frac{T}{M_W(\phi)}\sim g\frac{T}{\phi}
\sim \frac{\lambda}{g^2}\sim \frac{m_H^2}{m_W^2}
\label{betasm}
\end{equation}

Therefore the  validity of perturbation theory implies that
$m_H\simlt m_W$. This behaviour is supported by non-perturbative
calculations~\cite{Jansen}.

\section{Summary of physical results for the phase transition and
constrains on the Higgs mass}
\subsection{Standard Model results}

The effective potential for the Standard Model was analyzed in
Eqs.~(\ref{potsmtemp}), (\ref{potsmodel}) in the one-loop approximation, 
including leading order plasma effects. In this approximation, the
longitudinal components of the gauge bosons are screened by
plasma effects while the transverse components remain
unscreened. In this way a good approximation to the effective
potential including these plasma effects is provided by 
Eq.~(\ref{potfirst}), where the coefficient $E$ is given by
\begin{equation}
\label{escreened}
E_{\rm SM}=\frac{2}{3}\frac{2m_W^3+m_Z^3}{4\pi v^3}\sim 9.5\times 10^{-3}
\end{equation}
Now we can use Eq.~(\ref{minimotc}) and $m_h^2=2\lambda v^2$ to
write, 
\begin{equation}
\label{fiH}
\frac{\phi(T_c)}{T_c}\sim \frac{4Ev^2}{m_h^2}
\end{equation}
In this way the bound (\ref{boundfi}) translates into the bound
on the Higgs mass,
\begin{equation}
\label{boundH}
m_h\simlt \sqrt{\frac{4E}{1.3}}\sim 42\ GeV.
\end{equation}

The bound (\ref{boundH}) is excluded by LEP measurements~\cite{PDG}, 
and so the Standard Model is unable to keep any
previously generated baryon asymmetry. 
Including two-loop effects the bound is slightly increased
to $\sim 45$ GeV~\cite{AE}. Is it possible, in
extensions of the Standard Model, to overcome this difficulty?
We will see in the next section one typical example where the
Standard Model is extended: the well motivated
supersymmetric extension of the Standard Model. 

\subsection{MSSM results}

Among the extensions of the Standard Model, the physically
most motivated and phenomenologically most acceptable one is the
Minimal
Supersymmetric Standard Model. This model allows for extra
CP-violating phases besides the Kobayashi-Maskawa one, which could
help
in generating the observed baryon asymmetry~\cite{cn}. It is then
interesting to study whether in the MSSM the nature of the phase
transition can be significantly modified with respect to the 
Standard Model.

In this section we
extend the considerations of the previous section to the full 
parameter space,
characterizing the Higgs sector of the MSSM~\cite{eqz3,beqz,previous}. 
The main tool for our study is the one-loop, daisy-improved
finite-temperature
effective potential of the MSSM, $V_{\rm{eff}}(\phi,T)$. We are
actually
interested in the dependence of the potential on $\phi_1 \equiv {\rm
Re}
\, H_1^0$ and $\phi_2 \equiv {\rm Re} \,  H_2^0$ only, where $H_1^0$
and $H_2^0$ are the neutral
components of the Higgs doublets $H_1$ and $H_2$, thus $\phi$ will
stand
for $(\phi_1,\phi_2)$. Working in the 't~Hooft-Landau gauge and in
the
$\overline{DR}$-scheme, we can write
\be
\label{total}
V_{\rm{eff}}(\phi,T) = V_0(\phi)
+ V_1(\phi,0) + \Delta V_1(\phi,T)
+\Delta V_{\rm{daisy}}(\phi,T)+V_2(\phi,T) \, ,
\ee
where
\bea
\label{v0}
V_0(\phi) & = &  m_1^2 \phi_1^2 + m_2^2  \phi_2^2 + 2 m_3^2 \phi_1
\phi_2
+ {g^2+g'\,^2 \over 8} (\phi_1^2 -\phi_2^2)^2 \, ,
\\
\label{deltav}
V_1(\phi,0) & = & \sum_i {n_i \over
64 \pi^2} m_i^4 (\phi) \left[ \log {m_i^2 (\phi) \over Q^2} - {3
\over 2} \right]  \, ,
\\
\label{deltavt}
\Delta V_1(\phi,T) & = & {T^4 \over 2 \pi^2} \left\{ \sum_i
n_i \, J_i \left[ { m^2_i (\phi) \over T^2 } \right]  \right\} \, ,
\\
\label{dvdaisy}
\Delta V_{\rm{daisy}}(\phi,T) & = & - {T \over 12 \pi} 
\sum_{i={\rm bosons}} n_i
\left[ {\cal M}_i^3 (\phi, T ) - m_i^3 (\phi) \right] \, .
\eea
The masses ${\cal M}_i^2 (\phi,T)$ are obtained
from the ${m}_i^2 (\phi)$ by adding the leading $T$-dependent
self-energy
contributions, which are proportional to $T^2$. We recall that, in
the gauge
boson sector, only the longitudinal components ($W_L, Z_L, \gamma_L$)
receive
such contributions.

The relevant degrees of freedom for our calculation are:
$n_t = - 12$, 
$n_{\st_1} = n_{\st_2} = 6$,  
$n_W=6$, $n_Z=3$, 
$n_{W_L}=2$, $ n_{Z_L}=n_{\gamma_L}=1$.
The field-dependent top mass is
$m_t^2(\phi)=h_t^2 \phi_2^2$.
The entries of the field-dependent stop mass matrix are
\bea
\label{tlmass}
m_{\widetilde{t}_L}^2 (\phi) & = & m_{Q_3}^2 + m_t^2 (\phi) +
D_{\widetilde{t}_L}^2 (\phi)  \, ,
\\
\label{trmass}
m_{\widetilde{t}_R}^2 (\phi) & = & m_{U_3}^2 + m_t^2 (\phi) +
D_{\widetilde{t}_R}^2 (\phi)  \, ,
\\
\label{mixmass}
m_X^2 (\phi) & = & h_t (A_t \phi_2 - \mu \phi_1) \, ,
\eea
where $m_{Q}$, $m_{U}$ and $A_t$ are soft supersymmetry-breaking
mass
parameters, $\mu$ is a superpotential Higgs mass term, and
\bea
\label{dterms}
D_{\widetilde{t}_L}^2(\phi)& = & \left( {1 \over 2}-
{2 \over 3}\sin^2 \theta_W \right)
{g^2 + g'\,^2 \over 2}(\phi_1^2-\phi_2^2),\\
D_{\widetilde{t}_R}^2(\phi) &=&\left( {2 \over 3}\sin^2 \theta_W \right)
{g^2 + g'\,^2 \over 2}(\phi_1^2-\phi_2^2)
\eea
are the $D$-term contributions. The field-dependent stop masses are
then
\be
\label{mstop}
m_{\widetilde{t}_{1,2}}^2 (\phi) = {m^2_{\widetilde{t}_L} (\phi) +
m^2_{\widetilde{t}_R} (\phi) \over 2} \pm \sqrt{ \left[
{m^2_{\widetilde{t}_L} (\phi)- m^2_{\widetilde{t}_R} (\phi) \over 2}
\right]^2 + \left[ m_X^2(\phi) \right]^2  } \, .
\ee
The corresponding effective $T$-dependent masses,
${\cal M}^2_{\widetilde{t}_{1,2}}
(\phi,T)$, are given by expressions identical to (\ref{mstop}), apart
from the
replacement
\be
\label{repl}
m^2_{\widetilde{t}_{L,R}} (\phi) \, \rightarrow \,
{\cal M}^2_{\widetilde{t}_{L,R}}(\phi,T) \equiv
m^2_{\widetilde{t}_{L,R}} (\phi)+  \Pi_{\widetilde{t}_{L,R}}(T)  \,  .
\ee
The $\Pi_{\widetilde{t}_{L,R}}(T)$ are the leading parts of the
$T$-dependent
self-energies of $\widetilde{t}_{L,R}$\,,
\bea
\label{pistl}
\Pi_{\widetilde{t}_L}(T)& = &
{4 \over 9}g_s^2 T^2 +
{1 \over 4}g^2 T^2 +
{1 \over 108}g'\,^2  T^2 +
{1 \over 6}h_t^2 T^2 \, ,
\\
\label{pistr}
\Pi_{\widetilde{t}_R}(T) & =  &
{4 \over 9}g_s^2 T^2 +
{4 \over 27} g'\,^2 T^2 +
{1 \over 3}h_t^2 T^2 \, ,
\eea
where $g_s$ is the strong gauge coupling constant. Only loops of
gauge
bosons, Higgs bosons and third generation squarks have been included,
implicitly assuming that all remaining supersymmetric particles are
heavy and decouple.  

We shall work in the limit in which the left handed
stop is heavy, $m_Q \simgt 500$ GeV. In this limit,
the supersymmetric corrections to the precision electroweak
parameter $\Delta\rho$ become small and hence, this allows 
a good fit to the electroweak precision data coming from
LEP and SLD. 
Lower values of $m_Q$ make the phase transition
stronger and we are hence taking a conservative assumption
from the point of view of 
defining the region consistent with electroweak baryogenesis. The 
left handed stop decouples at finite temperature, but, at zero
temperature, it sets the scale of the Higgs masses 
as a function of $\tan\beta$. For 
right-handed stop masses below, or of order
of, the top quark mass, and for large values of the CP-odd 
Higgs mass, $m_A \gg M_Z$,
the one-loop improved Higgs
effective potential admits a high
temperature expansion,
\be
\label{potMSSM}
V_0 +V_1 = -\frac{m^2(T)}{2} \phi^2 - T \;
\left[E_{\rm SM}\; \phi^3 +  (2 N_c) \frac{\left(m_{\widetilde{t}}^{2} +
\Pi_{\st_R}(T)\right)^{3/2}}
{12 \pi} \right] + \frac{\lambda(T)}{8} \phi^4+\cdots
\ee
where
$N_c = 3$ is the number of colours and 
$E_{\rm SM}$
is the cubic term coefficient in the Standard Model case.

Within our approximation, the lightest stop mass is approximately
given by
\begin{equation}
m_{\widetilde{t}}^2 \simeq m_U^2 + 0.15 M_Z^2 \cos2\beta + m_t^2 
\left(1- \frac{\widetilde{A}_t^2}{m_Q^2}\right)
\end{equation}
where $\widetilde{A}_t = A_t - \mu/\tan\beta$ is the stop mixing 
parameter.
As was observed in Ref.~\cite{CQW0}, the phase transition strength
is maximized for values of the soft breaking parameter 
$m_U^2 = - \Pi_{\st_R}(T)$, for which the coefficient of the cubic
term in the effective potential,  
\be
E \simeq E_{\rm SM}+ \frac{h_t^3 \sin^3\beta \left(1 -
\widetilde{A}_t^2/m_Q^2\right)^{3/2}}{4 \sqrt{2} \pi},
\label{totalE}
\ee
can be one order of magnitude larger than $E_{\rm SM}$~\cite{CQW0}.
In principle, the above would allow a sufficiently strong first order
phase transition for Higgs masses as large as 100 GeV.
However, it was also noticed that such large negative values of
$m_U^2$ may induce the presence of color breaking minima at
zero or finite temperature~\cite{CQW0,CarWag}. Demanding the absence
of such dangerous minima, the one loop analysis leads to 
an upper bound on the lightest CP-even
Higgs mass of order 80 GeV. This bound was obtained
for values of $\widetilde{m}_U^2 = - m_U^2$ of order (80 GeV$)^2$.

The most important two loop corrections are of the form
$\phi^2 \log(\phi)$ and, as said above, are induced by 
the Standard Model weak gauge bosons as well as by the stop and
gluon loops~\cite{JoseR}. 
It was recently noticed that the coefficient
of these terms can be efficiently obtained by the study of the
three dimensional running mass of the scalar top and Higgs fields
in the dimensionally reduced
theory at high temperatures~\cite{Schmidt}. Equivalently, in a four
dimensional computation of the MSSM Higgs effective potential
with a heavy left-handed stop, we obtain~\cite{CQW0}
\begin{eqnarray}
V_2&\simeq & {\displaystyle \log\frac{\Lambda_H}{\phi}  }\\ 
&& {\displaystyle 
\frac{\phi^2 T^2}{32 \pi^2}
\left[\frac{51}{16}g^2  - 3 \left[ h_t^2 \sin\beta^2 \left(1- 
\frac{\widetilde{A}_t^2}{m_Q^2}\right)\right]^2
+  8 g_s^2 h_t^2 \sin^2\beta \left(1- \frac{\widetilde{A}_t^2}{m_Q^2}\right)
\right]   }\nonumber
\end{eqnarray}
where the first term comes from the Standard Model
gauge boson-loop contributions, 
while the second and third terms come from the
light supersymmetric particle loop contributions.
The scale 
$\Lambda_H$ depends on the finite corrections, which may be
obtained by the expressions given in~\cite{CQW}.
As mentioned above, the two-loop  
corrections are very important and, as has been shown in 
Ref.~\cite{JoseR}, they can make the phase transition strongly first
order even for $m_U \simeq 0$~\cite{CQW}. Concerning the validity of
the perturbative expansion, the $\beta$-parameter, similarly to
$\beta_{\rm SM}$, (\ref{betasm}), can be shown to be given by
\begin{equation}
\beta_{\rm MSSM}\sim \frac{m_h^2}{m_t^2}\ ,
\end{equation} 
which leads to a reliable perturbative expansion for a value of the
Higgs mass enhanced, with respect to its Standard Model value, by a
factor $\sim (m_t/m_h)^2$.

An analogous situation occurs in the $U$-direction
$(U\equiv \widetilde{t}_R)$. The one-loop 
expression is approximately given by
\begin{equation}
V_{0}(U) + V_1(U,T)  = 
\left(-\widetilde{m}_U^2 + \gamma_U T^2 \right) U^2 -
T E_U U^3 + \frac{\lambda_U}{2} U^4,
\label{upot}
\end{equation}
where $\gamma_U$ and $E_U$ were given in~\cite{CQW}.

Analogous to the case of the field $\phi$, the two loop corrections
to the $U$-potential are dominated by gluon and stop loops and are
approximately given by
\begin{equation}
V_2(U,T) = \frac{U^2 T^2}{16\pi^2}\left[  \frac{100}{9} 
g_s^4 - 2 h_t^2 \sin^2\beta \left(1-\frac{\widetilde{A}_t^2}{m_Q^2}
\right) \right] \log\left(\frac{\Lambda_U}{U}\right)
\end{equation}
where, as in the Higgs case, the scale $\Lambda_U$ may only
be obtained after the finite corrections to the effective
potential are computed~\cite{CQW}.

Once the effective potential in the $\phi$ and $U$ directions are
computed, one can study the strength of the electroweak phase
transition, as well as the 
presence of potential color breaking
minima. 
At one-loop, it was observed that requiring the 
stability of the physical vacuum at zero temperature was enough
to assure the absolute stability of the potential at finite temperature.
As has been first noticed in Ref.~\cite{Schmidt},
once two loop corrections are included, the situation is more 
complicated~\cite{CQW}.

At zero temperature the minimization of the 
effective potential for the fields $\phi$ and $U$ shows
that the true minima are located for vanishing
values of one of the two fields. The two set of minima
are connected through a family of saddle points for
which both fields acquire non-vanishing  
values. Due to the nature of the high
temperature corrections, we do not expect a modification
of this conclusion at finite temperature. 

Two parameters control the presence of color breaking minima:
$\widetilde{m}_U^c$, defined as the smallest value of $\widetilde{m}_U$
for which a color breaking minimum deeper than the electroweak
breaking minimum is present 
at $T = 0$, and $T_c^U$, the critical
temperature for the transition into a color breaking minimum in the
$U$-direction. The value of $\widetilde{m}_U^c$ may be obtained by
analysing the effective potential for the field $U$ at zero temperature,
and it is approximately given by~\cite{CQW0}
\begin{equation}
\widetilde{m}_U^c \simeq
\left( \frac{m_H^2 \; v^2 \; g_s^2}{12} \right)^{1/4}.
\label{boundmu}
\end{equation}

Defining the critical temperature as that at which the potential
at the symmetry preserving and broken minima  are degenerate,
four situations can happen in the comparison of the critical
temperatures along the $\phi$ ($T_c$) and $U$ ($T_c^U$) transitions:
{\bf a)} $T_c^U < T_c$;   $\widetilde{m}_U < \widetilde{m}_U^c$; 
{\bf b)} $T_c^U < T_c$;   $\widetilde{m}_U > \widetilde{m}_U^c$; 
{\bf c)} $T_c^U > T_c$;   $\widetilde{m}_U < \widetilde{m}_U^c$; 
{\bf d)} $T_c^U > T_c$;   $\widetilde{m}_U > \widetilde{m}_U^c$. 

In case a), as the universe cools down, a phase transition into
a color preserving minimum occurs, which remains stable until
$T = 0$. This situation, of absolute 
stability of the physical vacuum, is the most
conservative requirement to obtain electroweak baryogenesis.
In case b), at $T = 0$ the color breaking minimum is deeper than
the physical one implying that
the color preserving minimum becomes unstable
for finite values of the temperature, with $T<T_c$. A physically
acceptable situation may only occur if the lifetime of the 
physical vacuum is smaller than the age of the universe. 
We shall denote this situation 
as \lq\lq metastability''.
In case c), as the universe cools down, a color breaking minimum
develops which, however, becomes metastable as the temperature
approaches zero. A physically acceptable situation can only
take place if a two step phase transition occurs, that is
if the color breaking minimum has a lifetime lower than the 
age of the universe at some temperature $T < T_c$~\cite{Schmidt}. 
Finally, in case d) the color breaking
minimum is absolutely stable and hence, the situation becomes
physically unacceptable.
\begin{figure}[htb]
\centerline{
\psfig{figure=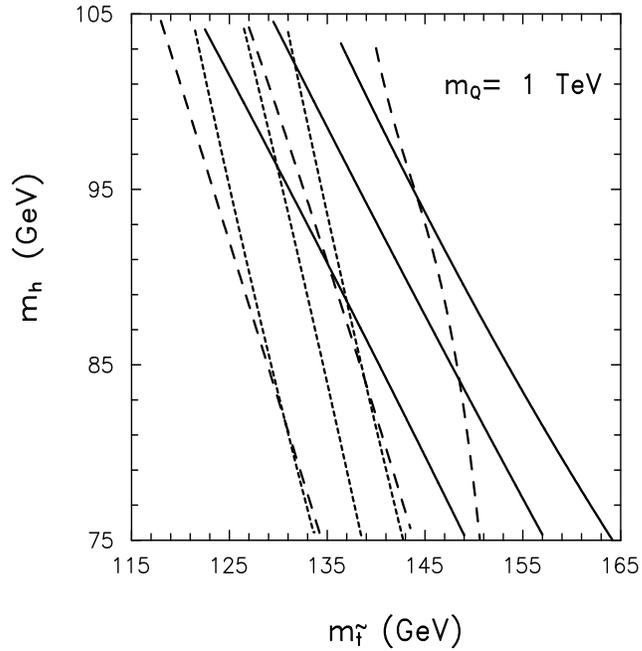,width=8.5cm,height=8.5cm,bbllx=6.cm,bblly=5.cm,bburx=20.cm,bbury=18cm}}
\caption{Values of $m_h$, $m_{\st}$ for which $v(T_c)/T_c = 1$ (solid
line), $T_c^U = T_c$ (dashed line), $\widetilde{m}_U = 
\widetilde{m}_U^c$ (short-dashed line), for $m_Q = 1$ TeV
and $\widetilde{A}_t=0,200,300$ GeV.
The region on the left of the solid line is consistent with a strongly
first order phase transition. A two step phase transition may occur in
the regions on the left of the
dashed line, while on the left of the short-dashed line, the physical
vacuum at $T = 0$
becomes metastable. The region on the left of both the dashed
and short-dashed lines leads to a stable color breaking vacuum state
at zero temperature and is hence physically unacceptable.}
\label{mexico}
\end{figure}

Figure~\ref{mexico} shows the region of parameter space consistent with
a sufficiently strong phase transition for 
$\widetilde{A}_t=0,200,300$ GeV. For low values of the
mixing, $\widetilde{A}_t \simlt 200$ GeV, 
case a) or c) may occur but, contrary to what
happens at one-loop, case b) is not realized. 
For the case of no mixing, this result
is in agreement with the analysis of~\cite{Schmidt}.
The region of absolute stability of the physical
vacuum for $\widetilde{A}_t \simeq 0$
is bounded to values of the Higgs mass of order
95 GeV. There is a small region at the right of the
solid line,
in which a two-step phase transition may
take place, for values of the parameters which
would lead to $v/T < 1$ for $T = T_c$, but may
evolve to larger values at some $T < T_c$ at which the 
second of the two
step phase transition into the physical vacuum takes place.
This region disappears
for larger values of the stop mixing mass parameter. 
For values of the mixing parameter $\widetilde{A}_t$ between
200 GeV and 300 GeV, both situations, cases b) and c) may
occur, depending on the value of $\tan\beta$.  
For large values of the stop mixing,
$\widetilde{A}_t > 300$ GeV,
a two-step phase transition does not take place. 

All together, and even demanding absolute stability
of the physical vacuum, electroweak baryogenesis seems to work
for a wide region of Higgs and stop mass values. Higgs masses
between the present experimental limit, of about 90 GeV, 
and around
105~GeV are consistent with this scenario. Similarly, the running
stop mass may vary from values of order 165 GeV (of the same
order as the top quark mass one) and 100 GeV. 
Observe that, due to the influence
of the D-terms, values of $m_{\widetilde{t}} \simeq 165$ GeV,
$\widetilde{A}_t \simeq 0$ and $m_h \simeq 75$ GeV, are achieved for
small positive values of $m_U$. Also observe that
for lower values of $m_Q$ the phase transition may become
more strongly first order and slightly larger values of the
stop masses may be obtained. Remind that these results are based on
the two-loop improved effective potential. Recent non-perturbative 
calculations~\cite{npmssm} confirm the validity of these
perturbative results.



\section*{References}
\begin{thebibliography}{99}

\bibitem{EHS} W. Heisenberg and H. Euler, {\em Z. Phys.} {\bf 98}
(1936) 714; J. Schwinger, \pr {\bf 82} (1951) 664.
%
\bibitem{GSWJL} J. Goldstone, A. Salam and S. Weinberg, \pr {\bf
127} (1962) 965; G. Jona-Lasinio, \nc {\bf 34} (1964) 1790.
%
\bibitem{CW} S. Coleman and E. Weinberg, \pr {\bf D7} (1973) 1888.
%
\bibitem{J} R. Jackiw, \pr {\bf D9} (1974) 1686.
%
\bibitem{IIM} J. Iliopoulos, C. Itzykson and A. Martin, \rmp
{\bf 47} (1975) 165.
%
\bibitem{BjD} J.D. Bjorken and S.D. Drell, {\em Relativistic Quantum
Mechanics} (McGraw-Hill, 1964); {\em Relativistic Quantum Fields}
(McGraw-Hill, 1965).
%
\bibitem{GR} I.S. Gradshteyn and I.M. Ryzhik, {\em Table of
Integrals Series and Products} (Academic Press, 1965).
%
\bibitem{DR} G. t'Hooft and M. Veltman, \np {\bf B44} (1972)
189; C.G. Bollini and J.J. Giambiagi, \pl {\bf B40} (1972) 366;
J.F. Ashmore, \ncl {\bf 4} (1972) 289. 
%
%
\bibitem{MS} G. t'Hooft and M. Veltman, \np {\bf B61} (1973)
455; W.A. Bardeen, A.J. Buras, D.W. Duke and T. Muta, \pr {\bf
D18} (1978) 3998.
%
\bibitem{COLL} See, {\it e.g.}: J. Collins, {\it Renormalization}
(Cambridge University Press, 1984).
%
\bibitem{DRED} W. Siegel, \pl {\bf B84} (1979) 193.
%
\bibitem{AH} G.W. Anderson and L.J. Hall, \pr {\bf D45} (1992)
2685. 
%
\bibitem{IEP} B. Kastening, \pl {\bf B283} (1992) 287; M. Bando,
T. Kugo, N. Maekawa and H. Nakano, \pl {\bf B301} (1993) 83; C.
Ford, D.R.T. Jones, P.W. Stephenson and M.B. Einhorn, \np {\bf
B395} (1993) 17.
%
\bibitem{CEQR} A. Casas, J.R. Espinosa, M. Quir\'os and A.
Riotto, \NPB{436}{95}{3}.

\bibitem{DJ} L. Dolan and R. Jackiw, \pr {\bf D9} (1974) 3320.
%
\bibitem{W} S. Weinberg, \pr {\bf D9} (1974) 3357.
%
\bibitem{B} R.H. Brandenberger, \rmp {\bf 57} (1985) 1.
%
\bibitem{LW} N.P. Landsman and Ch.G. van Weert, \prc {\bf 145}
(1987) 141.
%
\bibitem{HPA} M. Quir\'os, {\em Helv. Phys. Acta} {\bf 67}
(1994) 451.
%
\bibitem{Kapusta} J.I. Kapusta, {\em Finite-temperature field theory}
(Cambridge University Press, 1989).
%
\bibitem{KMS} R. Kubo, {\em J. Phys. Soc. Japan} {\bf 12} (1957)
570; P.C. Martin and J. Schwinger, \pr {\bf 115} (1959) 1342.
%
\bibitem{M} T. Matsubara, \ptp {\bf 14} (1955) 351.
%
\bibitem{MATSU} H. Matsumoto, Y. Nakano, H. Umezawa, F. Mancini
and M. Marinaro, \ptp {\bf 70} (1983) 599; H. Matsumoto, Y.
Nakano and H. Umezawa, \jmp {\bf 25} (1984) 3076.
%
\bibitem{KEL} L.V. Keldish, {\em Sov. Phys. JETP} {\bf 20}
(1964) 1018.
%
\bibitem{EA} T.S. Evans, \pr {\bf D47} (1993) R4196; T. Altherr,
CERN preprint, CERN-TH.6942/93.
%
\bibitem{RVI} R. Kobes, \pr {\bf D42} (1990) 562 and \prl {\bf
67} (1991) 1384; T.S. Evans, \pl {\bf B249} (1990) 286, \pl {\bf
B252} (1990) 108, \np {\bf B371} (1992) 340; P. Aurenche and T.
Becherrawy, \np {\bf B379} (1992) 259; M.A. van Eijck and Ch. G.
van Weert, \pl {\bf B278} (1992) 305.
%
\bibitem{K} D.A. Kirzhnits, {\it JETP Lett.} {\bf 15} (1972) 529.
%
\bibitem{KL} D.A. Kirzhnits and A.D. Linde, \pl {\bf 42B} (1972)
471; D.A. Kirzhnits and A.D. Linde, {\it JETP} {\bf 40} (1974) 628;
D.A. Kirzhnits and A.D. Linde, {\it Ann. Phys.} {\bf 101}
(1976) 195.
%
\bibitem{LINDE} A.D. Linde, {\it Rep. Prog. Phys.} {\bf 42}
(1979) 389; A.D. Linde, \pl {\bf 99B} (1981) 391; 
A.D. Linde, \pl {\bf 99B} (1981) 391; A.D. Linde, {\it Particle
Physics and Inflationary Cosmology} (Harwood, Chur, Switzerland,
1990).
%
\bibitem{INFLATION} A.D. Linde, \pl {\bf B108} (1982) 389;
A. Albrecht and P.J. Steinhardt, \prl {\bf 48} (1082) 1226.
%
%
\bibitem{DLHLL} M. Dine, R.G. Leigh, P. Huet, A. Linde and D.
Linde, \pl {\bf B283} (1992) 319; \pr {\bf D46} (1992) 550.
%
\bibitem{C} S. Coleman, \pr {\bf D15} (1977) 2929.
%
\bibitem{CC} C.G. Callan and S. Coleman, \pr {\bf D16} (1977)
1762.
%
\bibitem{CDL} S. Coleman and F. De Luccia, \pr {\bf D21} (1980)
3305. 
%
\bibitem{LTUN} A.D. Linde, \pl {\bf 70B} (1977) 306; {\bf 100B}
(1981) 37; \np {\bf B216} (1983) 421.
%
\bibitem{LSTV} L. McLerran, M. Shaposhnikov, N. Turok and M.
Voloshin, \pl {\bf B256} (1991) 451.
%
\bibitem{DHS} M. Dine, P. Huet and R. Singleton Jr., \np {\bf
B375} (1992) 625.
%
\bibitem{Marcos} J.M. Moreno, M. Quir{\'o}s and M. Seco, 
\NPB{526}{98}{489}.
%
\bibitem{KT} E.W. Kolb and M.S. Turner, {\it The Early Universe}
(Addison-Wesley, 1990).
%
\bibitem{SAKHA} A.D. Sakharov, {\em Zh. Eksp. Teor. Fiz. Pis'ma}
{\bf 5} (1967) 32; {\em JETP Lett.} {\bf 91B} (1967) 24.
%
\bibitem{PDG} Particle Data Group, Review of Particle
Properties, {\em Eur. Phys. J.} {\bf C3} (1998) 1.
%
\bibitem{KW} E.W. Kolb and S. Wolfram, \np {\bf B172} (1980)
224; \pl {\bf B91} (1980) 217.
%
\bibitem{KRS} V.A. Kuzmin, V.A. Rubakov and M.E. Shaposhnikov,
\pl {\bf B155} (1985) 36; \pl {\bf B191} (1987) 171.
%
\bibitem{AnnRev} A.G. Cohen, D.B. Kaplan and A.E. Nelson, {\em
Annu. Rev. Nucl. Part. Sci.} {\bf 43} (1993) 27.
%
\bibitem{Hooft} G. t' Hooft, \prl {\bf 37} (1976) 8; \pr {\bf
D14} (1976) 3432.
%
\bibitem{CKM} M. Kobayashi and M. Maskawa, \ptp {\bf 49} (1973) 652.
%
\bibitem{Sph} N.S. Manton, \pr {\bf D28} (1983) 2019; F.R.
Klinkhamer and N.S. Manton, \pr {\bf D30} (1984) 2212.
%
\bibitem{Sph2} J. Kunz, B. Kleihaus and Y. Brihaye, \pr {\bf
D46} (1992) 3587.
%
\bibitem{Sph3} Y. Brihaye and J. Kunz, \pr {\bf D48} (1993) 3884.
%
\bibitem{David} J.M. Moreno, D.H. Oaknin and M. Quir{\'o}s,
[hep-ph/9605387], \NPB{483}{97}{267}.
%
\bibitem{ratesym} P. Arnold and L. McLerran, \pr {\bf D36}
(1987) 581; S.Yu. Khlebnikov and M.E. Shaposhnikov, \np {\bf
B308} (1988) 885.
%
\bibitem{kappa} J. Ambjorn, M. Laursen and M. Shaposhnikov, \pl
{\bf B197} (1989)49; J. Ambjorn, T. Askaard, H. Porter and M.
Shaposhnikov, \pl {\bf B244} (1990) 479; \np {\bf B353} (1991)
346; J. Ambjorn and K. Farakos, \pl {\bf B294} (1992) 248;
J. Ambjorn and A. Krasnitz, \PLB{362}{95}{97}.
%
\bibitem{extra} P. Arnold, D. Son and L.G. Yaffe,\PRD{55}{97}{6264};
hep-ph/9810216; hep-ph/9810217;
G.D. Moore, \PRD{59}{99}{014503}; hep-ph/9810313; D. Bodeker,
\PLB{426}{98}{351}; hep-ph/9810265.
%
\bibitem{CLMW} L. Carson, Xu Li, L. McLerran and R.-T. Wang, \pr
{\bf D42} (1990) 2127. 
%
\bibitem{VB} M.E. Shaposhnikov, \np {\bf B287} (1987) 757; \np
{\bf B299} (1988) 797; A.I. Bochkarev and M.E. Shaposhnikov,
\mpl {\bf A2} (1987) 417.
%
\bibitem{boundsph} A.I. Bochkarev, S.V. Kuzmin and M.E.
Shaposhnikov, \pr {\bf D43} (1991) 369.
%
\bibitem{F} P. Fendley, \pl {\bf B196} (1987) 175.
%
\bibitem{EQZ1} J.R. Espinosa, M. Quir\'os and F. Zwirner, \pl
{\bf B291} (1992) 115.
%
\bibitem{BOOKS}
D.J.~Gross, R.D.~Pisarski and L.G.~Yaffe,  {\em Rev. Mod. Phys.} {\bf
53} (1981) 43. 
%
\bibitem{P}
R.R. Parwani, Phys. Rev. {\bf D45} (1992) 4695.
%
\bibitem{AE}
P.~Arnold and O.~Espinosa, \pr {\bf D47} (1993) 3546.
%
\bibitem{CARRINGTON}
M.E. Carrington, Phys. Rev. {\bf D45} (1992) 2933.
%
\bibitem{Jansen} K.~Jansen, {\it Nucl. Phys. (Proc. Supl.)} 
{\bf B47} (1996) 196  [hep-lat/9509018].
%
\bibitem{cn}
A.G.~Cohen and A.E.~Nelson, \pl {\bf B297 (1992)} 111.
%
\bibitem{eqz3}
J.R.~Espinosa, M.~Quir\'os and F.~Zwirner, \PLB{307}{93}{106}.
%
\bibitem{beqz} A. Brignole, J.R.~Espinosa, M.~Quir\'os and F.~Zwirner, 
\PLB{324}{94}{181}.
%
\bibitem{previous}
G.F.~Giudice, \pr {\bf D45} (1992) 3177;
S.~Myint, \pl {\bf B287} (1992) 325.
%
\bibitem{CQW0} M. Carena, M. Quiros and C.E.M. Wagner,
\PLB{380}{96}{81}
%
\bibitem{CarWag} M. Carena and C.E.M. Wagner, \NPB{452}{95}{45}
%
\bibitem{JoseR} J.R. Espinosa, \NPB{475}{96}{273};
B. de Carlos and J.R. Espinosa, [hep-ph/9703212], \NPB{503}{97}{24}.
%
\bibitem{Schmidt} D. Bodeker, P. John, M. Laine and M.G. Schmidt,
\NPB{497}{97}{387}
%
\bibitem{CQW} M. Carena, M. Quir{\'o}s and C.E.M. Wagner, [hep-ph/9710401]
{\em Nucl. Phys.} {\bf B524} (1998) 3.
%
\bibitem{npmssm} M. Laine and K. Rummukainen, \PRL{80}{98}{5259};
\NPB{535}{98}{423}.

\end{thebibliography}
\end{document}